\documentclass[11pt,letterpaper]{article}
\usepackage[letterpaper,margin=1in]{geometry}

\usepackage[T1]{fontenc}
\usepackage{amsmath,amssymb,amsthm}
\newtheorem{lemma}{Lemma}
\usepackage{graphicx}
\usepackage{subcaption}
\usepackage{url}
\usepackage[numbers,sort&compress]{natbib}

\def\defn#1{\textbf{\textit{\boldmath #1}}}

\usepackage{booktabs}
\usepackage{float}
\usepackage{tikz}
\usepackage{colortbl}
\counterwithin{figure}{section}
\counterwithin{table}{section}

\usetikzlibrary{shapes.geometric}
\usetikzlibrary{decorations.markings}
\newcommand{\wirey}[1]{\ifcase#1\or 7\or 6\or 5\or 4\or 1.7\or 0.7\or -0.3\or -1.3\fi}
\colorlet{faceR}{red!60!white}
\colorlet{faceO}{orange!90!white}
\colorlet{faceG}{green!55!white}
\colorlet{faceY}{yellow!80!white}
\colorlet{faceW}{white}
\colorlet{faceB}{blue!40!white}
\colorlet{networkFill}{violet!7}

\usepackage{xcolor}
\newcommand{\new}[1]{#1}

\definecolor{nnorange}{RGB}{235,150,45}
\definecolor{nnblue}{RGB}{40,70,200}
\newcommand{\nnorbit}[4]{%
  \begin{tikzpicture}[line join=round]
    \filldraw[fill=gray!18, draw=black, line width=0.8pt] (0,0) rectangle (2,2);
    \draw[gray!55, line width=0.5pt] (0.7,0)--(0.7,2);
    \draw[gray!55, line width=0.5pt] (1.3,0)--(1.3,2);
    \draw[gray!55, line width=0.5pt] (0,1.3)--(2,1.3);
    \draw[gray!55, line width=0.5pt] (0,0.7)--(2,0.7);
    \node at (0.7,2.32) {$1V$};
    \node at (1.3,2.32) {$2V$};
    \node at (-0.34,1.3) {$1H$};
    \node at (-0.34,0.7) {$2H$};
    \filldraw[fill=#1, draw=black, line width=0.5pt] (0.54,1.14) rectangle (0.86,1.46);
    \filldraw[fill=#2, draw=black, line width=0.5pt] (1.14,1.14) rectangle (1.46,1.46);
    \filldraw[fill=#3, draw=black, line width=0.5pt] (0.54,0.54) rectangle (0.86,0.86);
    \filldraw[fill=#4, draw=black, line width=0.5pt] (1.14,0.54) rectangle (1.46,0.86);
  \end{tikzpicture}%
}

\definecolor{nnyellow}{RGB}{255,232,20}
\definecolor{nnsideblue}{RGB}{30,80,220}
\definecolor{nnsidered}{RGB}{225,32,40}
\newcommand{\nnncubebody}[4]{%
  \foreach \i in {0,1,2,3}{
    \filldraw[fill=nnsideblue, draw=black!80, line width=0.8pt]
      ({0.82*(\i)},{0.47*(-\i)}) --
      ({0.82*(\i+1)},{0.47*(-\i-1)}) --
      ({0.82*(\i+1)},{0.47*(-\i-1)-0.945}) --
      ({0.82*(\i)},{0.47*(-\i)-0.945}) -- cycle;
  }
  \foreach \j in {0,1,2,3}{
    \filldraw[fill=nnsidered, draw=black!80, line width=0.8pt]
      ({0.82*(4+\j)},{0.47*(\j-4)}) --
      ({0.82*(4+\j+1)},{0.47*(\j+1-4)}) --
      ({0.82*(4+\j+1)},{0.47*(\j+1-4)-0.945}) --
      ({0.82*(4+\j)},{0.47*(\j-4)-0.945}) -- cycle;
  }
  \foreach \i in {0,1,2,3}{
    \foreach \j in {0,1,2,3}{
      \def\cellcol{white}%
      \ifnum\i=1\relax\ifnum\j=2\relax\def\cellcol{#1}\fi\ifnum\j=1\relax\def\cellcol{#3}\fi\fi
      \ifnum\i=2\relax\ifnum\j=2\relax\def\cellcol{#2}\fi\ifnum\j=1\relax\def\cellcol{#4}\fi\fi
      \filldraw[fill=\cellcol, draw=black!80, line width=0.8pt]
        ({0.82*(\i+\j)},{0.47*(\j-\i)}) --
        ({0.82*(\i+1+\j)},{0.47*(\j-\i-1)}) --
        ({0.82*(\i+\j+2)},{0.47*(\j-\i)}) --
        ({0.82*(\i+\j+1)},{0.47*(\j-\i+1)}) -- cycle;
    }
  }
}
\newcommand{\nnncube}[4]{%
  \begin{tikzpicture}[scale=0.5,line join=round]
    \nnncubebody{#1}{#2}{#3}{#4}
  \end{tikzpicture}%
}

\let\realbfseries=\bfseries
\def\bfseries{\realbfseries\boldmath}

\usepackage{hyperref}
\hypersetup{colorlinks=true, linkcolor=blue, citecolor=blue, urlcolor=blue}

\title{Solving Large Rubik's Cubes \\ within a Factor of $2.37+o(1)$ of Optimal}

\newlength{\authorrowwidth}
\newcommand{\authorblock}[3]{%
  \begin{tabular}[t]{@{}c@{}}
    \normalsize\textbf{#1}\footnotemark[#2]\\[2pt]
    \footnotesize\begin{tabular}[t]{@{}c@{}}#3\end{tabular}
  \end{tabular}}

\author{%
\begin{minipage}{\authorrowwidth}
\noindent
\authorblock{Jeremy Mrzyglocki}{1}
  {MIT CSAIL, Cambridge\\ Massachusetts, USA\\[1pt] TU Munich, Germany}\hfill
\authorblock{Tomas Rokicki}{2}
  {Palo Alto\\ California, USA}\hfill
\authorblock{Maxine Perroni-Scharf}{3}
  {MIT CSAIL, Cambridge\\ Massachusetts, USA}\hfill
\authorblock{Erik D. Demaine}{4}
  {MIT CSAIL, Cambridge\\ Massachusetts, USA}%
\end{minipage}%
}

\date{}

\begin{document}

\maketitle

\begingroup
\renewcommand{\thefootnote}{}
\footnotetext[1]{%
  \textsuperscript{*}\,\texttt{jeremy@mrzyglocki.com}\quad
  \textsuperscript{\dag}\,\texttt{rokicki@gmail.com}\quad
  \textsuperscript{\ddag}\,\texttt{max1@mit.edu}\quad
  \textsuperscript{\S}\,\texttt{edemaine@mit.edu}}
\endgroup
\setcounter{footnote}{0}

\begin{abstract}
We design optimized algorithms to solve the $n \times n \times n$ Rubik's Cube
within $30.5 \, n^2 / \lg n + o(n^2 / \lg n)$ moves.
This upper bound is asymptotically within a factor of $2.37$
of the known lower bound, which applies to almost all instances.
The constant $30.5$ is also the first explicit constant factor for the
known upper bound of $O(n^2 / \lg n)$, where the implicit constant of the previous algorithm was $1.01 \cdot 10^{40}$ (ten duodecillion).
Thus we improve the upper-bound constant by a factor of more than $3.30 \cdot 10^{38}$ (330 undecillion).
Our algorithm runs in time linear in the input size $N = \Theta(n^2)$,
and uses $O(\sqrt{N} \, \lg N)$ extra space
(beyond the state of the cube).
Our implementation is the first practical cube solver
to generate solutions of length $O(n^2/\lg n)$.
Our latest solver outperforms existing solvers for $n$ as small as $840$.

\end{abstract}

\clearpage

\section{Introduction}
\label{introduction}

The Rubik's Cube is perhaps the world's best-known puzzle,
which recently celebrated its 50th anniversary.
The Cube has also fascinated a wide range of research communities, including mathematicians, computer scientists, and engineers~\cite{khemani2018solving}.
In robotics, it serves as a benchmark for dexterous manipulation~\cite{OpenAIRubiks}.
In machine learning, it is a testbed for graph algorithms~\cite{mcaleer2018solving,agostinelli2019deepcubea,chervov2025machine}.
In complexity theory, Demaine, Eisenstat, and Rudoy~\cite{demaine2017solving} proved NP-completeness of solving an $n \times n \times n$ Rubik's Cube optimally from a given state.

A central question in cube theory is to determine the worst-case number of moves required to solve any configuration of the cube.
In other words (by symmetry), what is the diameter of the configuration space, where vertices represent configurations and edges correspond to valid moves?
For the $3 \times 3 \times 3$ cube, Rokicki, Kociemba, Davidson, and Dethridge~\cite{rokicki2014diameter} computed this number (sometimes called God's number) to be exactly 20 moves in the half-turn metric.
For the $n \times n \times n$ cube, at ESA 2011,
Demaine, Demaine, Eisenstat, Lubiw, and Winslow~\cite{DemaineEtAlRubik} showed that the number grows as $\Theta(n^2 / \lg n)$, a logarithmic factor fewer moves than standard algorithms ---
essentially solving $\Theta(\lg n)$ cubies instead of $\Theta(1)$ cubies within $\Theta(1)$ moves.

This asymptotic bound raises the natural theoretical question:
what is the correct leading constant?
And on the practical side, can the theoretical logarithmic speedup lead to shorter solutions to $n \times n \times n$ cubes for moderately large~$n$?

\subsection{Our Results}

In this paper, we make theoretical and practical progress on these two questions by designing new algorithms for solving $n \times n \times n$ Rubik's Cubes that are both theoretically and practically efficient.
We start from the $O(n^2 / \lg n)$ algorithmic outline,
and optimize many parts by developing several new ideas,
including better, more extensive commutator move sequences, a better selection of solving stages, improved parallelization techniques, and dynamic programming for
medium~$n$.

On the theoretical side, we design an algorithm that solves any configuration
of an $n \times n \times n$ Rubik's Cube in at most
$30.5 \, n^2 / \lg n + o(n^2/\lg n)$ moves.
This is the first explicit constant for an $O(n^2 / \lg n)$ upper bound.
The previous $O(n^2 / \lg n)$ upper bound \cite{DemaineEtAlRubik} just aimed
for the right asymptotic growth, so it considered any constant fast enough ---
but working through the algorithm (see Table~\ref{old analysis}),
it uses the rather large constant $24! \approx 6.2 \cdot 10^{23}$
in two places, resulting in an overall upper-bound constant of
$10 \big(\frac{24!}{4!^6} - 1\big) \big(\frac{24!}{2} - 1\big) =
10{,}071{,}\allowbreak 957{,}\allowbreak 728{,}\allowbreak 521{,}\allowbreak 484{,}\allowbreak 625{,}\allowbreak 393{,}\allowbreak 791{,}\allowbreak 240{,}\allowbreak 345{,}\allowbreak 432{,}\allowbreak 100{,}010
\approx 1.01 \cdot 10^{40}$
(ten duodecillion).
Our upper-bound constant improves on this constant by a factor of more than $3.30 \cdot 10^{38}$
(330 undecillion).
%
By contrast, the previous lower-bound constant \cite{DemaineEtAlRubik}
is the quite reasonable $\frac{1}{4} \lg\frac{24!}{(4!)^6} \approx 12.88$.
Our new upper bound is thus within a factor of $30.5 / 12.88 < 2.37$ of the lower bound.

\begin{figure}[t]
  \centering
  \includegraphics[width=\textwidth]{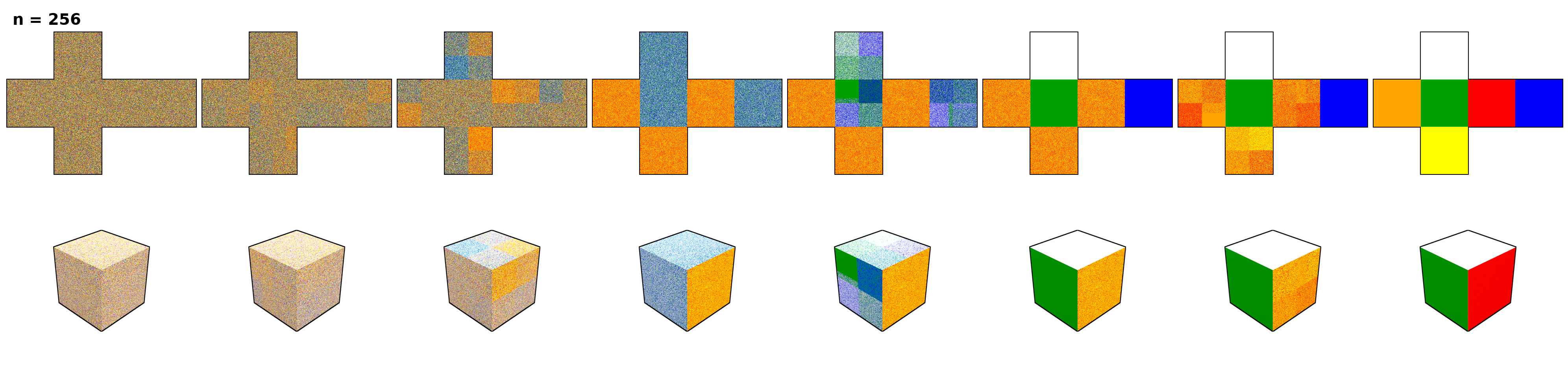}
  \caption{Eight evenly spaced frames of a render of an $n=256$ solve.} 
  \label{fig:video}
\end{figure}

On the practical side, we implement our algorithm in an open-source solver for large Rubik's Cubes.
Alongside the solver, we provide validation scripts that, for each permutation network underlying our solvers, exhaustively check every possible orbit configuration and confirm that the network solves it, giving an independently verifiable, complete proof of correctness for each solver. The solver and validation scripts are available as supplementary material.\footnote{Repository: \url{https://github.com/rokicki/brob-paper-data}.}
To the best of our knowledge, this implementation is the first with a practical worst-case guarantee matching that asymptotic bound.
Empirically, it also outperforms all previously described approaches for sufficiently large~$n$, roughly $n=840$.
We provide animations of the solving algorithm on an $n \times n \times n$ Rubik's Cube with $n=256$ and $n=65{,}536$; see Figure~\ref{fig:video}.

We also improve the best algorithms for the $n \times n \times 1$ Rubik's Cube.
This case is roughly two-dimensional, making it much easier to understand
figures and algorithms, so it serves as a helpful warmup
(as it did in \cite{DemaineEtAlRubik}).
Again the worst-case number of moves is $\Theta(n^2/\lg n)$,
and the constants in \cite{DemaineEtAlRubik} are implicit,
but working through the algorithm (see Table~\ref{old analysis nxnx1}),
it achieves a leading constant of $18$ as presented,
or $12$ if we just bound horizontal and vertical moves separately.
We reduce the upper-bound constant to $1.5$,
improving by a factor of either $12$ or $8$.
By contrast, the previous lower-bound constant \cite{DemaineEtAlRubik}
works out to $\frac{1}{4} \lg 6 \approx 0.646$.
Our new upper bound is thus within a factor of $1.5 / 0.646 < 2.322$ of the lower bound.

On this journey to determine the leading constant for $n \times n \times n$ Rubik's
Cubes, we encountered many entertaining problems and objects, including sorting
networks, permutation networks, a number of interesting combinatorial optimization
problems which we tackled with beam search, dynamic-programming optimization of
discrete strip widths, symbolic evaluation of asymptotes of algebraic expressions, and
plain old data-structure design and implementation.

\begin{table}
  \def\header#1{\textbf{#1}}
  \begin{minipage}{\textwidth}
  \small
  \setlength{\tabcolsep}{4pt}
  \begin{tabular}{lll}
    \header{Step} & \header{Action achieved} & \header{Move upper bound} \\
    \hline
    Lemma 8 & any permutation from Table 1 & 10 moves \\
    Lemma 9 & any permutation in $A_{24}$ & $|A_{24}|-1 = \frac{24!}{2}-1$ permutations from Table 1%
    \footnote{Lemma 9 \label{gap}%
      states that the permutations from Table 1 generate
      $A_{24}$, via a computation with GAP~\cite{GAP4},
      which uses the Schreier--Sims algorithm to compute a stabilizer chain
      and strong generating set, then computes the group order
      from the orbit sizes of successive stabilizers.
      But this computation gives no upper bound on the number of generators required.
      The obvious bound is that a shortest generator sequence corresponds
      to a shortest path in the Cayley graph of $A_{24}$, which consists of
      at most $|A_{24}| - 1$ steps because shortest paths do not repeat
      vertices.}
     \\
    Lemma 9 & any permutation in $A_{24}$ & $10 \big(\frac{24!}{2}-1\big)$ moves \\
    Lemma 10 & solve one config patterned $X \times Y$ & $10 \big(\frac{24!}{2} - 1\big) (X + Y + 1)$ moves \\
    Lemma 11 & solve one config within $X \times Y$ & $10 \big(\frac{24!}{2} - 1\big) \big((X + 1) 2^X + Y\big)$ moves \\
    Lemma 12 & solve one config within $\frac12 \lg Y \times Y$ & $10 \big(\frac{24!}{2} - 1\big) \Big(\big(\frac12 \lg Y + 1\big) 2^{\frac12 \lg Y} + Y\Big)$ \\&& $= 10 \big(\frac{24!}{2} - 1\big) Y + o(Y)$ moves \\
    Lemma 12 & solve one config within $X \times Y$ & $10 \big(\frac{24!}{2} - 1\big) X Y / \frac12 \lg Y + o(X Y / \lg Y)$ \\&& $= 20 \big(\frac{24!}{2} - 1\big) X Y / \lg Y + o(X Y / \lg Y)$ moves \\
    Theorem 3 & solve one config in $\sqrt{\frac{n}{2}} \times \sqrt{\frac{n}{2}}$ & $20 \big(\frac{24!}{2} - 1\big) \frac{n}{2} / \lg \sqrt{\frac{n}{2}} + o(n / \lg n)$ \\&& $= 20 \big(\frac{24!}{2} - 1\big) n / \lg n + o(n / \lg n)$ moves \\
    Theorem 3 & solve one config in $n \times n$ & $\big(\sqrt{\frac{n}{2}}\big)^2 \cdot 20 \big(\frac{24!}{2} - 1\big) n / \lg n + o(n^2 / \lg n)$ \\&& $= 10 \big(\frac{24!}{2} - 1\big) n^2 / \lg n + o(n^2 / \lg n)$ moves \\
    Theorem 3 & solves all configs in $n \times n$ & $10 \big(\frac{24!}{4!^6} - 1\big) \big(\frac{24!}{2} - 1\big) n^2 / \lg n + o(n^2 / \lg n)$ moves%
    \footnote{The number of sticker configurations for an orbit is $24! / 4!^6$. If we distinguish cubies as in the supercube, we would instead use $24! / 2$. One configuration is already solved, so needs no moves, hence the $-1$.}
    \\ \hline
  \end{tabular}
  \caption{Derivation of the implicit leading constant in the $O(n^2 / \lg n)$
    upper bound for solving the $n \times n \times n$ Rubik's Cube
    from \cite{DemaineEtAlRubik}:
    $10 \big(\frac{24!}{4!^6} - 1\big) \big(\frac{24!}{2} - 1\big) =
    10{,}071{,}\allowbreak 957{,}\allowbreak 728{,}\allowbreak 521{,}\allowbreak 484{,}\allowbreak 625{,}\allowbreak 393{,}\allowbreak 791{,}\allowbreak 240{,}\allowbreak 345{,}\allowbreak 432{,}\allowbreak 100{,}010
    \approx 1.00720 \cdot 10^{40}$.
    Lemma, theorem, and table numbers all refer to the arXiv version of \cite{DemaineEtAlRubik}.}
  \label{old analysis}
  \end{minipage}
\end{table}

\begin{table}
  \def\header#1{\textbf{#1}}
  \setlength{\tabcolsep}{2pt}
  \begin{minipage}{\textwidth}
  \begin{tabular}{llll}
    \header{Step} & \header{Pattern solved} & \header{Move upper bound} & \header{Decoupled upper bound} \\
    \hline
    Lemma 2 & 1 config & 6 & 4 horizontal + 4 vertical \\
    Lemma 4 & 1 config in $X \times Y$ pattern & $6(X+Y)$ & $4(X+Y)$ \\
    Lemma 5 & 1 config within $X \times Y$ & $6 (X 2^X + Y)$ & $4 (X 2^X + Y)$ \\
    Lemma 6 & 1 config within $\frac12 \lg Y \times Y$ & $6 \big(\frac12 \sqrt{Y} \lg Y + Y\big)$ & $4 \big(\frac12 \sqrt{Y} \lg Y + Y\big)$ \\
    Lemma 6 & 1 config within $X \times Y$ & $6 \big(X / \frac12 \lg Y\big) \big(\frac12 \sqrt{Y} \lg Y + Y\big)$ & $4 \big(X / \frac12 \lg Y\big) \big(\frac12 \sqrt{Y} \lg Y + Y\big)$ \\
    && $= 12 X Y / \lg Y + o(X Y)$ & $= 8 X Y / \lg Y + o(X Y)$ \\
    Theorem 1 & 1 config in $n \times n$ & $12 \big(\frac{n}{2}\big)^2 / \lg n + o(n^2/\lg n)$ & $8 \big(\frac{n}{2}\big)^2 / \lg n + o(n^2/\lg n)$ \\
    && $= 3 n^2 / \lg n + o(n^2/\lg n)$ & $= 2 n^2 / \lg n + o(n^2/\lg n)$ \\
    Theorem 1 & all configs in $n \times n$ & $18 n^2 / \lg n + o(n^2/\lg n)$ & $12 n^2 / \lg n + o(n^2/\lg n)$
    \\ \hline
  \end{tabular}
  \caption{Derivation of the implicit leading constant in the $O(n^2 / \lg n)$
    upper bound for solving the $n \times n \times 1$ Rubik's Cube
    from \cite{DemaineEtAlRubik}:
    $18$ or $12$.
    The two move upper bounds come from slightly different analyses depending on how we approximate a sum in Lemma~4.
    Lemma and theorem numbers all refer to the arXiv version of \cite{DemaineEtAlRubik}.}
  \label{old analysis nxnx1}
  \end{minipage}
\end{table}

\subsection{Prior Work}
\label{Prior_work}

Several $\Theta(n^2)$-move solvers have been developed
for the $n \times n \times n$ Rubik's Cube, although many are released just as GitHub code and lack detailed analysis.
RCube~\cite{RCubeGitHub} is a commutator-based solver,
with video demonstrations for $n=32{,}768$ and $n=65{,}536$~\cite{RCubeYouTube32768,RCubeYouTube65536}.
It provides a useful baseline for move counts and solution-rendering speed.
Other block-building solvers~\cite{RubikNxNxNSolver,dwalton} achieve low move counts for small $n$, but do not scale as well to large $n$.
These solvers are sequential, not obviously parallelizable, and explicitly move cubies, making them computationally slow; in particular, executing a single move already takes linear time. Our analysis shows that their worst-case move counts are $\Theta(n^2)$.

The $O(n^2/\lg n)$ upper bound of \cite{DemaineEtAlRubik} establishes that solutions exist but does not construct them.
Specifically, a computation with GAP
(described in Table~\ref{old analysis}, footnote \ref{gap})
showed that a certain set of 24 permutations generates $A_{24}$,
the alternating group which represents the reachable configurations
of one orbit of 24 center cubies.
This computation implies that any even permutation of a single orbit can in principle be expressed as a composition of these generators, but leaves open how to do so.
Because $A_{24}$ has the rather large size $24!/2 \approx 3.1 \cdot 10^{23}$,
it is difficult to brute-force for short solutions, even though it should
``just take constant time''.
Because of this gap, many in the community believed that an $O(n^2 / \lg n)$ algorithm would not be practical for reasonably small $n$.
Kaleem and Kaleem~\cite{onDemaineEtAlRubik} reduced the move count of specific 3-cycle subroutines (for example, from 10 moves to 8), but this work does not address the lack of a concrete overall upper bound.

\section{Common Definitions}
\label{commondefinitions}

An $n \times n \times n$ Rubik's Cube consists of
$6 n^2 - 12 n + 8$ \defn{cubies} ---
the mechanical pieces that move on the surface of the cube.
Corner, edge, and center cubies have 3, 2, and 1 colored stickers, respectively.
An \defn{orbit} (or ``cubie cluster'' in the terminology of~\cite{DemaineEtAlRubik}) consists of the up to 24 cubies whose positions are interchangeable by twisting moves.
(An orbit consists of fewer than 24 cubies at corners, middle edges, and the very central center when $n$ is odd.)
Cubies always remain within their orbit.

\begin{figure}[t]
    \centering
    \begin{tikzpicture}[x=0.85cm,y=0.85cm,line join=round]
      \foreach \ox/\oy/\fc/\lTL/\iTL/\lTR/\iTR/\lBR/\iBR/\lBL/\iBL in {
        2/2/faceW/A/0/B/1/C/2/D/3,
        0/0/faceO/E/4/F/5/G/6/H/7,
        2/0/faceG/I/8/J/9/K/10/L/11,
        4/0/faceR/M/12/N/13/O/14/P/15,
        6/0/faceB/Q/16/R/17/S/18/T/19,
        2/-2/faceY/U/20/V/21/W/22/X/23}{
        \fill[\fc] (\ox,\oy) rectangle ({\ox+2},{\oy+2});
        \draw[black,line width=0.4pt] ({\ox+1},\oy) -- ({\ox+1},{\oy+2});
        \draw[black,line width=0.4pt] (\ox,{\oy+1}) -- ({\ox+2},{\oy+1});
        \draw[black,line width=0.9pt] (\ox,\oy) rectangle ({\ox+2},{\oy+2});
        \foreach \lx/\ly/\ltr/\idx in {
          0.5/1.5/\lTL/\iTL, 1.5/1.5/\lTR/\iTR,
          1.5/0.5/\lBR/\iBR, 0.5/0.5/\lBL/\iBL}{
          \node at ({\ox+\lx},{\oy+\ly+0.17}) {\ltr};
          \node[font=\scriptsize] at ({\ox+\lx},{\oy+\ly-0.24}) {\idx};
        }
      }
    \end{tikzpicture}
    \caption{The Speffz sticker indexing scheme usually used for blindfolded solving,
    shown as a cube net (faces $U$, $L$, $F$, $R$, $B$, $D$ in white, orange, green,
    red, blue, yellow). \new{Each quadrant shows its Speffz letter and, below it, the
    0-indexed position ($0$--$23$) used throughout this paper.}
    Letter scheme from \url{https://www.speedsolving.com/wiki/index.php/Speffz}.}
    \label{fig:Speffz}
\end{figure}
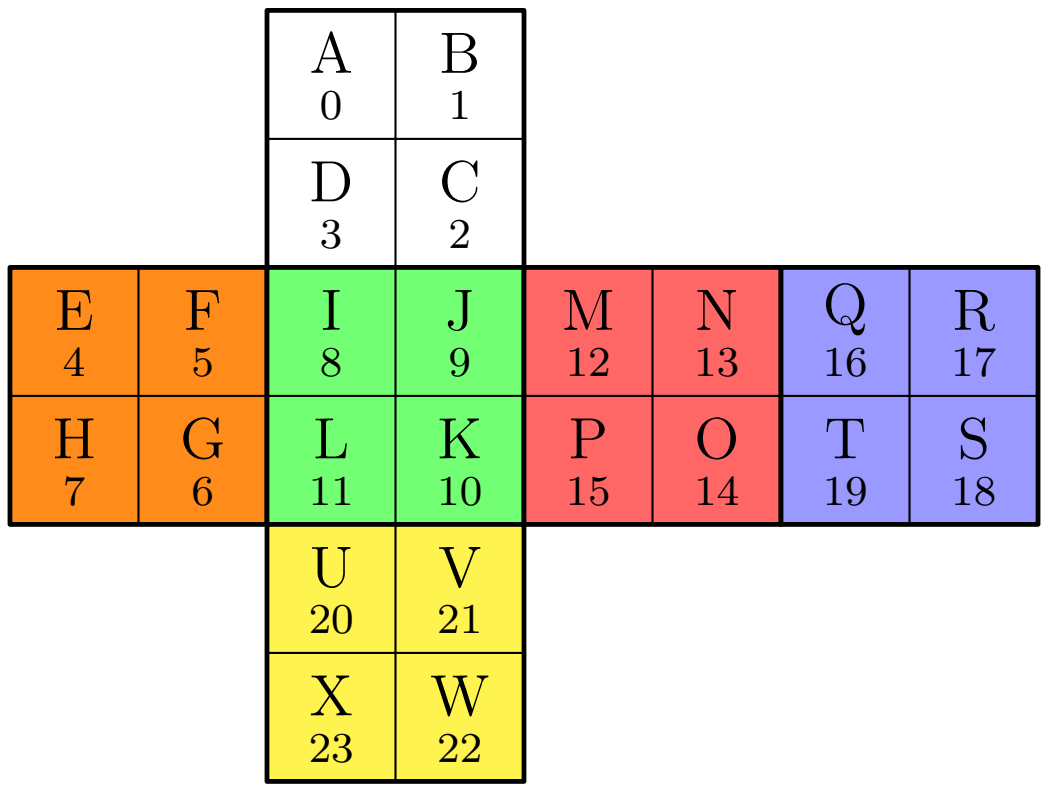

We use \defn{Speffz order} to index the fixed number of 24 orbit positions.
We start from the WCA (World Cube Association) standard orientation; the cube is held with the green face facing front and the white face facing up~\cite{speffz_wiki,WCApos}.
Speffz order iterates through all quadrants of the faces in the order U, L, F, R, B, and D, meaning Up, Left, Front, Right, Back, and Down
(Figure~\ref{fig:Speffz}).
The color order on those faces is white, orange, green, red, blue, yellow.

Our move metric is the \defn{Slice Turn Metric} (STM), where one move twists a single planar slice/layer by any multiple of 90 degrees. Because every move in this metric turns a single slice, we use the terms \emph{move} and \emph{slice move} interchangeably.
We denote such a move using \defn{Singmaster notation}~\cite{SingmasterNotation},
which consists of a sequence of up to three parts:
\begin{enumerate}
\item an optional numerical depth prefix between $1$ and $\lceil n/2 \rceil$,
      where $1$ (the default) denotes the outermost layer of the face and
      $\lceil n/2 \rceil$ denotes the innermost layer reachable from that face
      (the middle slice when $n$ is odd);
\item a letter U, L, F, R, B, or D indicating the Up, Left, Front, Right,
      Back, or Down face from which the slice is counted; and
\item an optional suffix indicating the degree of twist: no suffix denotes a
      $90^\circ$ clockwise twist, ``$'$'' denotes a $90^\circ$ counterclockwise
      twist, and ``$2$'' denotes a $180^\circ$ turn, as viewed from the
      perspective of the face being turned.
\end{enumerate}

An \defn{outer-turn move} is a move without a depth prefix, i.e., a twist of the outermost layer of a face. $X'$ denotes the inverse of a move sequence~$X$,
which is equivalent to reversing the order of the moves and inverting their rotations.
For example, $(3R' \; 4U \; 2L2)' = 2L2 \; 4U' \; 3R$. Every move sequence on the cube permutes pieces within their orbits. We use certain move sequences, called ``commutators'', that affect only a few local orbits and leave the remaining orbits unchanged.
Specifically, for two move sequences $A$ and $B$,
their \defn{commutator} is defined as $[A, B] = A\ B\ A'\ B'$,
whose inverse is $[A,\ B]' = [B,\ A] = B\ A\ B'\ A'$.

Throughout this paper, $\lg n$ denotes $\log_2 n$. For asymptotic analysis, we restrict attention to even $n$, because the additional cubies in odd cubes do not affect the leading term \cite{DemaineEtAlRubik}.
On the $n \times n \times n$ Rubik's Cubes we mostly use move sequences that displace only three cubies total, which we refer to as \defn{3-cycles}.

The methods for the $n \times n \times 1$ and $n \times n \times n$ cases follow the same outline. First we
search for short move sequences that permute only a few pieces within each orbit.
Then we construct a \defn{universal solving sequence}, or \defn{universal sequence} for short: a single sequence of such permutations,
of bounded length, from which an ordered selection (subsequence) solves any configuration
of an orbit. We lay the mathematical groundwork for this construction and relate the
length of the universal sequence to the leading constant. Finally, in the $n \times n \times n$ case, we examine several methods for reducing the number of permutations needed in the universal sequence.

\section{Calculating a Leading Constant for $n \times n \times 1$ Rubik's Cubes}
As a warmup, we consider algorithms for the $n \times n \times 1$ Rubik's Cube, explaining and analyzing the algorithm of \cite{DemaineEtAlRubik} and then improving its leading constant.

\begin{figure}[t]
    \centering
    \begin{tikzpicture}[scale=0.85,line join=round,font=\small]
      \nnncubebody{white}{white}{white}{white}
      \draw[orange!90!black, line width=1.6pt]
        ({0.82*(1+1)},{0.47*(1-1)}) --
        ({0.82*(3+1)},{0.47*(1-3)}) --
        ({0.82*(3+3)},{0.47*(3-3)}) --
        ({0.82*(1+3)},{0.47*(3-1)}) -- cycle;
      \draw[gray!70,line width=0.5pt] (4.92,1.41) -- (4.51,1.175);
      \draw[gray!70,line width=0.5pt] (5.74,0.94) -- (5.33,0.705);
      \node at (4.99,1.55) {$1V$};
      \node at (5.81,1.08) {$2V$};
      \draw[gray!70,line width=0.5pt] (1.23,1.645) -- (2.05,1.175);
      \draw[gray!70,line width=0.5pt] (0.41,1.175) -- (1.23,0.705);
      \node at (0.95,1.80) {$1H$};
      \node at (0.13,1.33) {$2H$};
      \node[orange!80!black] at (3.28,2.35) {\footnotesize one orbit (4 cubies)};
    \end{tikzpicture}
    \caption{A single center orbit on a physical $4 \times 4 \times 1$ cube. The four
    orbit cubies (outlined) sit at the intersections of the vertical slices $1V$, $2V$
    and the horizontal slices $1H$, $2H$, matching the labels $1V$, $2V$, $1H$, $2H$
    of the abstract diagram in Figure~\ref{fig:nn1}. The front (blue) and right (red)
    faces are shown only to convey the three-dimensional shape.}
    \label{fig:nn1-key}
\end{figure}
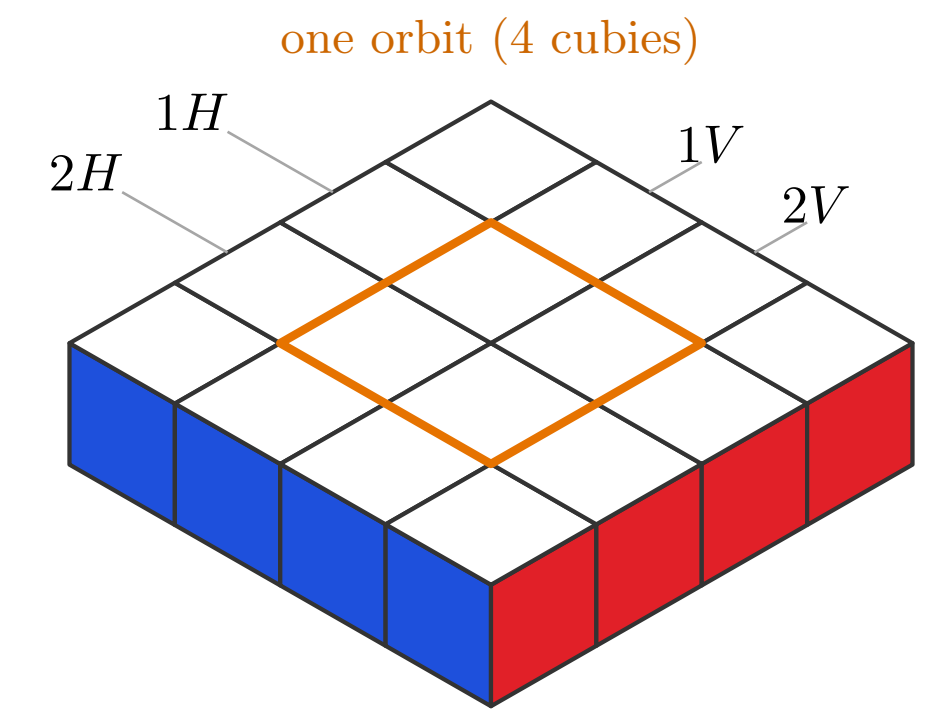

We define two vertical moves 1V and 2V, as well as two horizontal moves 1H and 2H, whose intersections define an orbit of an $n \times n \times 1$ Rubik's Cube. Each twist is $180^\circ$. The two vertical slices lie symmetrically on opposite sides of the cube, as do the two horizontal slices.
Demaine, Demaine, Eisenstat, Lubiw, and Winslow~\cite{DemaineEtAlRubik} show that, for the $n \times n \times 1$ Rubik's Cube, the center cubies can be partitioned into orbits of size~$4$. \new{For odd $n$, the cubies in the middle row and middle column (coordinate $(n+1)/2$) lie in degenerate orbits of size $1$ or $2$; we disregard them, because there are only $O(n)$ of them, so they do not contribute to the leading term of the move count.} Each of the size-$4$ orbits has six possible states, and any combination of states among the orbits is achievable through the use of commutators.
\begin{lemma}[Locality of commutators]
\label{lem:commutator-locality}
The commutator $[A,B] = A\ B\ A'\ B'$ moves only those cubies whose trajectory passes
through a slice of $A$ \emph{and} a slice of $B$; every other cubie is fixed. In
particular, for the center-solving commutators used in this paper the affected cubies lie within the
targeted center orbit---and, for parallelized commutators, within the union of the targeted
center orbits---so all other orbits, including the wing, edge and corner orbits, are left unchanged.
\end{lemma}
\begin{proof}
In $[A,B]$ each of $A$ and $B$ occurs twice, once in each direction. Consider a cubie
displaced by $B$ but whose positions before and after $B$ both lie outside the slices
turned by~$A$: then $A'$ does not displace it, and $B'$ returns it to its starting
position. The symmetric argument applies with the roles of $A$ and $B$ exchanged, and
cubies touched by neither $A$ nor $B$ are trivially fixed. Hence the net effect of
$[A,B]$ is confined to cubies whose trajectory passes through slices of both $A$
and~$B$. This observation is only a necessary condition for a cubie to move: the slice
intersections need not lie within the targeted center orbit. In $[3R,\ U\ 2L'\ U']$, for
example, the slices $3R$ and $U$ also meet at wing positions. That a commutator affects only
the targeted center orbit must therefore be checked directly. Tracking all $216$ stickers of a
$6\times6\times6$ cube, each of the six base commutators of Table~\ref{tab:3cycle-groups}
moves exactly three center stickers and fixes every other sticker, including all wings, edges
and corners; the parallelized examples of Figure~\ref{fig:parallelization-cases} likewise move
only center stickers of the targeted orbits. The permutation induced by each sequence we use for permnet62 and permnet61 is
listed in Appendices~\ref{app:permnet62-tables}--\ref{app:permnet61-tables} and verified by the
validation scripts of Section~\ref{introduction}.
\end{proof}
Figure~\ref{fig:nn1-key} illustrates a single orbit on a physical $4 \times 4 \times 1$ cube, which has exactly one center orbit, and shows how the moves $1V$, $2V$, $1H$, and $2H$ intersect at its four cubies.

\begin{figure}[t]
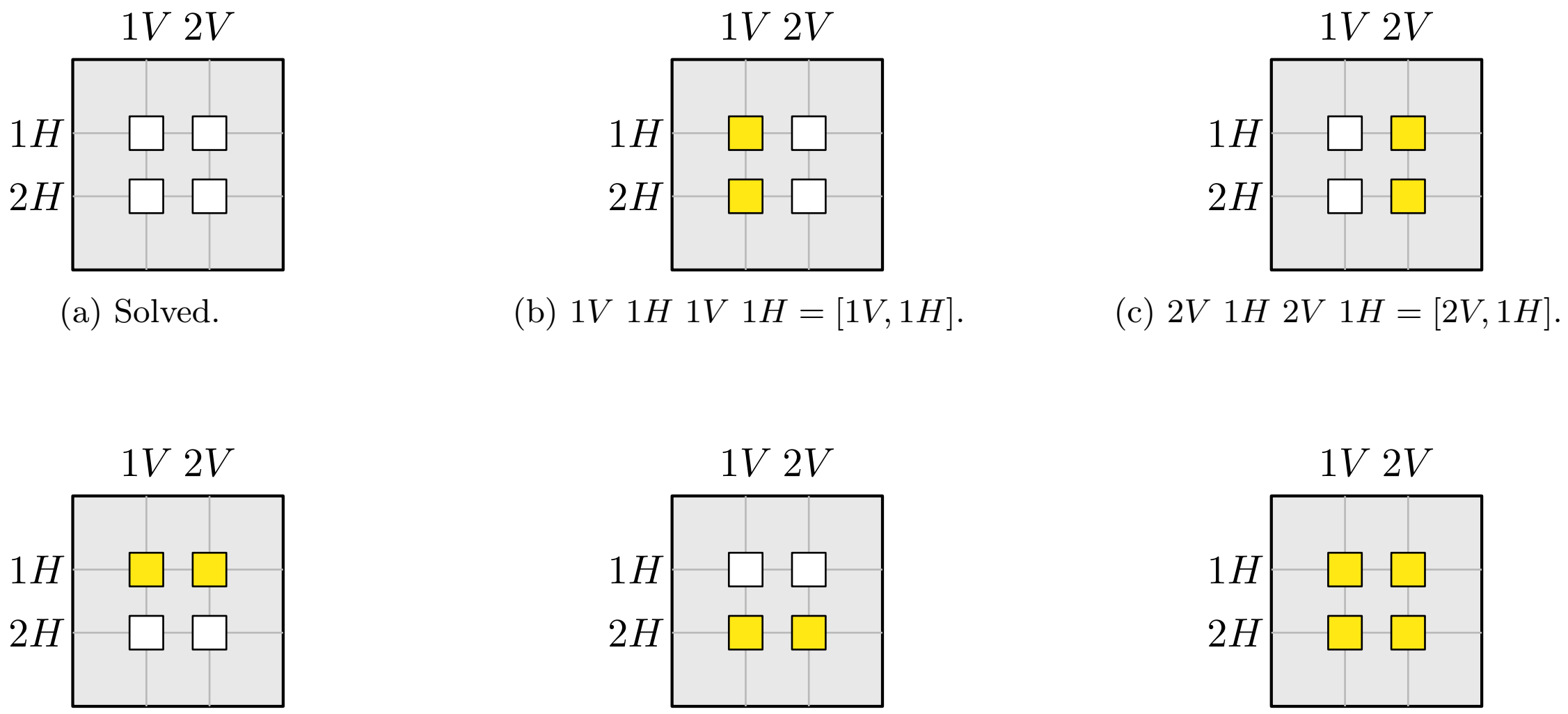

    \centering
    \begin{subfigure}[t]{0.31\textwidth}\centering
      \nnorbit{white}{white}{white}{white}
      \caption{Solved.\\\strut}
    \end{subfigure}\hfill
    \begin{subfigure}[t]{0.31\textwidth}\centering
      \nnorbit{nnyellow}{white}{nnyellow}{white}
      \caption{$1V\ 1H\ 1V\ 1H = \lbrack 1V, 1H\rbrack$.\\\strut}
    \end{subfigure}\hfill
    \begin{subfigure}[t]{0.31\textwidth}\centering
      \nnorbit{white}{nnyellow}{white}{nnyellow}
      \caption{$2V\ 1H\ 2V\ 1H = \lbrack 2V, 1H\rbrack$.\\\strut}
    \end{subfigure}

    \vspace{1.4em}

    \begin{subfigure}[t]{0.31\textwidth}\centering
      \nnorbit{nnyellow}{nnyellow}{white}{white}
      \caption{$1H\ 1V\ 1H\ 1V = \lbrack 1H, 1V\rbrack$.\\\strut}
    \end{subfigure}\hfill
    \begin{subfigure}[t]{0.31\textwidth}\centering
      \nnorbit{white}{white}{nnyellow}{nnyellow}
      \caption{$2H\ 1V\ 2H\ 1V = \lbrack 2H, 1V\rbrack$.\\\strut}
    \end{subfigure}\hfill
    \begin{subfigure}[t]{0.31\textwidth}\centering
      \nnorbit{nnyellow}{nnyellow}{nnyellow}{nnyellow}
      \caption{$1H\ 2H\ 1V\ 1H\ 2H\ 1V = \lbrack 1H\ 2H, 1V\rbrack$.}
    \end{subfigure}
    \caption{All possible states of an orbit of an $n \times n \times 1$ Rubik's Cube
    and their solutions. Adapted from~\cite[Figure 1]{DemaineEtAlRubik}. A white cell shows the top
    color ($0$); a yellow cell has been flipped to show its bottom color ($1$).}
    \label{fig:nn1}
\end{figure}

\begin{figure}[t]
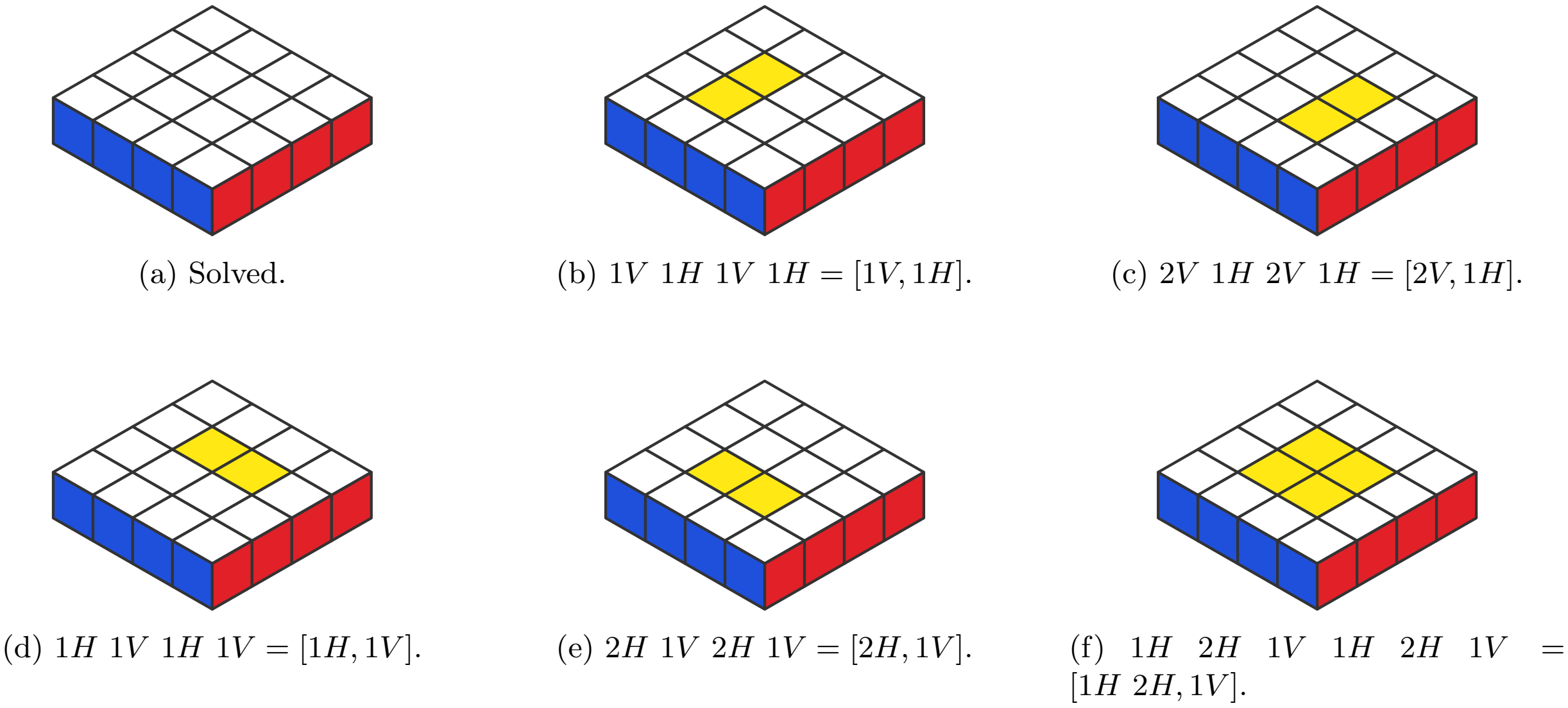

    \centering
    \begin{subfigure}[t]{0.31\textwidth}\centering
      \nnncube{white}{white}{white}{white}
      \caption{Solved.\\\strut}
    \end{subfigure}\hfill
    \begin{subfigure}[t]{0.31\textwidth}\centering
      \nnncube{nnyellow}{white}{nnyellow}{white}
      \caption{$1V\ 1H\ 1V\ 1H = \lbrack 1V, 1H\rbrack$.\\\strut}
    \end{subfigure}\hfill
    \begin{subfigure}[t]{0.31\textwidth}\centering
      \nnncube{white}{nnyellow}{white}{nnyellow}
      \caption{$2V\ 1H\ 2V\ 1H = \lbrack 2V, 1H\rbrack$.\\\strut}
    \end{subfigure}

    \vspace{1.2em}

    \begin{subfigure}[t]{0.31\textwidth}\centering
      \nnncube{nnyellow}{nnyellow}{white}{white}
      \caption{$1H\ 1V\ 1H\ 1V = \lbrack 1H, 1V\rbrack$.\\\strut}
    \end{subfigure}\hfill
    \begin{subfigure}[t]{0.31\textwidth}\centering
      \nnncube{white}{white}{nnyellow}{nnyellow}
      \caption{$2H\ 1V\ 2H\ 1V = \lbrack 2H, 1V\rbrack$.\\\strut}
    \end{subfigure}\hfill
    \begin{subfigure}[t]{0.31\textwidth}\centering
      \nnncube{nnyellow}{nnyellow}{nnyellow}{nnyellow}
      \caption{$1H\ 2H\ 1V\ 1H\ 2H\ 1V = \lbrack 1H\ 2H, 1V\rbrack$.}
    \end{subfigure}
    \caption{The six orbit states of Figure~\ref{fig:nn1} shown on a physical
    $4 \times 4 \times 1$ cube (one center orbit). A white center cubie shows the
    top color ($0$); a yellow one has been flipped to show its bottom color ($1$).
    The states and their solving move sequences correspond one-to-one to
    Figure~\ref{fig:nn1}.}
    \label{fig:nn1-3d}
\end{figure}

Then we expand these move sequences to simultaneously cover multiple orbits within rows of the puzzle in order to reduce the move count. Next, we extend the move sequences to two dimensions using combinatorial rectangles, yielding an algorithm with asymptotically better complexity. Evaluating the overall asymptotic complexity mathematically, we see that the number of stages of applying a single commutator is dominant. Finally, we focus on finding a short sequence of commutators that, when applied \emph{selectively}---each orbit performing only the subset of the sequence that its own state requires---suffices to solve all possible orbital states.
If we denote the colors of the cubies with the numbers $0$ and $1$, and list the visible colors on the top face in row-major order (see Figure~\ref{fig:nn1}), we can enumerate the states as follows:

\begin{center}
\begin{tabular}{lll}
\textbf{State} & \textbf{Solution commutator} & \textbf{case in Figure~\ref{fig:nn1}}\\
\hline
0000 & solved & (a) \\
1010 & $[1V, 1H]$ & (b)  \\
0101 & $[2V, 1H]$ & (c)  \\
1100 & $[1H, 1V]$ & (d)  \\
0011 & $[2H, 1V]$ & (e)  \\
1111 & $[1H\ 2H, 1V]$ & (f)  \\
\end{tabular}
\end{center}
States $(b)$--$(e)$ can be solved with a four-move commutator, while
the state $(f)$ requires a six-move commutator. Figure~\ref{fig:nn1-3d} shows the six states on a physical $4 \times 4 \times 1$ cube.
\new{There are only $O(n)$ non-center cubies, and each of
their orbits likewise has a bounded number of states and can be solved independently with
a constant number of moves, similarly to the centers. They therefore contribute only
lower-order terms to the total move count, and we ignore them from here on.} We assume
there are $n^2$ centers partitioned into $n^2/4$ orbits.

\subsection{0-Dimensional Approach}

One straightforward approach is to solve the puzzle orbit-by-orbit,
using the sequences above. In the worst case, every orbit is in state
$1111$, requiring $6$ moves per orbit, for a total of
\[
6 \cdot \frac{n^2}{4}.
\]
We refer to this strategy as a \defn{0-dimensional} (0D) approach, where
each orbit is solved independently.

\subsection{Row-by-Row (1D) Approach}

Another approach is to solve the puzzle row-by-row; this strategy is the \emph{1-dimensional} (1D) approach, in which orbits sharing a row (or column) are solved together (cf.\ Section~\ref{strip2d}).
For the first $n/2$ rows, we process orbits by state in five stages; because each orbit spans two rows symmetric about the center, processing these $n/2$ rows already reaches all orbits.
In each stage, we apply a commutator, but extend the vertical move
to all relevant columns simultaneously.
For example, if columns $3,4,8$ are in state $0011$, we apply
\[
[2H, 3V\ 4V\ 8V],
\]
solving all three orbits in one sequence. In the worst case, each row requires one commutator applied to $n/2$
columns, costing $(4+2+2+2+2) + 2\cdot\frac{n}{2} = 12 + n$ moves in the worst case. Over $n/2$ rows, the total worst-case move count is
\[
\frac{n}{2}(12 + n) = \frac{n^2}{2} + 6n.
\]
This row-by-row approach reduces the factor on the leading term from $1.5$ to $0.5$.

\subsection{Strip-Based (2D) Approach}
\label{strip2d}

\begin{figure}[t]
    \centering
    \includegraphics[width=0.6\textwidth]{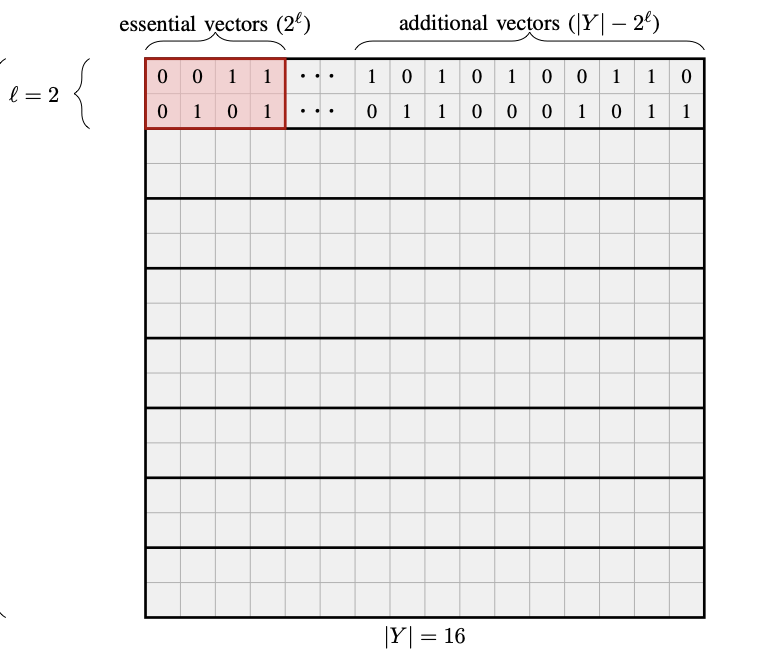}
    \caption{Partition of the grid of orbits into strips of width $w$ and
    essential/additional vectors, shown for a $16 \times 16$ grid and $w=2$.}
    \label{fig:ess}
\end{figure}

Following~\cite{DemaineEtAlRubik}, we can improve further by applying
commutators to combinatorial rectangles and by solving strips of width $w$,
where the \defn{width} $w$ is the number of rows a strip spans.
In the context of both commutator-based and 3-cycle-based solving algorithms (the latter on the $n \times n \times n$ Rubik's Cube, Section~\ref{algorithmicanalysis}), we use the terms \defn{0D parallelization}, \defn{1D parallelization}, and \defn{2D parallelization} to refer to solving only one orbit at a time, solving several orbits in rows or columns at a time, and solving several orbits in combinatorial rectangles across both rows and columns at a time, respectively.
We group orbits into combinatorial rectangles spanning multiple rows
and columns, where each column is labeled by five length-$w$ Boolean
vectors, one per state, indicating which of its orbits are in that state,
i.e., whether they need the commutator for that particular case applied
to them. Now we compare the labelings for each of the five stages.
Per vector, there are $2^w$ possible labelings.
Columns with the same labeling can be solved together using one commutator.
For states $(b)$--$(e)$, each commutator requires $2$ moves per column
and $2$ moves per active row.
For state $(f)$, the row cost becomes $4$ moves.
\new{Each stage spends $2$ moves on each of the $n/2$ columns of the strip,
so the five stages contribute $5 \cdot 2 \cdot \tfrac{n}{2} = 5n$ column
moves in total.
For the row moves, in the worst case every one of the $2^w$ labelings
occurs in each stage. We call the first occurrence of each distinct
labeling within a strip an \defn{essential vector}; every further (2nd,
3rd, \ldots) occurrence of the same labeling is an \defn{additional
vector}. An additional vector joins the combinatorial rectangle of its
essential vector and therefore costs only column moves, whereas each
essential vector additionally incurs the row moves of its rectangle
(Figure~\ref{fig:ess}). Because each of the $w$ bits is true in exactly half
of the $2^w$ labelings, the essential vectors of one stage activate
$w \cdot 2^{w-1}$ rows in total; summing the per-row cost over the five
stages ($2+2+2+2+4 = 12$ moves) gives}
$12 \cdot w \cdot 2^{w-1} = 6 w 2^w$ row moves.
Thus the cost per strip is at most
\[
5n + 6 w 2^w.
\]

With $n/(2w)$ strips, the worst-case total cost becomes
\[
\frac{(5n + 6 w 2^w)n}{2w}.
\]
\new{In~\cite{DemaineEtAlRubik}, the width is chosen as
$\tfrac{1}{2}\lg n$ (so that $2^w \approx \sqrt{n}$), which suffices for the
asymptotic bound $O(n^2/\lg n)$ but, in our accounting, doubles the
leading constant to $5$. Choosing instead}
\[
w = \lg n - 2 \lg \lg n
\]
\new{keeps the row-move term $6 w 2^w = O(n/\lg n) = o(n)$ while halving the
column-move term, giving a total cost of}
\[
2.5 \frac{n^2}{\lg n} + o\!\left(\frac{n^2}{\lg n}\right).
\]

\subsection{Improved Multi-State Approach}
\new{As established above, the total number of stages directly affects the
leading constant: each stage performs its own pass over the columns, so
five state-specific stages cost $5n$ column moves per strip. Our next
goal is therefore to reduce the number of stages. The key observation is
that a four-move commutator does not merely solve one state --- it
permutes all six states simultaneously. Hence, instead of dedicating one
stage to each state, we can chain commutators so that every state is
mapped to the solved state by some subsequence of the chain, allowing a
single pass to handle multiple states at once.} To this end, we consider all
eight four-move commutators:

\begin{center}
\small
\begin{tabular}{c|cccccccc}
 & $[1H,1V]$ & $[1H,2V]$ & $[2H,1V]$ & $[2H,2V]$ & $[1V,1H]$ & $[1V,2H]$ & $[2V,1H]$ & $[2V,2H]$ \\
\hline
0000 & 1010 & 0101 & 1010 & 0101 & 1100 & 0011 & 1100 & 0011 \\
0011 & 1111 & 1111 & 0000 & 0000 & 0101 & 1010 & 1010 & 0101 \\
0101 & 0011 & 1100 & 1100 & 0011 & 1111 & 1111 & 0000 & 0000 \\
1010 & 1100 & 0011 & 0011 & 1100 & 0000 & 0000 & 1111 & 1111 \\
1100 & 0000 & 0000 & 1111 & 1111 & 1010 & 0101 & 0101 & 1010 \\
1111 & 0101 & 1010 & 0101 & 1010 & 0011 & 1100 & 0011 & 1100 \\
\end{tabular}
\end{center}
Consider applying subsequences of
\[
[1H,1V]\ [1H,2V]\ [2V,1H]
\]
to the solved state $0000$.

\begin{center}
\begin{tabular}{ll}
\textbf{Subsequence} & \textbf{Final state} \\
\hline
-- & 0000 \\
$[1H,1V]$ & 1010 \\
$[1H,2V]$ & 0101 \\
$[2V,1H]$ & 1100 \\
$[1H,1V][1H,2V]$ & 0011 \\
$[1H,1V][2V,1H]$ & 1111 \\
$[1H,2V][2V,1H]$ & 0000 \\
$[1H,1V][1H,2V][2V,1H]$ & 1010 \\
\end{tabular}
\end{center}

\noindent The first six entries generate all states.
Reversing the order of the move sequences and inverting each one restores each of those states to solved. This reversal and inversion yield a three-stage algorithm:
\begin{itemize}
\item Stage 1: $[1H,2V]$
\item Stage 2: $[2V,1H]$
\item Stage 3: $[1V,1H]$
\end{itemize}
Each stage reduces the number of remaining states.
For one $n/2 \times w$ strip, the cost per stage is $n + w 2^w$.
Thus the total cost is bounded by:
\[
3(n + w 2^w)\cdot \frac{n}{2w}.
\]
Using the same optimized $w$, we obtain:
\[
1.5 \frac{n^2}{\lg n} + o\!\left(\frac{n^2}{\lg n}\right).
\]
By reducing the number of stages from $5$ to $3$, we reduce the
leading constant proportionally.

\section{$n\times n\times n$ Algorithm \& Mathematical Analysis}
\label{algorithmicanalysis}
The state space of an orbit is much larger for an \( n \times n \times n \) Rubik's Cube (\(6\) vs.\ \(\approx 3.25 \cdot 10^{15}\)), so we cannot perform a simple enumeration of states or construct a simple solution graph as we did for the \( n \times n \times 1 \) case. This larger state space makes the problem more challenging, and we need to develop different methods.

An even $n\times n\times n$ Rubik's Cube can be solved starting with the corners,
then the edges, and finally the center cubies.
Corners can be solved in at most $14$ moves (Section~\ref{midedges}); such constant bounds can be verified via exhaustive search with tools like twsearch~\cite{twsearch}.

\new{Our solving strategy places cubies using 3-cycles, which are even permutations; a sequence of 3-cycles can therefore only solve an orbit whose permutation is even. A single quarter-turn slice move applies a 4-cycle---an odd permutation---to an edge-cubie orbit, flipping that orbit's parity. Because there are $n/2-1$ edge-cubie orbits, all parities can be fixed with at most $n/2-1$ such moves, after which every remaining orbit permutation is even and can be solved by 3-cycles alone.} There are $O(n)$ edge cubies, so they contribute only $O(n)$ moves.
The number of center cubies scales as $O(n^2)$, and thus dominates the overall move count.
(See \cite{DemaineEtAlRubik} for details.)
Therefore the rest of the paper focuses only on centers.

\subsection{Deriving an $O(n^2)$ upper bound}
\label{upper_n2}

First we derive a function $F_{0\mathrm{D}}(n)$ for a low quadratic upper bound on move count. Demaine, Demaine, Eisenstat, Lubiw, and Winslow~\cite{DemaineEtAlRubik} show how 24 3-cycles can be constructed and that those cases are enough to solve any permutation in $A_{24}$ (cf.\ Section~\ref{Prior_work}). In the following, we use move sequences for all $4000$ usable center 3-cycles, which we obtain from six base commutators (Lemma~\ref{lem:threecycles}). They have several important properties that we introduce successively. One such property is that the maximal move count per 3-cycle is 12 moves.

\begin{table}[t]
  \centering
  \setlength{\tabcolsep}{4pt}
  \begin{tabular}{p{0.42\linewidth} p{0.25\linewidth} c}
    \toprule
    \textbf{Class} & \textbf{Representative commutator} & \textbf{Parallelization behavior} \\
    \midrule
    $2{+}1$ cubies where the two faces are adjacent and the two cubies on the same face are adjacent
      & \href{https://alpha.twizzle.net/explore/?puzzle=6x6x6&alg=%5B3R%2C+U+2L%27+U%27%5D}{[3R, U 2L' U']} & $+2/+2$ \\[3pt]

    $2{+}1$ cubies where the two faces are opposite and the two cubies on the same face are adjacent
      & \href{https://alpha.twizzle.net/explore/?puzzle=6x6x6&alg=%5B3R2%2C+U+2L2+U%27%5D}{[3R2, U 2L2 U']} & $+2/+2$ \\[3pt]

    $2{+}1$ cubies where the two faces are adjacent and the two cubies on the same face are diagonal
      & \href{https://alpha.twizzle.net/explore/?puzzle=6x6x6&alg=%5B3L+2U+3L%27%2C+U2%5D}{[3L 2U 3L', U2]} & $+2/+4$ \\[3pt]

    $2{+}1$ cubies where the two faces are opposite and the two cubies on the same face are diagonal
      & \href{https://alpha.twizzle.net/explore/?puzzle=6x6x6&alg=%5B3R+F2+2U2+F2+3R%27%2C+U2%5D}{[3R F2 2U2 F2 3R', U2]} & $+2/+4$ \\[3pt]

    $1{+}1{+}1$ cubies on three different faces sharing a corner (e.g.\ the $U$, $F$, and $R$ faces)
      & \href{https://alpha.twizzle.net/explore/?puzzle=6x6x6&alg=%5B2U%2C+B+3L+B%27%5D}{[2U, B 3L B']} & $+2/+2$ \\[3pt]

    $1{+}1{+}1$ cubies on three different faces including an opposite pair (e.g.\ the $U$, $F$, and $B$ faces)
      & \href{https://alpha.twizzle.net/explore/?puzzle=6x6x6&alg=%5B3R%2C+U+2L+U%27%5D}{[3R, U 2L U']} & $+2/+2$ \\
    \bottomrule
  \end{tabular}
  \caption{Equivalence classes of 3-cycles on a 24-cubie orbit, together with representative commutators and their parallelization behavior.}
  \label{tab:3cycle-groups}
\end{table}

\begin{lemma}[Realizing all 3-cycles]
\label{lem:threecycles}
Every one of the $4000$ usable center 3-cycles---that is, all $2\binom{24}{3} = 4048$
3-cycles of the $24$ orbit positions except the $48$ whose three cubies lie on a single
face---can be realized by a move sequence that fixes all other orbits. All $4000$ move
sequences are generated from the six base commutators of Table~\ref{tab:3cycle-groups} by
mirroring, rotating, inverting, and applying outer-turn setup moves.
\end{lemma}
\begin{proof}
The three cubies of a usable 3-cycle are distributed across the faces in one of six
ways---$2{+}1$ cubies with the two faces adjacent or opposite and the two same-face
cubies adjacent or diagonal (four classes), and $1{+}1{+}1$ cubies on three faces sharing
a corner or including an opposite pair (two classes)---which are exactly the rows of
Table~\ref{tab:3cycle-groups}. Table~\ref{tab:3cycle-groups} exhibits a representative
commutator for each class; by Lemma~\ref{lem:commutator-locality} each such commutator
affects only its targeted orbit. Applying the cube's rotational and mirror symmetries and
inversion to a representative, together with outer-turn setup moves, reaches every 3-cycle
in that class, so the six representatives generate all $4000$ cases. The $48$ excluded
single-face 3-cycles are not needed, as a single face's four center stickers share a
color.
\end{proof}

A center cubie orbit contains 24 cubies with 6 groups of 4 same-colored cubies each. The number of states of an orbit of cubie centers is $24!/(4!^6)$. We write $(a,b,c)$---or $\operatorname{rot}(a,b,c)$---for the 3-cycle that sends the sticker at position $a$ to position $b$, $b$ to $c$, and $c$ to $a$, where positions are indexed $0$--$23$ in Speffz order (Figure~\ref{fig:Speffz}).

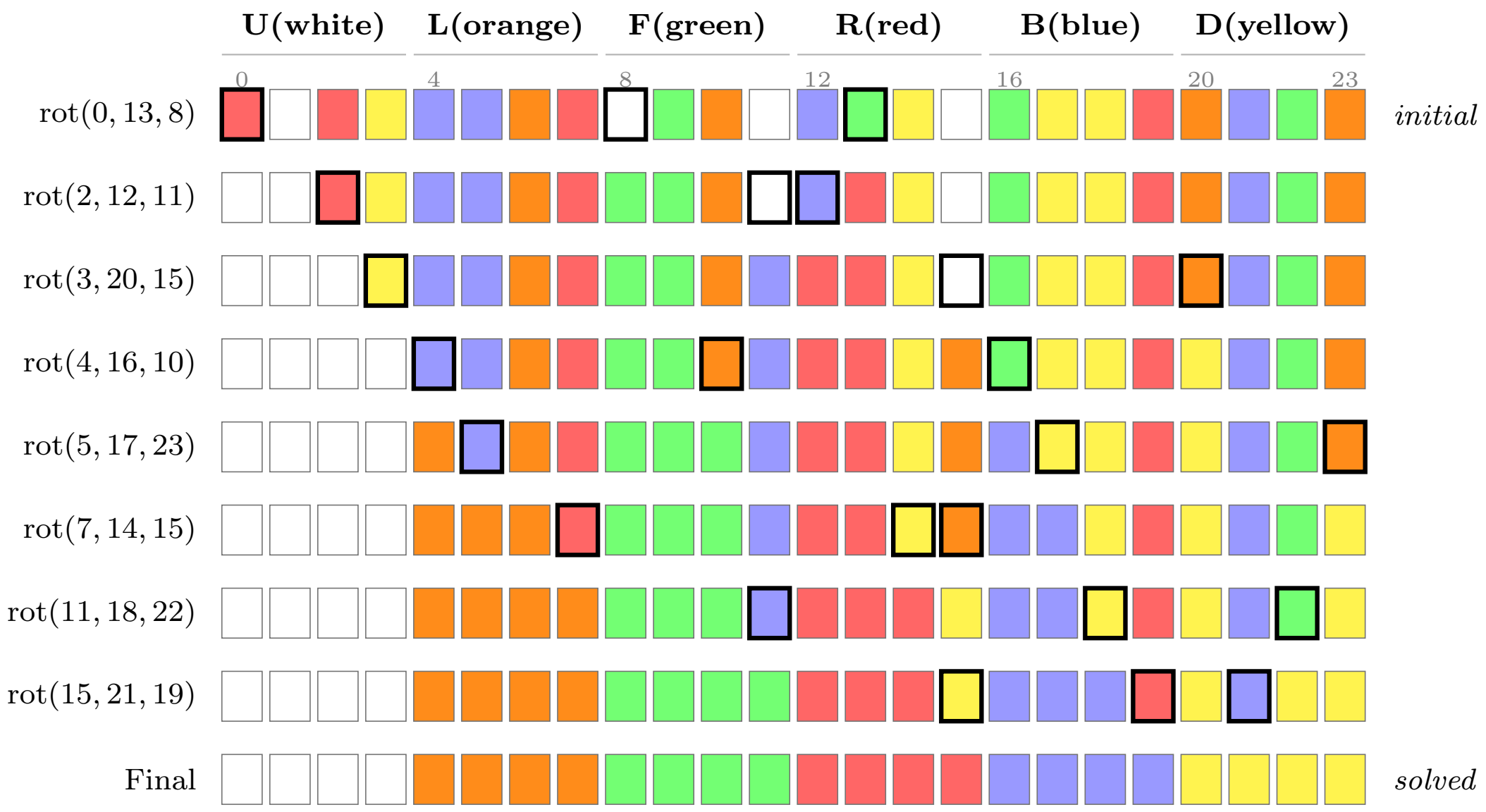
\begin{figure}[t]
  \centering
  \begin{tikzpicture}[x=0.45cm,y=0.78cm]
    \newcommand{\grow}[2]{%
      \foreach \cl [count=\c] in {#2} {
        \filldraw[fill=\cl, draw=black!55, line width=0.3pt]
          ({\c-0.42},{#1-0.3}) rectangle ({\c+0.42},{#1+0.3});
      }}
    \newcommand{\ghl}[2]{%
      \foreach \c in {#2} {
        \draw[black, line width=1.2pt]
          ({\c-0.42},{#1-0.3}) rectangle ({\c+0.42},{#1+0.3});
      }}
    \foreach \f/\lbl in {0/U(white),1/L(orange),2/F(green),3/R(red),4/B(blue),5/D(yellow)} {
      \node[font=\scriptsize\bfseries] at ({\f*4+2.5},1.05) {\lbl};
      \draw[gray!60, line width=0.4pt] ({\f*4+0.58},0.72) -- ({\f*4+4.42},0.72);
    }
    \foreach \c/\l in {1/0,5/4,9/8,13/12,17/16,21/20,24/23} \node[font=\tiny,gray] at (\c,0.42) {\l};
    \grow{0}{faceR,faceW,faceR,faceY,faceB,faceB,faceO,faceR,faceW,faceG,faceO,faceW,faceB,faceG,faceY,faceW,faceG,faceY,faceY,faceR,faceO,faceB,faceG,faceO}
    \ghl{0}{1,9,14}
    \grow{-1}{faceW,faceW,faceR,faceY,faceB,faceB,faceO,faceR,faceG,faceG,faceO,faceW,faceB,faceR,faceY,faceW,faceG,faceY,faceY,faceR,faceO,faceB,faceG,faceO}
    \ghl{-1}{3,12,13}
    \grow{-2}{faceW,faceW,faceW,faceY,faceB,faceB,faceO,faceR,faceG,faceG,faceO,faceB,faceR,faceR,faceY,faceW,faceG,faceY,faceY,faceR,faceO,faceB,faceG,faceO}
    \ghl{-2}{4,16,21}
    \grow{-3}{faceW,faceW,faceW,faceW,faceB,faceB,faceO,faceR,faceG,faceG,faceO,faceB,faceR,faceR,faceY,faceO,faceG,faceY,faceY,faceR,faceY,faceB,faceG,faceO}
    \ghl{-3}{5,11,17}
    \grow{-4}{faceW,faceW,faceW,faceW,faceO,faceB,faceO,faceR,faceG,faceG,faceG,faceB,faceR,faceR,faceY,faceO,faceB,faceY,faceY,faceR,faceY,faceB,faceG,faceO}
    \ghl{-4}{6,18,24}
    \grow{-5}{faceW,faceW,faceW,faceW,faceO,faceO,faceO,faceR,faceG,faceG,faceG,faceB,faceR,faceR,faceY,faceO,faceB,faceB,faceY,faceR,faceY,faceB,faceG,faceY}
    \ghl{-5}{8,15,16}
    \grow{-6}{faceW,faceW,faceW,faceW,faceO,faceO,faceO,faceO,faceG,faceG,faceG,faceB,faceR,faceR,faceR,faceY,faceB,faceB,faceY,faceR,faceY,faceB,faceG,faceY}
    \ghl{-6}{12,19,23}
    \grow{-7}{faceW,faceW,faceW,faceW,faceO,faceO,faceO,faceO,faceG,faceG,faceG,faceG,faceR,faceR,faceR,faceY,faceB,faceB,faceB,faceR,faceY,faceB,faceY,faceY}
    \ghl{-7}{16,20,22}
    \grow{-8}{faceW,faceW,faceW,faceW,faceO,faceO,faceO,faceO,faceG,faceG,faceG,faceG,faceR,faceR,faceR,faceR,faceB,faceB,faceB,faceB,faceY,faceY,faceY,faceY}
    \node[anchor=east,font=\scriptsize] at (0.35,0) {$\operatorname{rot}(0,13,8)$};
    \node[anchor=east,font=\scriptsize] at (0.35,-1) {$\operatorname{rot}(2,12,11)$};
    \node[anchor=east,font=\scriptsize] at (0.35,-2) {$\operatorname{rot}(3,20,15)$};
    \node[anchor=east,font=\scriptsize] at (0.35,-3) {$\operatorname{rot}(4,16,10)$};
    \node[anchor=east,font=\scriptsize] at (0.35,-4) {$\operatorname{rot}(5,17,23)$};
    \node[anchor=east,font=\scriptsize] at (0.35,-5) {$\operatorname{rot}(7,14,15)$};
    \node[anchor=east,font=\scriptsize] at (0.35,-6) {$\operatorname{rot}(11,18,22)$};
    \node[anchor=east,font=\scriptsize] at (0.35,-7) {$\operatorname{rot}(15,21,19)$};
    \node[anchor=east,font=\scriptsize] at (0.35,-8) {Final};
    \node[anchor=west,font=\scriptsize\itshape] at (24.7,0) {initial};
    \node[anchor=west,font=\scriptsize\itshape] at (24.7,-8) {solved};
  \end{tikzpicture}
  \caption{Example run of the greedy algorithm on one center orbit. Each row shows
    the 24 orbit positions in Speffz order, grouped by face. The 3-cycle
    $\operatorname{rot}(a,b,c)$ on the left---which sends the sticker at position
    $a$ to $b$, $b$ to $c$, and $c$ to $a$---is about to be applied to that row;
    its three affected cells are outlined in bold, and the next row shows the
    resulting state. After eight 3-cycles the orbit is solved.}
  \label{fig:greedy-example}
\end{figure}

\begin{lemma}[Greedy orbit solving]
\label{lem:greedy}
Any state of a $24$-cubie center orbit can be solved using at most $12$ 3-cycles.
\end{lemma}
\begin{proof}
We index the 24 orbit positions in Speffz order (Figure~\ref{fig:Speffz}) and
call a position \emph{solved} if its cubie has the color of its face. Each step
applies a 3-cycle $\operatorname{rot}(a,b,c)$. Position $a$ is the lowest-indexed
unsolved position; let $X$ be the color of $a$'s face and $Y$ the color of the
cubie at $a$. Position $b$ is an unsolved position on the face of color~$Y$, and
position $c$ is an unsolved position holding a cubie of color~$X$, so both $a$ and $b$ become solved. If
possible, we choose $b$ and $c$ such that the cubie at $b$ has the color of
$c$'s face, thereby solving all three positions. If instead $a$ and the only
unsolved position on face~$Y$ hold each other's colors, we take $b$ to be a solved
position on face~$Y$ and $c$ the unsolved one; again $a$ and $c$ become solved and
$b$ stays solved. In all cases, $b$ and $c$ have larger index than $a$, and previously
solved positions remain solved. Every 3-cycle used exists by
Lemma~\ref{lem:threecycles}, because $a$ and $b$ lie on different faces. Because
each 3-cycle solves at least two of the 24 cubies, a greedy solve needs at most
$12$ 3-cycles per orbit. Figure~\ref{fig:greedy-example} shows an example run.
\end{proof}

\noindent This bound on the number of 3-cycles gives a single-orbit move bound of
\[
F_{\text{single orbit}} = 12 \cdot 12 = 144.
\]

There are $(n/2 - 1)^2$ center orbits, so this orbit-by-orbit approach leads to an upper bound of $144 \cdot (n/2 - 1)^2 = 36n^2 - 144n + 144$ moves to solve all centers. This move count is $O(n^2)$.

Another property of this set of 4000 3-cycle sequences is their systematic structure: each has the form \(A [B, C] A'\), where \(A\) consists of at most two outer-turn setup moves---we call such moves \defn{conjugation moves}. If \(K\) such 3-cycles that share the same conjugation moves are applied consecutively, then \(K-1\) cancellations of \(A\) and \(A'\) occur. In our algorithms this happens because all applications of one stage of a universal sequence share its conjugation moves and can be executed consecutively without changing the order of operations within any orbit (Lemma~\ref{lem:cancellation}). For example:
\[
  U' ~ [4F, \, U' ~ 3B' ~ U] ~ U, ~  U' ~ [U' ~ 3L2 ~ U, \, 3R2] ~ U, ~  F' ~ [4R, \, U' ~ 3L' ~ U] ~ F, ~ F' ~ [5R, \, U' ~ 2L' ~ U] ~ F.
\]
These 40 moves are reduced by cancellation to
\[
  U' ~ [4F, \, U' ~ 3B' ~ U] ~ [U' ~ 3L2 ~ U, \, 3R2] U \quad F' ~ [4R, \, U' ~ 3L' ~ U] ~ [5R, \, U' ~ 2L' ~ U] F
\]
(36 moves).

\begin{lemma}[Conjugation-move cancellation]
\label{lem:cancellation}
Fix a stage of a universal sequence, with representative move sequence $A\,[B,C]\,A'$. If the
stage is applied $K$ times in succession---to different orbits or to different combinatorial
rectangles---then these $K$ applications cost at most $\sum_i t_i + O(1)$ moves, where $t_i$ is the
length of the $i$th application without conjugation moves. Hence, if every stage is executed in
this batched way, in the order of the universal sequence, a universal sequence of $L$ stages
incurs only $O(L)$ conjugation moves in total.
\end{lemma}
\begin{proof}
The $i$th application has the form $A\,[B_i,C_i]\,A'$, where the inner commutator has
length $t_i$. The conjugation moves $A$ are outer-turn moves and are the same for every
application of a given stage; only the inner slice moves depend on the orbits involved. When the
$K$ applications of one stage are executed consecutively, the $A'$ ending one application cancels
with the $A$ beginning the next, so only the first $A$ and the last $A'$ remain, and the $K$
applications cost at most $\sum_i t_i + O(1)$ moves. This batching preserves the order of the
universal sequence: all applications of a stage finish before the next stage begins, so each
orbit still applies its subsequence in order. The applications within one stage act on distinct
orbits, which have disjoint support by Lemma~\ref{lem:commutator-locality} and therefore commute.
\end{proof}
This cancellation lemma means that for any algorithm with many consecutive $A\,[B,C]\,A'$-type of move sequences, the dominant factor will not be influenced by $A$ and $A'$.

\subsection{Using parallelization with 3-cycles to lower the $O(n^2)$ bound}\label{greedy-example}
\begin{figure}[t]
  \centering
  \begin{subfigure}[t]{0.3\textwidth}
    \centering
    \includegraphics[width=\textwidth]{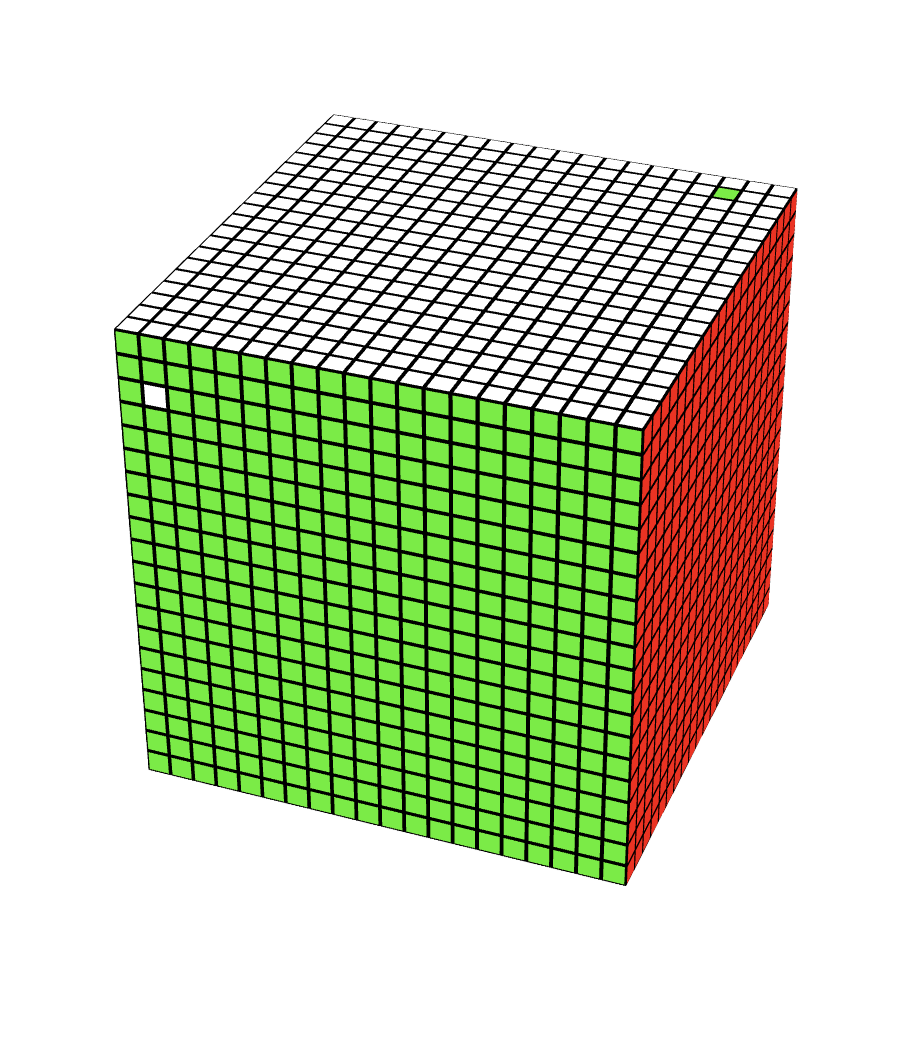}
    \caption{0D: \href{https://alpha.twizzle.net/explore/?puzzle=20x20x20&alg=\%5B3R\%2C+U\%27+2L\%27+U\%5D}{\texttt{[3R, U' 2L' U]}}}
    \label{fig:case-0d}
  \end{subfigure}
  \hfill
  \begin{subfigure}[t]{0.3\textwidth}
    \centering
    \includegraphics[width=\textwidth]{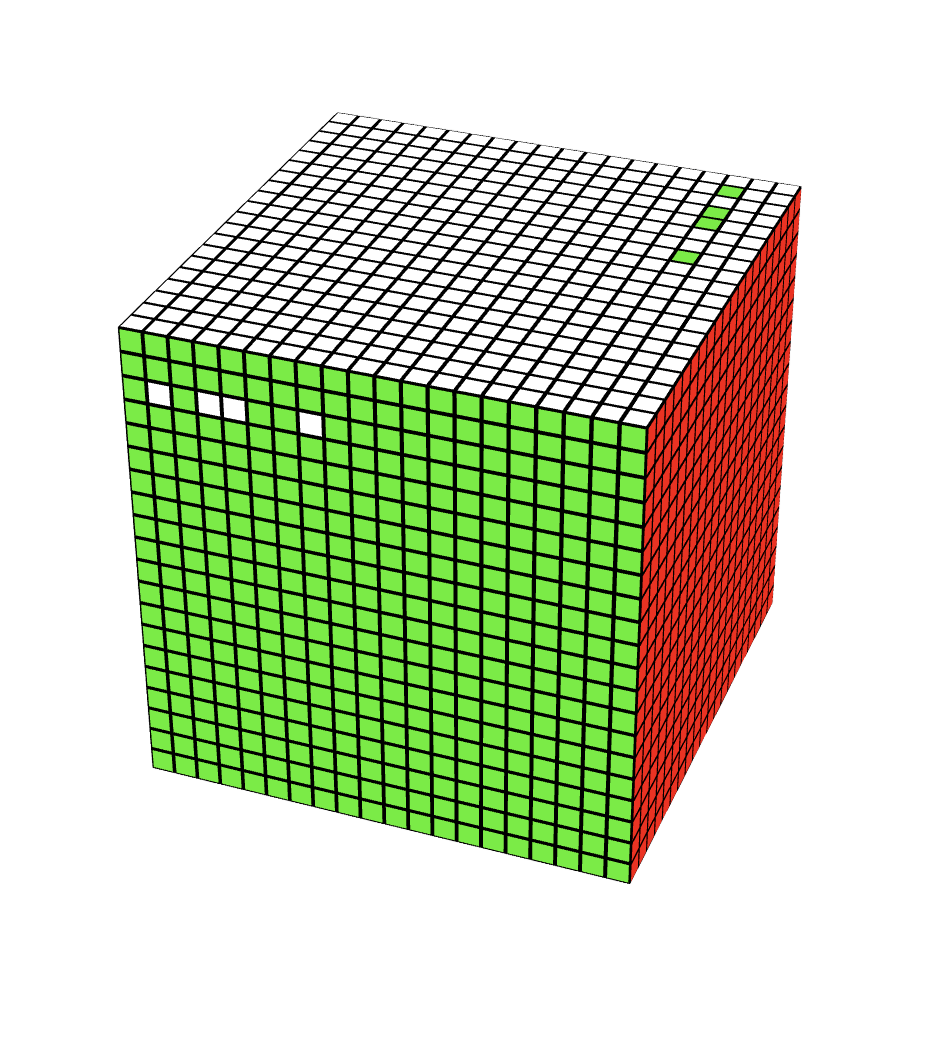}
    \caption{1D: \href{https://alpha.twizzle.net/explore/?puzzle=20x20x20&alg=\%5B3R\%2C+U\%27+8L\%27+5L\%27+4L\%27+2L\%27+U\%5D}{\texttt{[3R, U' 8L' 5L' 4L' 2L' U]}}}
    \label{fig:case-1d}
  \end{subfigure}
  \hfill
  \begin{subfigure}[t]{0.3\textwidth}
    \centering
    \includegraphics[width=\textwidth]{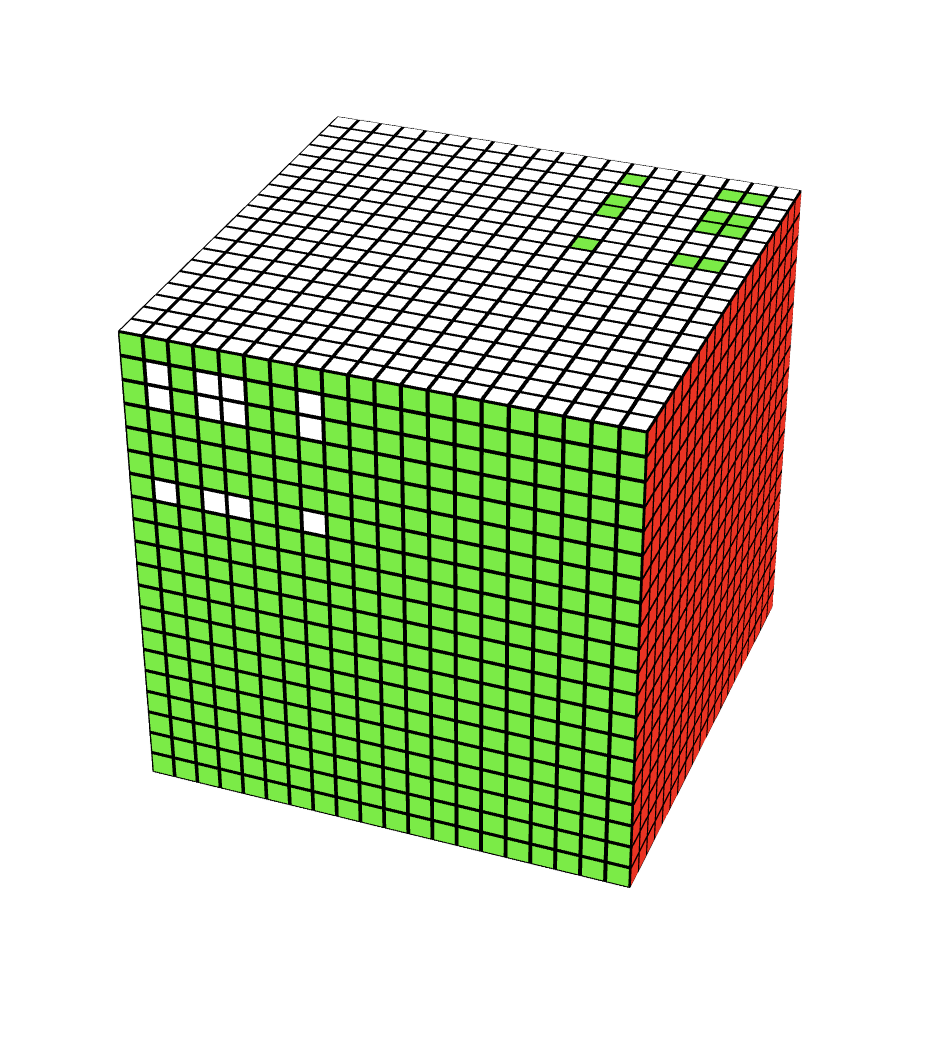}
    \caption{2D: \href{https://alpha.twizzle.net/explore/?puzzle=20x20x20&alg=\%5B2R+3R+7R\%2C+U\%27+8L\%27+5L\%27+4L\%27+2L\%27+U\%5D}{\texttt{[2R 3R 7R, U' 8L' 5L' 4L' 2L' U]}}}
    \label{fig:case-2d}
  \end{subfigure}
  \caption{Examples of 0D, 1D, and 2D parallelization of a 3-cycle on a $20\times20\times20$ Rubik's Cube, influencing one or several cubies simultaneously.}
  \label{fig:parallelization-cases}
\end{figure}

Parallelization is the key concept for an $O(n^2 / \lg n)$ move count. For the $n \times n \times n$ cube, we follow the definitions of 0D, 1D, and 2D parallelization introduced for the $n \times n \times 1$ cube in Section~\ref{strip2d}.
Figure~\ref{fig:parallelization-cases} shows three examples on an $n\times n\times n$ Rubik's Cube: (1) a 3-cycle without parallelization (the 0D case), (2) several orbits parallelized in rows or columns (the 1D case), and (3) several orbits parallelized in combinatorial rectangles across rows and columns (the 2D case).

We distinguish between $+2$/$+2$ and $+2$/$+4$ parallelization behavior for each 3-cycle case by the number of slice moves that must be inserted for parallelization in each direction, i.e., by how much the move count increases. For example, $+2{/}+4$ means that along one axis the algorithm incurs a $+2$ increase in move count per extra orbit, while along the other axis it incurs $+4$. In Table~\ref{tab:3cycle-groups}, we list 3-cycle move sequences that are all parallelizable with $+2$ in at least one direction, enabling the calculation above. This parallelization behavior improves on that of the 10-move sequence in~\cite{DemaineEtAlRubik} that has $+4{/}+4$ behavior.

Inserting slice moves into the different 3-cycle algorithms produces the 1D-parallelization effect defined above. Because all 4000 cases are either $+2$/$+2$- or $+2$/$+4$-parallelizable (Table~\ref{tab:3cycle-groups}), at least one direction can be chosen so that an additional permutation of three cubies in the same row or column costs only $+2$ moves, resulting in
\[
F_{1\mathrm{D}}(n) = 12 \cdot 2 \cdot \left(\frac{n}{2} - 1\right)^2 + O(n) = 6n^2 + O(n).
\]

\subsection{Solving a single orbit with a universal sequence of fixed-order 3-cycles (greedy approach)}
\label{greedy}

By grouping 3-cycles from different orbits into buckets and then executing the same 3-cycle types collectively across orbits, we can increase the probability $p$ that several orbits participate in the same permutation at once. This grouping assumes that we fix an order in which 3-cycle types may occur, even though a given orbit need not participate every time---just as in the $n \times n \times 1$ case (Section~\ref{strip2d}). We call this fixed multi-stage order a \defn{universal sequence}. The universal sequence is applied \defn{selectively}: it is a fixed schedule of stages, and a given orbit applies a stage only if its current state calls for it, skipping the rest, so that every orbit runs a subsequence of the same global order. For any orbit configuration, the universal sequence contains at least one subsequence of permutations that solves the full orbit. The concept of universal sequences allows a high degree of parallelization with 3-cycles. Furthermore, it allows outer-turn-move cancellations as described above.
The maximum depth of the problem of permuting the orbit with 3-cycles is 12 3-cycles, because, as in the 0D approach of Section~\ref{upper_n2}, each 3-cycle solves at least two pieces. Thus a trivial approach, in which we iterate over all 4000 move sequences 12 times, yields a universal sequence of length
\[
  L = 12 \cdot 4000 = 48{,}000.
\]
Because there are many 3-cycle types, the hit probability $p$ is low, so the average cost per solved cubie remains high: commutator 3-cycles often affect only one orbit and thus do not benefit from parallelizable algorithms. However, we can instead use the greedy algorithm of Section~\ref{upper_n2}, which is single-pass by design: it always solves the lowest unsolved position. If we order the 4000 3-cycle types by ascending first (lowest) index, then the at most 12 3-cycles that the greedy algorithm applies to any orbit state appear in exactly this order, giving a universal sequence of length
\[
  L = 4000.
\]
This shorter universal sequence reduces both the total move count and the upper bound in the $O(n^2 / \lg n)$ case.
\subsection{Analysis of existing 3-cycle-based solver}
\label{existing_solver}
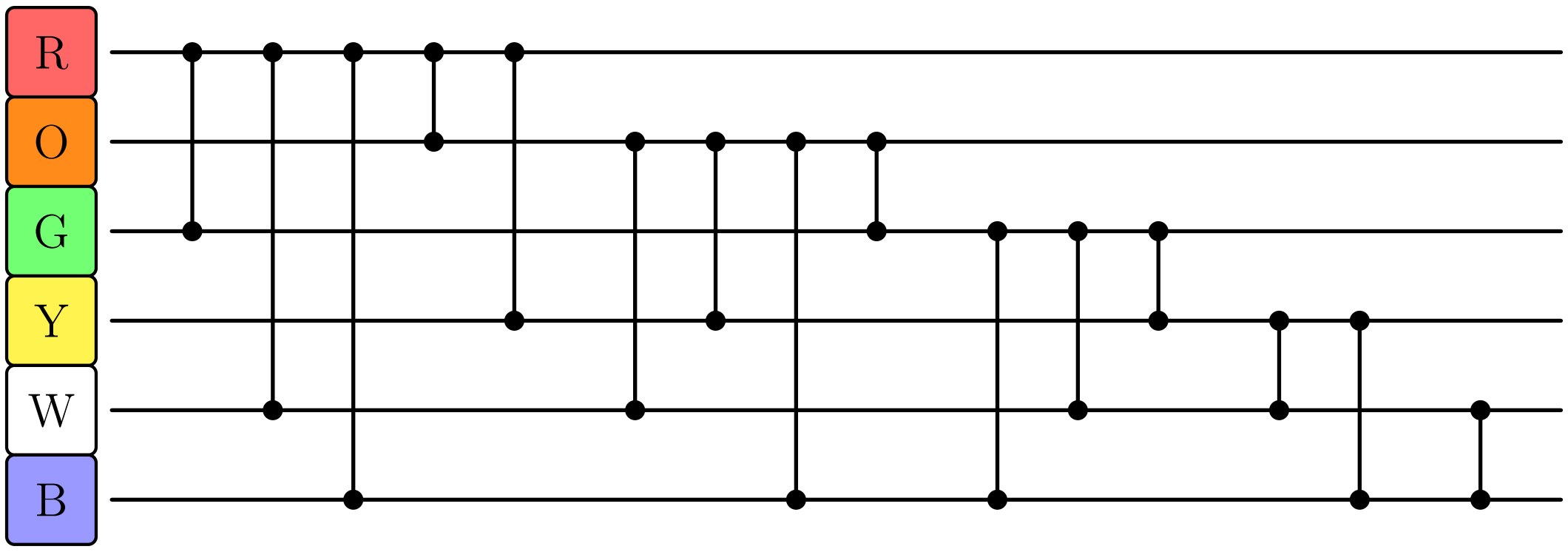
\begin{figure}[t]
    \centering
    \begin{tikzpicture}[x=0.9cm,y=0.8cm,line cap=round,line join=round]
        \tikzset{
            facebox/.style={draw=black, thick, rounded corners=2pt, minimum width=0.8cm, minimum height=0.8cm},
            wire/.style={black, line width=1pt},
            comparator/.style={black, line width=1pt},
            dot/.style={circle, fill=black, inner sep=1.8pt}
        }

        \foreach \name/\y/\bgcol in {R/5/faceR, O/4/faceO, G/3/faceG, Y/2/faceY, W/1/faceW, B/0/faceB} {
            \node[facebox, fill=\bgcol] at (-0.6,\y) {\large \name};
            \draw[wire] (0,\y) -- (14.4,\y);
        }

        \foreach \x/\yone/\ytwo in {
            0.8/5/3,
            1.6/5/1,
            2.4/5/0,
            3.2/5/4,
            4.0/5/2,
            5.2/4/1,
            6.0/4/2,
            6.8/4/0,
            7.6/4/3,
            8.8/3/0,
            9.6/3/1,
            10.4/3/2,
            11.6/2/1,
            12.4/2/0,
            13.6/1/0
        } {
            \draw[comparator] (\x,\yone) -- (\x,\ytwo);
            \node[dot] at (\x,\yone) {};
            \node[dot] at (\x,\ytwo) {};
        }

    \end{tikzpicture}
    \caption{Sorting-network-style diagram for sorting across six cube faces. Each horizontal wire represents one face, and each vertical connector indicates a scheduled comparison between two faces. The order of comparisons (left to right) follows the order used by RCube, rather than a simple regular schedule.}
    \label{fig:rcube_12_6}
\end{figure}

The existing 3-cycle-based solver RCube~\cite{RCubeGitHub} consists of 15 phases where, for each face, a certain color is transferred toward its corresponding side. The pattern is: $G \to R, W \to R, B \to R, O \to R, Y \to R, W \to O, Y \to O, B \to O, G \to O, B \to G, W \to G, Y \to G, W \to Y, B \to Y, B \to W$. Figure~\ref{fig:rcube_12_6} shows the face-comparison diagram. It highlights the six cube faces and the comparison order between them. 

We call each such phase a \defn{two-face swap}. Each two-face swap consists of several 3-cycle permutations applied one after another. While the implementation of RCube uses some further optimizations, we ignore those in the analysis, as they do not improve the leading term of the upper bound. The two-face swap consists of two nested for-loops, resulting in $4 \times 4 = 16$ 3-cycles listed here:

\[
\boxed{
\begin{aligned}
&(0,4,5)\; (0,5,6)\; (0,6,7)\; (0,7,4)\; (1,4,5)\; (1,5,6)\; (1,6,7)\; (1,7,4)\\
&(2,4,5)\; (2,5,6)\; (2,6,7)\; (2,7,4)\;(3,4,5)\; (3,5,6)\; (3,6,7)\; (3,7,4)
\end{aligned}
}
\]

\begin{figure}[t]
    \centering
    \resizebox{\textwidth}{!}{%
    \begin{tikzpicture}[x=0.78cm,y=0.72cm,line cap=round,line join=round]
        \tikzset{
            wire/.style={black, line width=0.9pt},
            cyc/.style={black, line width=0.9pt,
                postaction={decorate, decoration={markings,
                    mark=at position 0.5 with {\arrow{>}}}}},
            dot/.style={circle, fill=black, inner sep=1.6pt},
            feed/.style={gray!55, line width=0.5pt}
        }

        \foreach \i in {1,...,8} {
            \pgfmathtruncatemacro{\wlbl}{\i-1}
            \node[left] at (-0.55,{\wirey{\i}}) {\small \wlbl};
            \draw[wire] (0,{\wirey{\i}}) -- (14.6,{\wirey{\i}});
        }

        \foreach \x/\a/\b/\c/\s in {
            0.8/1/5/6/1, 1.6/1/6/7/-1, 2.4/1/7/8/1, 3.2/1/8/5/-1,
            4.4/2/5/6/-1, 5.2/2/6/7/1, 6.0/2/7/8/-1, 6.8/2/8/5/1,
            8.0/3/5/6/1, 8.8/3/6/7/-1, 9.6/3/7/8/1, 10.4/3/8/5/-1,
            11.6/4/5/6/-1, 12.4/4/6/7/1, 13.2/4/7/8/-1, 14.0/4/8/5/1
        } {
            \pgfmathsetmacro{\dx}{0.42*\s}
            \path (\x,{\wirey{\a}}) coordinate (A)
                  (\x-\dx,{\wirey{\b}}) coordinate (B)
                  (\x+\dx,{\wirey{\c}}) coordinate (C);
            \draw[cyc] (A) -- (B);
            \draw[cyc] (B) -- (C);
            \draw[cyc] (C) -- (A);
            \node[dot] at (A) {};
            \node[dot] at (B) {};
            \node[dot] at (C) {};
        }

        \fill[gray!22] (-5,0.5) -- (-5,3.5) -- (-3.7,4.8) -- (-3.7,1.8) -- cycle; 
        \fill[white]   (-8,3.5) -- (-5,3.5) -- (-3.7,4.8) -- (-6.7,4.8) -- cycle; 
        \fill[white]   (-8,0.5) -- (-5,0.5) -- (-5,3.5) -- (-8,3.5) -- cycle;     
        \foreach \k in {1,2,3,4,5,6}{
          \draw[gray!35,line width=0.3pt] ({-8+3*\k/7},0.5) -- ({-8+3*\k/7},3.5);
          \draw[gray!35,line width=0.3pt] (-8,{0.5+3*\k/7}) -- (-5,{0.5+3*\k/7});
          \draw[gray!35,line width=0.3pt] ({-8+3*\k/7},3.5) -- ({-8+3*\k/7+1.3},4.8);
          \draw[gray!35,line width=0.3pt] ({-8+1.3*\k/7},{3.5+1.3*\k/7}) -- ({-5+1.3*\k/7},{3.5+1.3*\k/7});
          \draw[gray!45,line width=0.3pt] (-5,{0.5+3*\k/7}) -- (-3.7,{1.8+3*\k/7});
          \draw[gray!45,line width=0.3pt] ({-5+1.3*\k/7},{0.5+1.3*\k/7}) -- ({-5+1.3*\k/7},{3.5+1.3*\k/7});
        }
        \draw[black,line width=0.9pt] (-8,0.5) -- (-5,0.5) -- (-5,3.5) -- (-8,3.5) -- cycle;
        \draw[black,line width=0.9pt] (-8,3.5) -- (-6.7,4.8) -- (-3.7,4.8) -- (-5,3.5);
        \draw[black,line width=0.9pt] (-5,0.5) -- (-3.7,1.8) -- (-3.7,4.8);
        \foreach \s/\t/\w in {0.357/0.643/1, 0.643/0.643/2, 0.357/0.357/3, 0.643/0.357/4}{
          \filldraw[fill=yellow, draw=black, line width=0.4pt]
            ({-8+3*(\s-0.06)+1.3*(\t-0.06)},{3.5+1.3*(\t-0.06)}) --
            ({-8+3*(\s+0.06)+1.3*(\t-0.06)},{3.5+1.3*(\t-0.06)}) --
            ({-8+3*(\s+0.06)+1.3*(\t+0.06)},{3.5+1.3*(\t+0.06)}) --
            ({-8+3*(\s-0.06)+1.3*(\t+0.06)},{3.5+1.3*(\t+0.06)}) -- cycle;
          \draw[feed] ({-8+3*\s+1.3*\t},{3.5+1.3*\t}) to[out=6,in=180] (-1.15,{\wirey{\w}});
        }
        \foreach \s/\t/\w in {0.357/0.643/5, 0.643/0.643/6, 0.357/0.357/7, 0.643/0.357/8}{
          \filldraw[fill=green!60!black, draw=black, line width=0.4pt]
            ({-8+3*(\s-0.06)},{0.5+3*(\t-0.06)}) rectangle ({-8+3*(\s+0.06)},{0.5+3*(\t+0.06)});
          \draw[feed] ({-8+3*\s},{0.5+3*\t}) to[out=0,in=180] (-1.15,{\wirey{\w}});
        }
        \node[align=center,font=\small] at (-6.35,-0.35) {$n\times n\times n$ cube};
    \end{tikzpicture}%
    }
    \caption{3-cycle permutation network on wires $0$--$7$, with a schematic
    $n \times n \times n$ cube showing how the wires map to four orbit positions on
    two faces (lower and upper in hierarchy) each. Each triangle represents one 3-cycle
    following the permutation network used in RCube.}
    \label{fig:rcube-3cycle-network}
\end{figure}
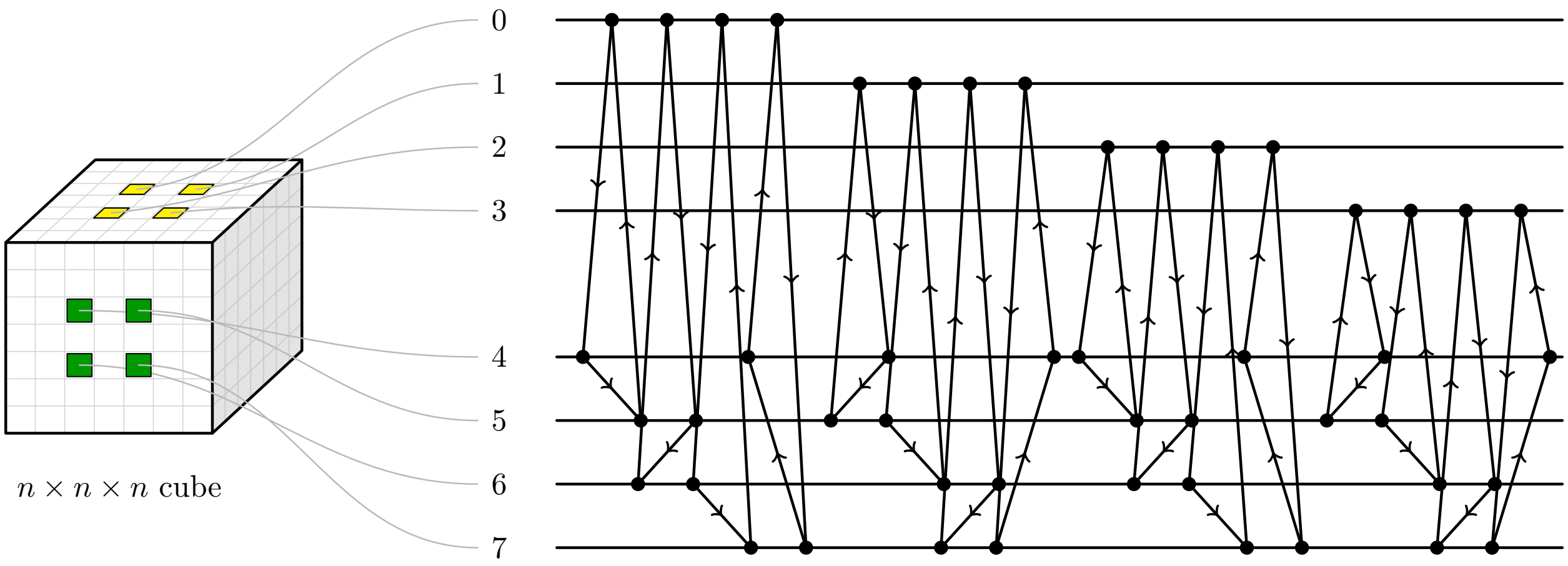
\begin{figure}[t]
    \centering
    \resizebox{\textwidth}{!}{%
    \begin{tikzpicture}[x=0.78cm,y=0.72cm,line cap=round,line join=round]
        \tikzset{
            wire/.style={black, line width=0.9pt},
            skip/.style={gray!45, line width=0.7pt},
            act/.style={orange!90!black, line width=1.3pt,
                postaction={decorate, decoration={markings,
                    mark=at position 0.5 with {\arrow{>}}}}},
            skipdot/.style={circle, fill=gray!45, inner sep=1.2pt},
            actdot/.style={circle, fill=orange!90!black, inner sep=1.6pt},
            feed/.style={gray!55, line width=0.5pt}
        }

        \newcommand{\scube}[1]{%
          \fill[gray!22] ({#1+3},0.5) -- ({#1+3},3.5) -- ({#1+4.3},4.8) -- ({#1+4.3},1.8) -- cycle;
          \fill[white]   (#1,3.5) -- ({#1+3},3.5) -- ({#1+4.3},4.8) -- ({#1+1.3},4.8) -- cycle;
          \fill[white]   (#1,0.5) -- ({#1+3},0.5) -- ({#1+3},3.5) -- (#1,3.5) -- cycle;
          \foreach \k in {1,2,3,4,5,6}{
            \draw[gray!35,line width=0.3pt] ({#1+3*\k/7},0.5) -- ({#1+3*\k/7},3.5);
            \draw[gray!35,line width=0.3pt] (#1,{0.5+3*\k/7}) -- ({#1+3},{0.5+3*\k/7});
            \draw[gray!35,line width=0.3pt] ({#1+3*\k/7},3.5) -- ({#1+3*\k/7+1.3},4.8);
            \draw[gray!35,line width=0.3pt] ({#1+1.3*\k/7},{3.5+1.3*\k/7}) -- ({#1+3+1.3*\k/7},{3.5+1.3*\k/7});
            \draw[gray!45,line width=0.3pt] ({#1+3},{0.5+3*\k/7}) -- ({#1+4.3},{1.8+3*\k/7});
            \draw[gray!45,line width=0.3pt] ({#1+3+1.3*\k/7},{0.5+1.3*\k/7}) -- ({#1+3+1.3*\k/7},{3.5+1.3*\k/7});
          }
          \draw[black,line width=0.9pt] (#1,0.5) -- ({#1+3},0.5) -- ({#1+3},3.5) -- (#1,3.5) -- cycle;
          \draw[black,line width=0.9pt] (#1,3.5) -- ({#1+1.3},4.8) -- ({#1+4.3},4.8) -- ({#1+3},3.5);
          \draw[black,line width=0.9pt] ({#1+3},0.5) -- ({#1+4.3},1.8) -- ({#1+4.3},4.8);
        }

        \scube{-8}
        \foreach \s/\t/\w/\col in {0.357/0.643/1/green!60!black, 0.643/0.643/2/yellow,
                                   0.357/0.357/3/green!60!black, 0.643/0.357/4/yellow}{
          \filldraw[fill=\col, draw=black, line width=0.4pt]
            ({-8+3*(\s-0.06)+1.3*(\t-0.06)},{3.5+1.3*(\t-0.06)}) --
            ({-8+3*(\s+0.06)+1.3*(\t-0.06)},{3.5+1.3*(\t-0.06)}) --
            ({-8+3*(\s+0.06)+1.3*(\t+0.06)},{3.5+1.3*(\t+0.06)}) --
            ({-8+3*(\s-0.06)+1.3*(\t+0.06)},{3.5+1.3*(\t+0.06)}) -- cycle;
          \draw[feed] ({-8+3*\s+1.3*\t},{3.5+1.3*\t}) to[out=6,in=180] (-1.15,{\wirey{\w}});
        }
        \foreach \s/\t/\w/\col in {0.357/0.643/5/yellow, 0.643/0.643/6/green!60!black,
                                   0.357/0.357/7/yellow, 0.643/0.357/8/green!60!black}{
          \filldraw[fill=\col, draw=black, line width=0.4pt]
            ({-8+3*(\s-0.06)},{0.5+3*(\t-0.06)}) rectangle ({-8+3*(\s+0.06)},{0.5+3*(\t+0.06)});
          \draw[feed] ({-8+3*\s},{0.5+3*\t}) to[out=0,in=180] (-1.15,{\wirey{\w}});
        }
        \node[align=center,font=\small\itshape] at (-6.35,-0.35) {scrambled};

        \scube{16.4}
        \foreach \s/\t/\w in {0.357/0.643/1, 0.643/0.643/2, 0.357/0.357/3, 0.643/0.357/4}{
          \filldraw[fill=yellow, draw=black, line width=0.4pt]
            ({16.4+3*(\s-0.06)+1.3*(\t-0.06)},{3.5+1.3*(\t-0.06)}) --
            ({16.4+3*(\s+0.06)+1.3*(\t-0.06)},{3.5+1.3*(\t-0.06)}) --
            ({16.4+3*(\s+0.06)+1.3*(\t+0.06)},{3.5+1.3*(\t+0.06)}) --
            ({16.4+3*(\s-0.06)+1.3*(\t+0.06)},{3.5+1.3*(\t+0.06)}) -- cycle;
          \draw[feed] ({16.4+3*\s+1.3*\t},{3.5+1.3*\t}) to[out=174,in=0] (15.75,{\wirey{\w}});
        }
        \foreach \s/\t/\w in {0.357/0.643/5, 0.643/0.643/6, 0.357/0.357/7, 0.643/0.357/8}{
          \filldraw[fill=green!60!black, draw=black, line width=0.4pt]
            ({16.4+3*(\s-0.06)},{0.5+3*(\t-0.06)}) rectangle ({16.4+3*(\s+0.06)},{0.5+3*(\t+0.06)});
          \draw[feed] ({16.4+3*\s},{0.5+3*\t}) to[out=180,in=0] (15.75,{\wirey{\w}});
        }
        \node[align=center,font=\small\itshape] at (18.05,-0.35) {solved};

        \foreach \i in {1,...,8} {
            \pgfmathtruncatemacro{\wlbl}{\i-1}
            \node[left] at (-0.55,{\wirey{\i}}) {\small \wlbl};
            \node[right] at (15.15,{\wirey{\i}}) {\small \wlbl};
            \draw[wire] (0,{\wirey{\i}}) -- (14.6,{\wirey{\i}});
        }

        \foreach \x/\a/\b/\c/\s in {
            0.8/1/5/6/1, 1.6/1/6/7/-1, 2.4/1/7/8/1,
            4.4/2/5/6/-1, 5.2/2/6/7/1, 6.0/2/7/8/-1, 6.8/2/8/5/1,
            8.0/3/5/6/1, 9.6/3/7/8/1, 10.4/3/8/5/-1,
            11.6/4/5/6/-1, 12.4/4/6/7/1, 13.2/4/7/8/-1, 14.0/4/8/5/1
        } {
            \pgfmathsetmacro{\dx}{0.42*\s}
            \path (\x,{\wirey{\a}}) coordinate (A)
                  (\x-\dx,{\wirey{\b}}) coordinate (B)
                  (\x+\dx,{\wirey{\c}}) coordinate (C);
            \draw[skip] (A) -- (B) -- (C) -- cycle;
            \node[skipdot] at (A) {}; \node[skipdot] at (B) {}; \node[skipdot] at (C) {};
        }

        \foreach \x/\a/\b/\c/\s in {
            3.2/1/8/5/-1, 8.8/3/6/7/-1
        } {
            \pgfmathsetmacro{\dx}{0.42*\s}
            \path (\x,{\wirey{\a}}) coordinate (A)
                  (\x-\dx,{\wirey{\b}}) coordinate (B)
                  (\x+\dx,{\wirey{\c}}) coordinate (C);
            \draw[act] (A) -- (B); \draw[act] (B) -- (C); \draw[act] (C) -- (A);
            \node[actdot] at (A) {}; \node[actdot] at (B) {}; \node[actdot] at (C) {};
        }
    \end{tikzpicture}%
    }
    \caption{Selective application of the fixed permutation network of
    Figure~\ref{fig:rcube-3cycle-network}. For one example orbit, the highlighted
    3-cycles (orange) form the subsequence actually applied---the orbit's ``path''
    through the network---while the grayed-out 3-cycles are skipped. The left cube
    shows the scrambled state of the eight selected orbit positions (top-face cubies
    on wires $1,3,4,6$); for this state two 3-cycles suffice: $(0,7,4)$ routes the
    cubie on wire $4$ to wire $0$, then $(2,5,6)$ routes the cubie on wire $6$ to
    wire $2$. The right cube shows the same positions after this subsequence has
    been applied. A different orbit state selects a different subsequence of the
    \emph{same} fixed universal sequence.}
    \label{fig:selective-network}
\end{figure}
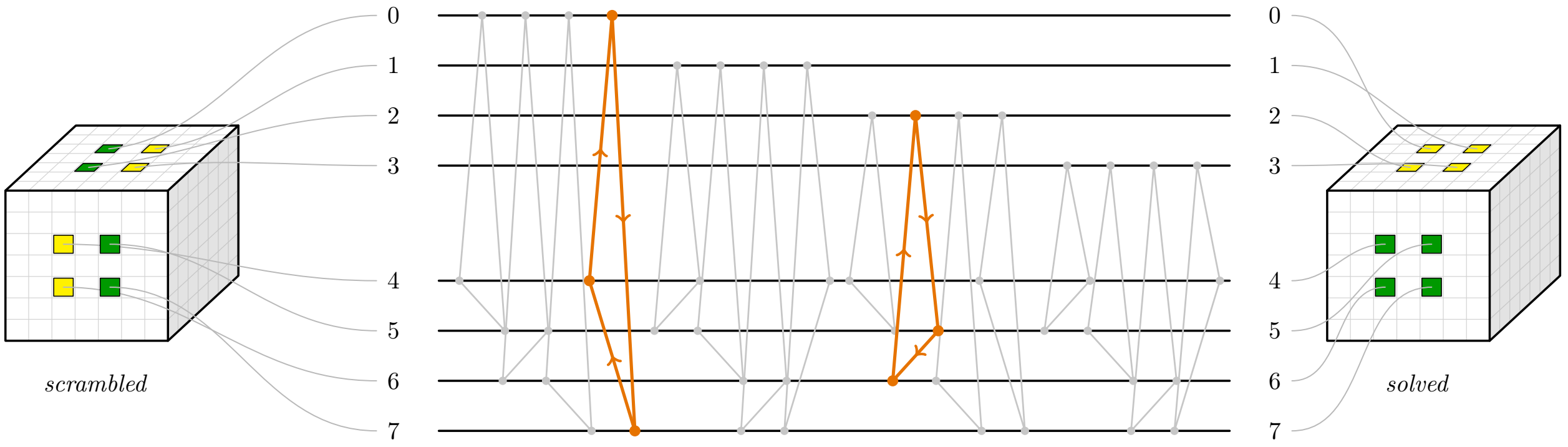

Figure~\ref{fig:rcube-3cycle-network} shows the 3-cycle permutation network on wires $0$ through $7$. The upper block uses wires $0$--$3$, located in the four quadrants of one face of the Rubik's Cube; the lower block uses wires $4$--$7$, located in the four quadrants of the other face. The larger vertical gap between wires $3$ and $4$ separates the two groups.
From this network structure it follows that the universal sequence is $15 \times 16$ 3-cycles long, leading to
\[
  L = 15 \cdot 16 = 240.
\]
Figure~\ref{fig:selective-network} illustrates the selective application of this
fixed network for one example orbit.

\subsection{Achieving $O(n^2 / \lg n)$ move count performance (2D parallelization)}
\label{2D_par}

By using $({+2}/{+2})$ 3-cycle algorithms, the constant in front of $n^2 / \lg n$ translates to the number of 3-cycles in the universal sequence times $1/2$.
As we have shown in Section~\ref{upper_n2}, there exist specific 3-cycle types with the $+2/+2$-parallelization behavior. For the solving algorithms we present and the calculations in this paper, we use only 3-cycles for which $+2{/}{+}2$-parallelization behavior exists. In the following, we calculate how the number of 3-cycle types in the ordered universal sequence influences our leading constant. For this calculation, it is important to note that the problem of 2D parallelizing 3-cycles of the same type on several orbits at once can be abstracted into finding combinatorial rectangles in a square 2D grid with side length $n/2 - 1$, where each hit or combination of hits via a combinatorial rectangle is associated with a cost. The way we achieve a low cost of $n^2 / \lg n$ is by finding combinatorial rectangles in strips of a certain width $w$, exactly as in the $n \times n \times 1$ case (Section~\ref{strip2d}) and as
described in~\cite[Lemmas~4--6]{DemaineEtAlRubik}.
We further improve that process to lower the leading constant by a factor of two.

As in Section~\ref{strip2d}, we partition the grid into $Y/w$ strips of constant width $w = w(Y)$, where $Y = n/2 - 1$ is the grid side length, batch columns with equal Boolean labelings into combinatorial rectangles, and account for the worst case by covering the cost of at least one of each essential vector while assuming that all remaining columns contain additional vectors (Figure~\ref{fig:ess}).

To calculate the cost of solving the worst case, we separate the costs into three
parts: outer-turn moves, vertical moves, and horizontal moves. The general structure of the commutators used is $[V,\, OHO]$, where the conjugation moves are already stripped because of the cancellation trick.
Each strip contains $2^w - 1$ combinatorial rectangles for all cases of non-zero essential bit vectors. Given that there are $Y/w$ strips, the total cost of outer-turn moves is at most $4 \cdot (Y/w) \cdot (2^w - 1)$,
the total cost of vertical moves is at most $2Y^2/w$,
and the total cost of horizontal moves is at most $Y \cdot 2^w$, because while we need to iterate over all ``1'' bits only (which occur exactly half of the time),
the factor of 2 cancels out with the 2 arising from the commutator structure.
Thus the total cost for solving all strips is
\[
  4 \cdot \frac{Y}{w} \cdot (2^w - 1) + \frac{2Y^2}{w} + Y \cdot 2^w.
\]
While~\cite{DemaineEtAlRubik} uses $w = \frac{1}{2}\lg Y$, we use
$w = \lg Y - 2\lg(\lg Y)$, which will lead to a better leading constant:
\begin{align*}
F_{\mathrm{move count}}
  &= \frac{4Y}{w}\!\left(\frac{Y}{(\lg Y)^2} - 1\right) + \frac{2Y^2}{w} + \frac{Y^2}{(\lg Y)^2} \\[4pt]
  &\;=\;
  \frac{2Y^2}{w}
  \;+\;
  \frac{Y^2}{(\lg Y)^2}
  \;+\;
  \frac{4Y^2}{w\,(\lg Y)^2}
  \;-\;
  \frac{4Y}{w}.
\end{align*}
Substituting $w = \lg Y - 2\lg(\lg Y)$ and ordering by asymptotic dominance:
\[
  F_{\mathrm{move count}}
  \;=\;
  \underbrace{\frac{2Y^2}{\lg Y - 2\lg\lg Y}}_{\displaystyle\Theta\!\left(\frac{Y^2}{\lg Y}\right)\,\text{(leading)}}
  \;+\;
  \underbrace{\frac{Y^2}{(\lg Y)^2}}_{\displaystyle\Theta\!\left(\frac{Y^2}{\lg^2 Y}\right)}
  \;+\;
  \underbrace{\frac{4Y^2}{\bigl(\lg Y - 2\lg\lg Y\bigr)(\lg Y)^2}}_{\displaystyle\Theta\!\left(\frac{Y^2}{\lg^3 Y}\right)}
  \;-\;
  \underbrace{\frac{4Y}{\lg Y - 2\lg\lg Y}}_{\displaystyle\Theta\!\left(\frac{Y}{\lg Y}\right)}.
\]
Now substituting $Y = n/2 - 1$ to get a function expressing the upper bound move count for solving one 3-cycle type of all centers of an $n \times n \times n$ Rubik's Cube:
\begin{multline*}
  F_{2\mathrm{D}}(n)
  =
  \frac{2\!\left(\frac{n}{2}-1\right)^{\!2}}%
       {\lg\!\left(\frac{n}{2}-1\right)-2\lg\lg\!\left(\frac{n}{2}-1\right)}
  \;+\;
  \frac{\left(\frac{n}{2}-1\right)^{\!2}}%
       {\left[\lg\!\left(\frac{n}{2}-1\right)\right]^{2}}
  \;+\;
  \frac{4\!\left(\frac{n}{2}-1\right)^{\!2}}%
       {\left[\lg\!\left(\frac{n}{2}-1\right)-2\lg\lg\!\left(\frac{n}{2}-1\right)\right]%
        \left[\lg\!\left(\frac{n}{2}-1\right)\right]^{2}} \\
  \;-\;
  \frac{4\!\left(\frac{n}{2}-1\right)}%
       {\lg\!\left(\frac{n}{2}-1\right)-2\lg\lg\!\left(\frac{n}{2}-1\right)}.
\end{multline*}
\[
  \lim_{n\to\infty}\frac{F_{2\mathrm{D}}(n)}{n^{2}/\lg n}
  =\lim_{n\to\infty}\frac{2\!\left(\tfrac{n}{2}-1\right)^{\!2}\lg n}
       {\bigl(\lg\!\left(\tfrac{n}{2}-1\right)-2\lg\lg\!\left(\tfrac{n}{2}-1\right)\bigr)\,n^{2}}
  = \tfrac{1}{2}.
\]
The leading constant for using one 3-cycle type on all center orbits is therefore $\mathbf{1/2}$. This calculation generalizes to the following tool, which we use to read off the leading constant of every solver in the remainder of the paper.

\begin{lemma}[Universal-sequence leading constant]
\label{lem:stageconstant}
Let a universal sequence consist of $L$ stages, each a move sequence with $+2/+X$
parallelization behavior for some constant $X$, such that every center orbit
configuration is selectively solved by some subsequence. Then the sequence solves all
centers of an $n \times n \times n$ Rubik's Cube in
\[
  \frac{L}{2}\cdot\frac{n^2}{\lg n} + o\!\left(\frac{n^2}{\lg n}\right)
\]
moves; that is, its leading constant is $L/2$.
\end{lemma}
\begin{proof}
The computation above shows that a single stage costs
$\tfrac12\,n^2/\lg n + o(n^2/\lg n)$: with the strip width $w = \lg Y - 2\lg\lg Y$ and
$Y = n/2-1$, the $+2$ (vertical) moves contribute the leading term $2Y^2/w \sim \tfrac12
n^2/\lg n$, while the $+X$ (horizontal) and outer-turn moves contribute only
$O(Y\cdot 2^w) + O\!\big(\tfrac{Y}{w}2^w\big) = o(Y^2/w)$; hence a constant $X$ affects
only lower-order terms. Summing over the $L$ stages multiplies the leading term by $L$,
giving $L/2$. The selective-solving hypothesis guarantees that these $L$ stages suffice to
solve every orbit.
\end{proof}

\noindent This lemma simplifies the assessment of asymptotic algorithm performance: to bound a solver's leading constant it suffices to count the stages of its universal sequence and verify the selective-solving property.

\subsection{Lowering the upper bound with better two-face swaps using 8-to-4 permutation networks}

\begin{figure}[t]
    \centering
    \begin{tikzpicture}[x=0.95cm,y=0.72cm,line cap=round,line join=round]
        \tikzset{
            wire/.style={black, line width=0.9pt},
            cyc/.style={black, line width=0.9pt,
                postaction={decorate, decoration={markings,
                    mark=at position 0.5 with {\arrow{>}}}}},
            dot/.style={circle, fill=black, inner sep=1.6pt}
        }

        \foreach \i in {1,...,8} {
            \pgfmathtruncatemacro{\wlbl}{\i-1}
            \node[left] at (-0.55,{\wirey{\i}}) {\small \wlbl};
            \draw[wire] (0,{\wirey{\i}}) -- (8.8,{\wirey{\i}});
        }

        \foreach \x/\a/\b/\c/\s in {
            0.8/2/3/5/ 1,
            1.8/2/5/8/-1,
            2.8/2/5/3/ 1,
            3.8/2/3/8/-1,
            4.8/1/8/5/ 1,
            5.8/2/3/7/-1,
            6.8/3/5/6/ 1,
            7.8/1/4/5/-1
        } {
            \pgfmathsetmacro{\dx}{0.42*\s}
            \path (\x,{\wirey{\a}}) coordinate (A)
                  (\x-\dx,{\wirey{\b}}) coordinate (B)
                  (\x+\dx,{\wirey{\c}}) coordinate (C);
            \draw[cyc] (A) -- (B);
            \draw[cyc] (B) -- (C);
            \draw[cyc] (C) -- (A);
            \node[dot] at (A) {};
            \node[dot] at (B) {};
            \node[dot] at (C) {};
        }
    \end{tikzpicture}
    \caption{3-cycle permutation network on wires $0$--$7$ for the new universal sequence, the building block of the sortnet96 algorithm (Section~\ref{6inputsortingnetworks}). Each triangle represents one 3-cycle.}
    \label{fig:new-permutation-network}
\end{figure}
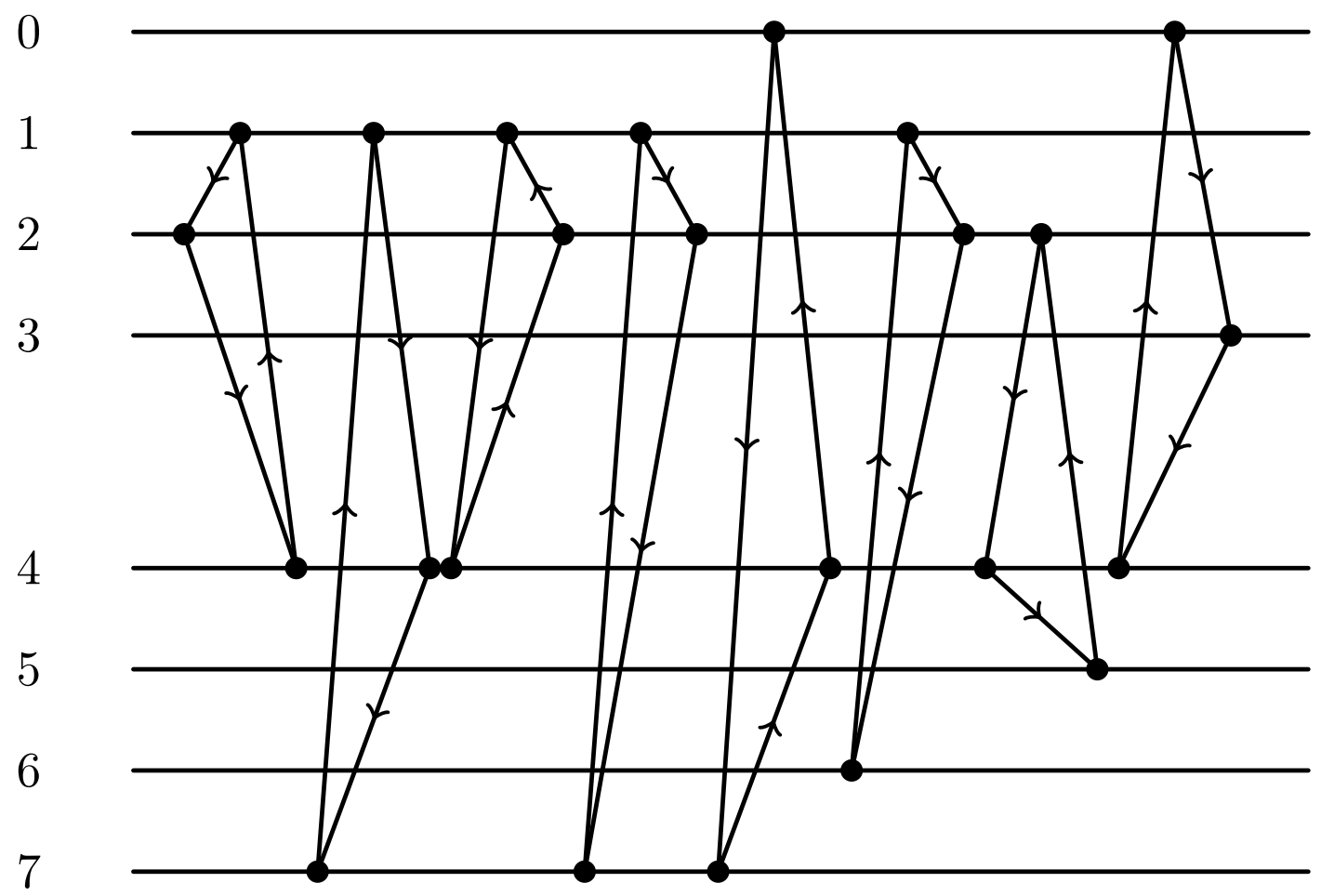

We can implement a new type of solver that is $O(n^2/\lg n)$ using the method described in the previous subsection, a chain of two-face swaps and a sorting network like the one in Section~\ref{existing_solver}. However, to increase the degree of parallelization, we use a permutation network to lower the number of 3-cycles needed for all $\binom{8}{4}
=70$ combinations for swapping the 4+4 center cubies. Such a network is not unique; many exist. Figure~\ref{fig:new-permutation-network} shows one of those 3-cycle permutation networks with wires $0$ through $7$ for the new universal sequence.

\[
\boxed{
  \text{3-cycles used: }(1,2,4)\;(1,4,7)\;(1,4,2)\;(1,2,7)\;(0,7,4)\;(1,2,6)\;(2,4,5)\;(0,3,4)
}
\]
We thereby reduce the leading constant to $60$:
\[
\begin{aligned}
F_{\mathrm{move count}}
  &= 15 \cdot 8 \cdot \frac{1}{2} \cdot \frac{n^2}{\lg n} + o\!\left(\frac{n^2}{\lg n}\right) = 60 \cdot \frac{n^2}{\lg n} + o\!\left(\frac{n^2}{\lg n}\right).
\end{aligned}
\]

These eight 3-cycles must solve every one of the $\binom{8}{4}=70$ two-face-swap
configurations of an orbit: for each configuration there must exist an ordered
subsequence of the eight 3-cycles that routes the four top-group cubies onto wires
$0$--$3$. We verified this solving property exhaustively by enumerating all $2^8$ subsequences over all
$70$ configurations; every configuration is solvable, and the minimal solving
subsequence uses at most six of the eight 3-cycles. Table~\ref{tab:permnet8-cases}
lists it for each configuration. For instance, configuration \texttt{1110\,0001}---in
which wires $3$ and $7$ hold each other's cubies, a single two-sticker swap---is solved
by exactly two 3-cycles: applying $(1,4,7)$ and then $(0,3,4)$ moves the top-group set
$\{0,1,2,7\}\to\{0,1,2,4\}\to\{0,1,2,3\}$ (columns $2$ and $8$ of its row).
We do not implement this solver as a standalone algorithm; it serves as a stepping stone
for the optimization in the next subsection.

\begin{table}[t]
  \centering \scriptsize \setlength{\tabcolsep}{3pt}
  \begin{tabular}{l|cccccccc @{\hspace{1.4em}} l|cccccccc}
    \toprule
    state & 1 & 2 & 3 & 4 & 5 & 6 & 7 & 8 & state & 1 & 2 & 3 & 4 & 5 & 6 & 7 & 8 \\
    \midrule
    \texttt{0000\,1111} &  &  &  & $\bullet$ & $\bullet$ & $\bullet$ & $\bullet$ & $\bullet$ & \texttt{1000\,0111} &  & $\bullet$ &  &  &  & $\bullet$ & $\bullet$ & $\bullet$ \\
    \texttt{0001\,0111} &  & $\bullet$ & $\bullet$ &  & $\bullet$ & $\bullet$ & $\bullet$ &  & \texttt{1000\,1011} &  &  &  & $\bullet$ &  & $\bullet$ &  & $\bullet$ \\
    \texttt{0001\,1011} &  &  &  & $\bullet$ & $\bullet$ & $\bullet$ &  &  & \texttt{1000\,1101} & $\bullet$ &  &  & $\bullet$ &  &  & $\bullet$ & $\bullet$ \\
    \texttt{0001\,1101} &  &  &  & $\bullet$ & $\bullet$ &  & $\bullet$ &  & \texttt{1000\,1110} & $\bullet$ &  &  &  &  & $\bullet$ & $\bullet$ & $\bullet$ \\
    \texttt{0001\,1110} &  &  &  &  & $\bullet$ & $\bullet$ & $\bullet$ &  & \texttt{1001\,0011} &  & $\bullet$ &  &  &  & $\bullet$ &  &  \\
    \texttt{0010\,0111} &  & $\bullet$ & $\bullet$ &  & $\bullet$ & $\bullet$ & $\bullet$ & $\bullet$ & \texttt{1001\,0101} &  & $\bullet$ &  &  &  &  & $\bullet$ &  \\
    \texttt{0010\,1011} & $\bullet$ &  &  &  & $\bullet$ & $\bullet$ &  & $\bullet$ & \texttt{1001\,0110} &  &  &  &  &  & $\bullet$ & $\bullet$ &  \\
    \texttt{0010\,1101} & $\bullet$ &  &  & $\bullet$ & $\bullet$ &  & $\bullet$ & $\bullet$ & \texttt{1001\,1001} & $\bullet$ &  &  & $\bullet$ &  &  &  &  \\
    \texttt{0010\,1110} & $\bullet$ &  &  &  & $\bullet$ & $\bullet$ & $\bullet$ & $\bullet$ & \texttt{1001\,1010} & $\bullet$ &  &  &  &  & $\bullet$ &  &  \\
    \texttt{0011\,0011} &  & $\bullet$ & $\bullet$ &  & $\bullet$ & $\bullet$ &  &  & \texttt{1001\,1100} & $\bullet$ &  &  &  &  &  & $\bullet$ &  \\
    \texttt{0011\,0101} &  & $\bullet$ & $\bullet$ &  & $\bullet$ &  & $\bullet$ &  & \texttt{1010\,0011} &  & $\bullet$ & $\bullet$ &  &  & $\bullet$ &  & $\bullet$ \\
    \texttt{0011\,0110} & $\bullet$ &  &  &  & $\bullet$ & $\bullet$ & $\bullet$ &  & \texttt{1010\,0101} &  & $\bullet$ &  &  &  &  & $\bullet$ & $\bullet$ \\
    \texttt{0011\,1001} & $\bullet$ &  &  & $\bullet$ & $\bullet$ &  &  &  & \texttt{1010\,0110} &  &  & $\bullet$ &  &  & $\bullet$ & $\bullet$ & $\bullet$ \\
    \texttt{0011\,1010} & $\bullet$ &  &  &  & $\bullet$ & $\bullet$ &  &  & \texttt{1010\,1001} & $\bullet$ &  &  & $\bullet$ &  &  &  & $\bullet$ \\
    \texttt{0011\,1100} & $\bullet$ &  &  &  & $\bullet$ &  & $\bullet$ &  & \texttt{1010\,1010} & $\bullet$ &  &  &  &  & $\bullet$ &  & $\bullet$ \\
    \texttt{0100\,0111} &  & $\bullet$ &  &  & $\bullet$ & $\bullet$ & $\bullet$ & $\bullet$ & \texttt{1010\,1100} &  &  & $\bullet$ &  &  &  & $\bullet$ & $\bullet$ \\
    \texttt{0100\,1011} &  &  &  &  & $\bullet$ & $\bullet$ &  & $\bullet$ & \texttt{1011\,0001} &  & $\bullet$ &  &  &  &  &  &  \\
    \texttt{0100\,1101} &  &  &  & $\bullet$ & $\bullet$ &  & $\bullet$ & $\bullet$ & \texttt{1011\,0010} &  &  & $\bullet$ &  &  & $\bullet$ &  &  \\
    \texttt{0100\,1110} &  &  &  &  & $\bullet$ & $\bullet$ & $\bullet$ & $\bullet$ & \texttt{1011\,0100} &  &  & $\bullet$ &  &  &  & $\bullet$ &  \\
    \texttt{0101\,0011} &  & $\bullet$ &  &  & $\bullet$ & $\bullet$ &  &  & \texttt{1011\,1000} &  &  & $\bullet$ &  &  &  &  &  \\
    \texttt{0101\,0101} &  & $\bullet$ &  &  & $\bullet$ &  & $\bullet$ &  & \texttt{1100\,0011} &  & $\bullet$ &  &  &  & $\bullet$ &  & $\bullet$ \\
    \texttt{0101\,0110} &  & $\bullet$ &  &  & $\bullet$ & $\bullet$ & $\bullet$ &  & \texttt{1100\,0101} &  &  &  & $\bullet$ &  &  & $\bullet$ & $\bullet$ \\
    \texttt{0101\,1001} &  &  &  & $\bullet$ & $\bullet$ &  &  &  & \texttt{1100\,0110} &  &  &  &  &  & $\bullet$ & $\bullet$ & $\bullet$ \\
    \texttt{0101\,1010} &  &  &  &  & $\bullet$ & $\bullet$ &  &  & \texttt{1100\,1001} &  &  &  & $\bullet$ &  &  &  & $\bullet$ \\
    \texttt{0101\,1100} &  &  &  &  & $\bullet$ &  & $\bullet$ &  & \texttt{1100\,1010} &  &  &  &  &  & $\bullet$ &  & $\bullet$ \\
    \texttt{0110\,0011} &  &  & $\bullet$ &  & $\bullet$ & $\bullet$ &  & $\bullet$ & \texttt{1100\,1100} & $\bullet$ &  &  &  &  &  & $\bullet$ & $\bullet$ \\
    \texttt{0110\,0101} &  & $\bullet$ &  &  & $\bullet$ &  & $\bullet$ & $\bullet$ & \texttt{1101\,0001} &  &  &  & $\bullet$ &  &  &  &  \\
    \texttt{0110\,0110} &  &  & $\bullet$ &  & $\bullet$ & $\bullet$ & $\bullet$ & $\bullet$ & \texttt{1101\,0010} &  &  &  &  &  & $\bullet$ &  &  \\
    \texttt{0110\,1001} &  &  &  &  & $\bullet$ &  &  & $\bullet$ & \texttt{1101\,0100} &  &  &  &  &  &  & $\bullet$ &  \\
    \texttt{0110\,1010} &  & $\bullet$ &  & $\bullet$ & $\bullet$ & $\bullet$ &  & $\bullet$ & \texttt{1101\,1000} & $\bullet$ &  &  &  &  &  &  &  \\
    \texttt{0110\,1100} &  &  &  &  & $\bullet$ &  & $\bullet$ & $\bullet$ & \texttt{1110\,0001} &  & $\bullet$ &  &  &  &  &  & $\bullet$ \\
    \texttt{0111\,0001} &  & $\bullet$ &  &  & $\bullet$ &  &  &  & \texttt{1110\,0010} &  &  & $\bullet$ &  &  & $\bullet$ &  & $\bullet$ \\
    \texttt{0111\,0010} &  &  & $\bullet$ &  & $\bullet$ & $\bullet$ &  &  & \texttt{1110\,0100} &  &  &  &  &  &  & $\bullet$ & $\bullet$ \\
    \texttt{0111\,0100} &  &  & $\bullet$ &  & $\bullet$ &  & $\bullet$ &  & \texttt{1110\,1000} &  &  &  &  &  &  &  & $\bullet$ \\
    \texttt{0111\,1000} &  &  &  &  & $\bullet$ &  &  &  & \texttt{1111\,0000} &  &  &  &  &  &  &  &  \\
    \bottomrule
  \end{tabular}
  \caption{Selective application of the 8-to-4 permutation network for all $\binom{8}{4}=70$ two-face-swap states. Each state is the $8$-bit pattern of which wires hold a top-group cubie (left nibble $=$ wires $0$--$3$, right nibble $=$ wires $4$--$7$); the solved state is \texttt{1111\,0000}. A filled dot in column $k$ marks that the $k$th 3-cycle $(1,2,4)\,(1,4,7)\,(1,4,2)\,(1,2,7)\,(0,7,4)\,(1,2,6)\,(2,4,5)\,(0,3,4)$ belongs to the minimal solving subsequence. All $70$ states are solvable.}
  \label{tab:permnet8-cases}
\end{table}

\subsection{Lowering the upper bound with better 6-input sorting networks}
\label{6inputsortingnetworks}

RCube uses 15 two-face comparisons. The optimal sorting network on 6 inputs (one per face) has 12 comparators; that no sorting network on 6 inputs with fewer comparators exists was proved by Floyd and Knuth~\cite{floyd1973bose} (see also~\cite[Section~5.3.4]{KnuthTAOCP3}). There exist many such sorting networks. For a two-face comparison to work, we need to define a decision rule. For RCube, this decision rule is to prioritize one color and move all of those cubies to one side. For our optimal sorting network introduced here, our decision rule is to sort high- and low-hierarchy colors onto higher- and lower-hierarchy faces equally, which will lead to a fully sorted cube at the end.
Figure~\ref{fig:optimal-sorting-network} shows one such optimal 6-input sorting network with 12 comparators. Our universal sequence is then $12 \times 8$ 3-cycles long, reducing the constant to $48$:

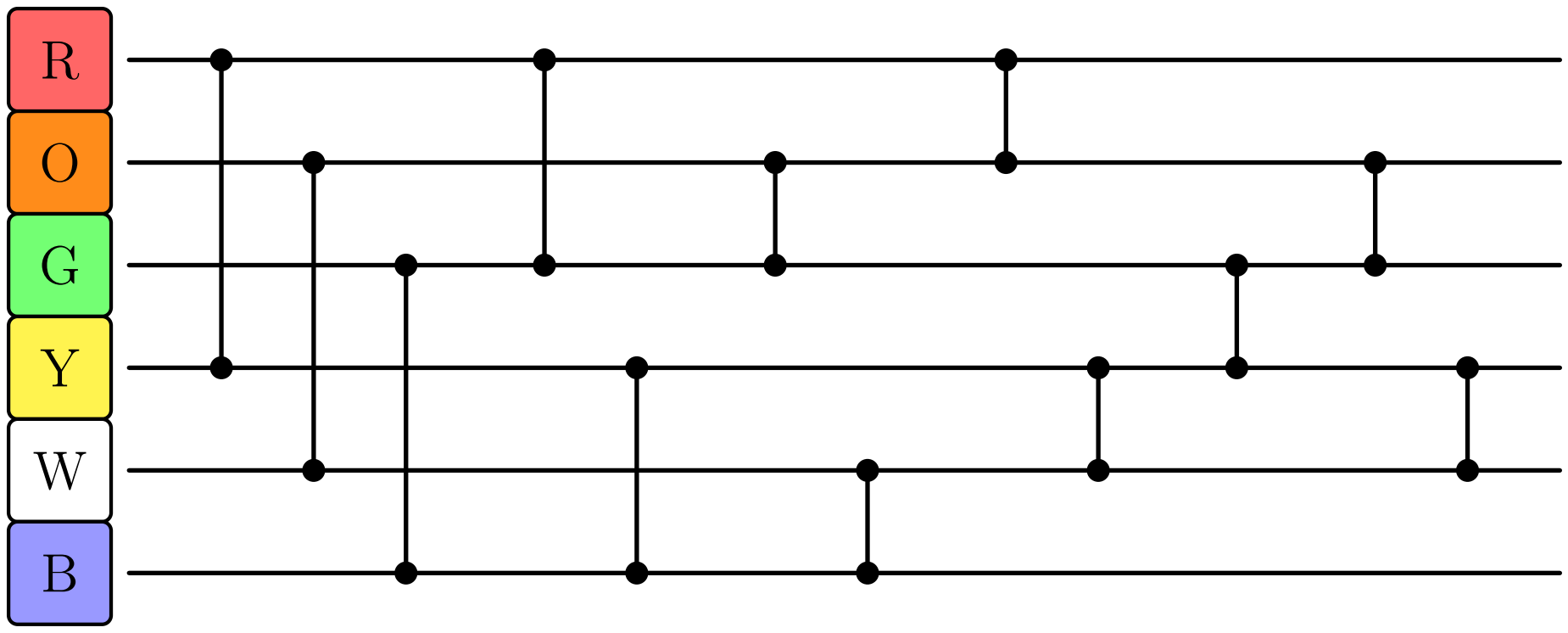
\begin{figure}[t]
    \centering
    \begin{tikzpicture}[x=0.9cm,y=0.8cm,line cap=round,line join=round]
        \tikzset{
            facebox/.style={draw=black, thick, rounded corners=2pt, minimum width=0.8cm, minimum height=0.8cm},
            wire/.style={black, line width=1pt},
            comparator/.style={black, line width=1pt},
            dot/.style={circle, fill=black, inner sep=1.8pt}
        }

        \foreach \name/\y/\bgcol in {R/5/faceR, O/4/faceO, G/3/faceG, Y/2/faceY, W/1/faceW, B/0/faceB} {
            \node[facebox, fill=\bgcol] at (-0.6,\y) {\large \name};
            \draw[wire] (0,\y) -- (12.4,\y);
        }

        \foreach \x/\yone/\ytwo in {
             0.8/5/2,
             1.6/4/1,
             2.4/3/0,
             3.6/5/3,
             4.4/2/0,
             5.6/4/3,
             6.4/1/0,
             7.6/5/4,
             8.4/2/1,
             9.6/3/2,
            10.8/4/3,
            11.6/2/1
        } {
            \draw[comparator] (\x,\yone) -- (\x,\ytwo);
            \node[dot] at (\x,\yone) {};
            \node[dot] at (\x,\ytwo) {};
        }

    \end{tikzpicture}
    \caption{A 6-input, 12-comparator sorting network, as used by the sortnet96 algorithm. The faces are shown top to bottom; each vertical connector is a comparator between two faces.}
    \label{fig:optimal-sorting-network}
\end{figure}
\[
\begin{aligned}
F_{\mathrm{move count}}
  &= 12 \cdot 8 \cdot \frac{1}{2} \cdot \frac{n^2}{\lg n} + o\!\left(\frac{n^2}{\lg n}\right) = 48 \cdot \frac{n^2}{\lg n} + o\!\left(\frac{n^2}{\lg n}\right).
\end{aligned}
\]
Expanding each of the 12 comparators of the optimal sorting network into the 8 3-cycles of the permutation network gives all $96$ 3-cycles of the universal sequence on 24 wires; we call this algorithm \defn{sortnet96}. The full combined network is shown in Figure~\ref{fig:full-combined-network} in Appendix~\ref{app:networks}.

\subsection{Upper bound using two sequential permutation networks}

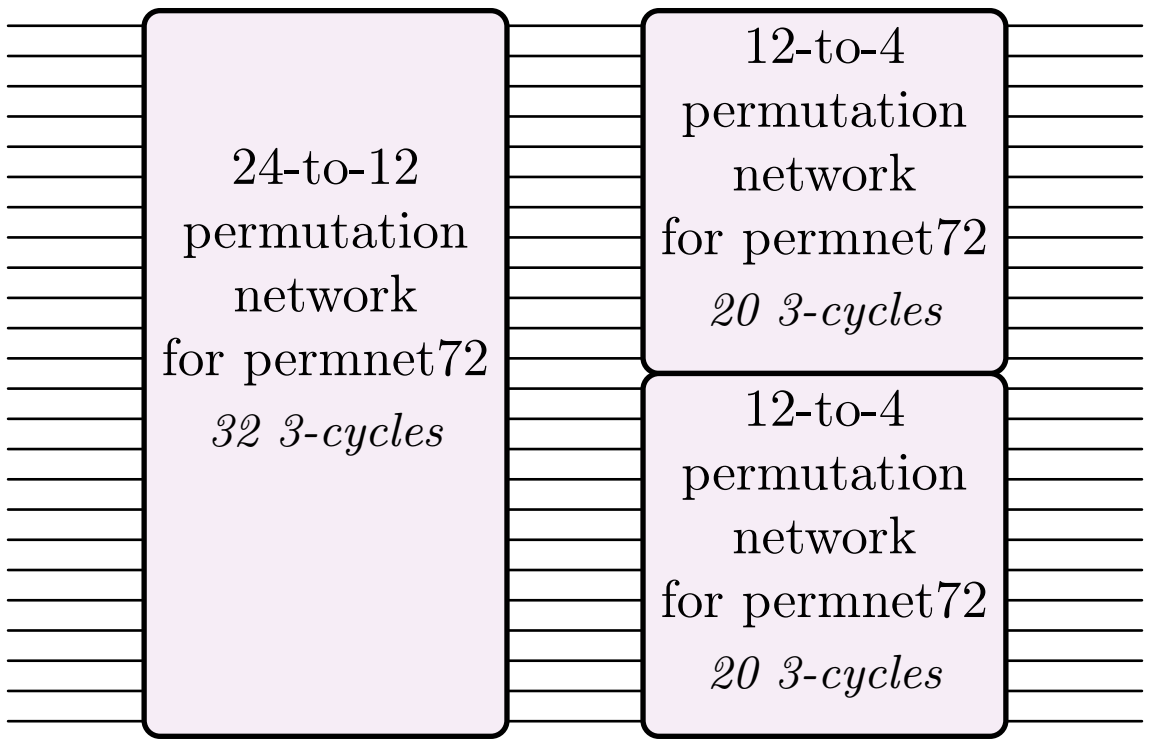
\begin{figure}[t]
  \centering
  \begin{tikzpicture}[x=0.6cm,y=0.2cm,line cap=round,line join=round]
    \tikzset{
      wire/.style={black, line width=0.5pt},
      netbox/.style={draw=black, line width=1pt, fill=networkFill, rounded corners=3pt}
    }
    \foreach \i in {1,...,24} {
      \draw[wire] (0,\i) -- (1.5,\i);
    }
    \draw[netbox] (1.5,0.5) rectangle (5.5,24.5);
    \node[align=center, font=\small] at (3.5,16) {24-to-12\\permutation\\network\\for permnet72};
    \node[align=center, font=\footnotesize\itshape] at (3.5,10.5) {32 3-cycles};
    \foreach \i in {1,...,24} {
      \draw[wire] (5.5,\i) -- (7,\i);
    }
    \draw[netbox] (7,12.5) rectangle (11,24.5);
    \node[align=center, font=\small] at (9,20) {12-to-4\\permutation\\network\\for permnet72};
    \node[align=center, font=\footnotesize\itshape] at (9,14.5) {20 3-cycles};
    \draw[netbox] (7,0.5) rectangle (11,12.5);
    \node[align=center, font=\small] at (9,8) {12-to-4\\permutation\\network\\for permnet72};
    \node[align=center, font=\footnotesize\itshape] at (9,2.5) {20 3-cycles};
    \foreach \i in {1,...,24} {
      \draw[wire] (11,\i) -- (12.5,\i);
    }
  \end{tikzpicture}
  \caption{Structure of the two sequential permutation networks of the permnet72 algorithm. All 24 wires first pass through the 24-to-12 permutation network (32 3-cycles), then split into two groups of 12, each passing through a 12-to-4 permutation network (20 3-cycles).}
  \label{fig:two-perm-networks}
\end{figure}

By increasing the size of cubie groups and thereby decreasing both the number of groups and the input size of the sorting networks, we further improve the leading constant. Ultimately, the lowest-order arrangement splits the full 24 cubies into two groups of 12 and then sorts each group into sets of 4.
Using these \defn{24-to-12} and \defn{12-to-4} permutation networks, we can further reduce the constant. Via hill climbing with random restarts, we find universal sequences of length 32 3-cycles for the 24-to-12 permutation network and length 20 3-cycles for the 12-to-4 permutation network, leading to a total number of 3-cycles of
\[
  L = 32+2\cdot20 = 72 .
\]
Figure~\ref{fig:two-perm-networks} illustrates the structure of the two sequential permutation networks. We call this algorithm \defn{permnet72}. The full combined network---the 32 3-cycles of the 24-to-12 permutation network on all 24 wires, followed by the 20 3-cycles of each 12-to-4 permutation network---is shown in Figure~\ref{fig:full-perm-network} in Appendix~\ref{app:networks}.

\begin{lemma}[permnet72]
\label{lem:permnet72}
The permnet72 universal sequence of $L = 32 + 2\cdot 20 = 72$ 3-cycles, each with $+2/+2$
parallelization behavior, selectively solves every center orbit and therefore solves all
centers of an $n \times n \times n$ Rubik's Cube in
$36\,n^2/\lg n + o(n^2/\lg n)$ moves; that is, permnet72 has leading constant $\mathbf{36}$.
\end{lemma}
\begin{proof}
The $24\to12{+}12$ network followed by the two $12\to4{+}4{+}4$ networks reduces any orbit
to the solved configuration; every configuration is selectively solved by some subsequence,
which we verify exhaustively (the $8$-to-$4$ building block in
Table~\ref{tab:permnet8-cases}, and the full networks by the validation scripts of
Section~\ref{introduction}). Each of the $72$ 3-cycles has a $+2/+2$ representative by
Table~\ref{tab:3cycle-groups}. Applying Lemma~\ref{lem:stageconstant} with $L = 72$ and
$X = 2$ gives leading constant $72/2 = 36$.
\end{proof}

\noindent\textbf{A universal sequence for the 24-to-12 permutation network:}
\[
\boxed{
\begin{aligned}
&(10,17,13)\;(1,12,17)\;(4,13,21)\;(6,15,21)\;(11,22,12)\;(10,14,18)\;(6,22,14)\;(11,19,18)\\
&(9,22,18)\;(9,12,15)\;(2,14,15)\;(8,13,19)\;(8,22,17)\;(6,19,12)\;(8,21,20)\;(11,13,16)\\
&(11,12,17)\;(11,14,19)\;(10,22,14)\;(2,17,21)\;(4,17,14)\;(1,14,13)\;(6,12,20)\;(10,15,16)\\
&(7,17,13)\;(1,17,18)\;(11,17,23)\;(8,17,15)\;(9,14,15)\;(0,21,12)\;(5,12,15)\;(3,23,15)
\end{aligned}
}
\]

\noindent\textbf{A universal sequence for the 12-to-4 permutation network:}
\[
\boxed{
\begin{aligned}
&(0,1,11)\;(2,3,11)\;(0,7,6)\;(1,6,2)\;(1,7,4)\;(6,11,10)\;(2,3,6)\;(1,2,5)\;(0,8,9)\;(4,10,11)\\
&(6,7,9)\;(5,9,8)\;(0,1,8)\;(2,9,10)\;(4,8,9)\;(2,3,8)\;(0,10,11)\;(2,3,6)\;(1,7,2)\;(0,4,5)
\end{aligned}
}
\]

\subsection{Lower bound constant}
\label{lower}

We derive the lower-bound constant from the state size and branching factor of the $n \times n \times n$ Rubik's Cube.
For even $n$, the cubies whose count scales as $O(n^2)$ are the non-diagonal, non-cross-center cubies. There are $\bigl(\tfrac{n}{2}-1\bigr)^2$ orbits with $\frac{24!}{(4!)^6}$ configurations each. So the leading term of the state size of the $n \times n \times n$ Rubik's Cube is
\[
  \left(\frac{24!}{(4!)^6}\right)^{\!\left(\frac{n}{2}-1\right)^2}.
\] In Slice Turn Metric the $n \times n \times n$ Rubik's Cube has $9n$ moves ($3$ axes, $n$ layers, $3$ turns each); turning the layer just turned again is never useful, so after the first move at most $9n-3$ moves are useful, and we use $9n-3$ as the branching factor (the exact effective value does not affect the leading constant). This counting argument implies that the number of states reachable in at most $L$ moves is at most
\[
  1 + 9n\sum_{j=0}^{L-1}(9n-3)^j \;=\; 1 + \frac{9n\bigl((9n-3)^L-1\bigr)}{9n-4}
  \;\le\; \frac{9n}{9n-4}\,(9n-3)^L ,
\]
so the lower bound $L$ satisfies
\[
  \frac{9n}{9n-4}\,(9n-3)^L \;\geq\; \left(\frac{24!}{(4!)^6}\right)^{\!\left(\frac{n}{2}-1\right)^2},
\]
which gives
\[
  L \;\geq\; \frac{\left(\dfrac{n}{2}-1\right)^2 \cdot \lg\dfrac{24!}{(4!)^6}
                   \;-\; \lg\dfrac{9n}{9n-4}}{\lg(9n-3)}.
\]
Denoting the right-hand side by $f(n)$, the lower bound constant is determined by
\[
  \lim_{n\to\infty} \frac{f(n)}{n^2/\lg n}
  \;=\; \lim_{n\to\infty}
    \frac{\left(\dfrac{n}{2}-1\right)^{\!2} \cdot \lg\!\dfrac{24!}{(4!)^6} \cdot \lg n}
         {\lg(9n-3) \cdot n^2}
  \;=\; \frac{1}{4} \lg\!\frac{24!}{(4!)^6} \;\approx\; 12.88.
\] We call this constant $C_{\mathrm{lb}} \approx 12.88$, so that $L \geq C_{\mathrm{lb}} \cdot \frac{n^2}{\lg n}$ asymptotically.

\subsection{Introducing permutation move sequences beyond 3-cycles}
\label{beyond3cycles}
{\color{black}
So far we have permuted the pieces of an orbit using only 3-cycles, and we showed
(Table~\ref{tab:3cycle-groups}) that every 3-cycle type we use admits a representative
with $+2/+2$ parallelization behavior. Now we allow move sequences that permute more than
three pieces at once, such as $3{+}3$ double 3-cycles and $5$-cycles; most of these are
easily recognized as commutator variations. As with 3-cycles, each candidate must pass a
set of conditions that reject sequences disturbing pieces in other orbits or interfering
across the diagonal center orbits. A single such sequence can realize a larger permutation
per stage, so that the universal sequence needs fewer stages overall---and, as we argue next,
without changing the leading-term factor.

Crucially, Lemma~\ref{lem:stageconstant} already covers this case: it applies to stages
with any constant $+2/+X$ behavior, not just $+2/+2$ 3-cycles. The leading constant is
therefore half the number of stages regardless of how large a permutation each stage
performs, so reducing the number of stages---by letting each stage perform a larger
permutation---lowers the constant directly.

Applying this idea to the two sequential permutation networks of
Figure~\ref{fig:two-perm-networks}, multi-piece sequences carry out the same
$24\to12{+}12$ and $12\to4{+}4{+}4$ splits as permnet72, but the $24\to12{+}12$ network now
uses only $26$ stages (down from $32$), while each $12\to4{+}4{+}4$ network uses $18$
stages (down from $20$). The total universal sequence therefore has
\[
  L = 26 + 2 \cdot 18 = 62
\]
stages. We call this algorithm \defn{permnet62}; by Lemma~\ref{lem:stageconstant}
its leading constant is $62/2 = 31$.
The search for the $24\to12{+}12$ network is computationally expensive, but the resulting
move sequences are easily verified on a $6\times6\times6$ Rubik's Cube or a virtual cube
simulator.

The two $12\to4{+}4{+}4$ networks of permnet62 are closely related: one is obtained from the
other by relabeling the face letters of its moves to act on the respective three-face
group. Tables~\ref{tab:permnet62-2412} and~\ref{tab:permnet62-124a}
(Appendix~\ref{app:permnet62-tables}) list, for each stage of the
$24\to12{+}12$ network and of the first $12\to4{+}4{+}4$ network, the representative move
sequence together with the permutation it induces on the orbit (written as a product of
disjoint cycles on the wire labels $0$--$23$). We omit the second $12\to4{+}4{+}4$
network, as it is the face-relabeled mirror of the first.

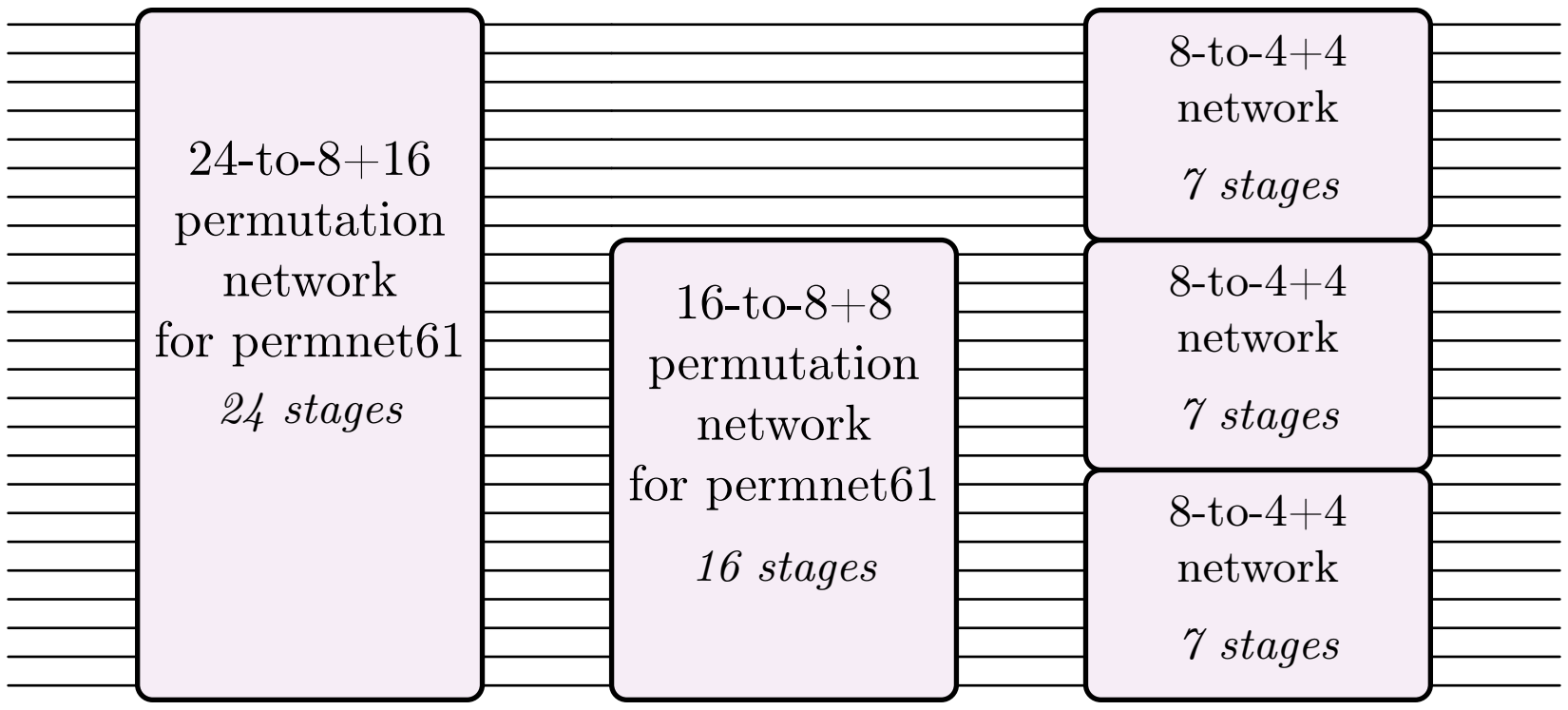
\begin{figure}[t]
  \centering
  \begin{tikzpicture}[x=0.6cm,y=0.2cm,line cap=round,line join=round]
    \tikzset{
      wire/.style={black, line width=0.5pt},
      netbox/.style={draw=black, line width=1pt, fill=networkFill, rounded corners=3pt}
    }
    \foreach \i in {1,...,24} { \draw[wire] (0,\i) -- (1.5,\i); }
    \draw[netbox] (1.5,0.5) rectangle (5.5,24.5);
    \node[align=center, font=\small] at (3.5,16) {24-to-8+16\\permutation\\network\\for permnet61};
    \node[align=center, font=\footnotesize\itshape] at (3.5,10.5) {24 stages};
    \foreach \i in {1,...,24} { \draw[wire] (5.5,\i) -- (7,\i); }
    \draw[netbox] (7,0.5) rectangle (11,16.5);
    \node[align=center, font=\small] at (9,11) {16-to-8+8\\permutation\\network\\for permnet61};
    \node[align=center, font=\footnotesize\itshape] at (9,5) {16 stages};
    \foreach \i in {17,...,24} { \draw[wire] (7,\i) -- (12.5,\i); }
    \foreach \i in {1,...,16} { \draw[wire] (11,\i) -- (12.5,\i); }
    \foreach \lo/\hi in {0.5/8.5, 8.5/16.5, 16.5/24.5} {
      \draw[netbox] (12.5,\lo) rectangle (16.5,\hi);
    }
    \foreach \y in {4.5,12.5,20.5} {
      \node[align=center, font=\footnotesize] at (14.5,\y+1.6) {8-to-4+4\\network};
      \node[align=center, font=\footnotesize\itshape] at (14.5,\y-2.2) {7 stages};
    }
    \foreach \i in {1,...,24} { \draw[wire] (16.5,\i) -- (18,\i); }
  \end{tikzpicture}
  \caption{Structure of the permutation networks of the permnet61 algorithm. All 24 wires
  first pass through the 24-to-8+16 permutation network (24 stages). The resulting group of
  16 then passes through the 16-to-8+8 permutation network (16 stages), while the group of 8
  passes through unchanged. Each of the three resulting groups of 8 finally passes through an
  8-to-4+4 permutation network (7 stages), giving $24 + 16 + 3\cdot 7 = 61$ stages in total.}
  \label{fig:permnet61-networks}
\end{figure}

The permutation network approach can be improved further by expanding our search to include
different decompositions into permutation networks. We performed a more general search,
including 24-to-12+12, 24-to-8+16, 16-to-8+8, 16-to-4+4+4+4, 12-to-4+4+4,
and 8-to-4+4 networks across all face configurations, attempting to find the best permutation
networks for each decomposition, and then using a simple dynamic programming approach
to optimize network selection for each individual cube size.  Our proof of concept
solver, brobdicube, performs this optimization at runtime from a library of algorithm sets.

The best network we found asymptotically when optimizing for the worst-case
move count we call
\defn{permnet61}.
This variant first splits the $24$ orbit positions into a group of $8$ and
a group of $16$ using a $24\to8{+}16$ network of $24$ stages, then splits the $16$-group
into two groups of $8$ using a $16\to8{+}8$ network of $16$ stages, and finally applies a
$7$-stage $8\to4{+}4$ network to each of the three resulting groups of $8$. The total
universal sequence therefore has
\[
  L = 24 + 16 + 3 \cdot 7 = 61
\]
stages. Figure~\ref{fig:permnet61-networks} shows this architecture, and the corresponding
move sequences are listed in Tables~\ref{tab:permnet61-2416}--\ref{tab:permnet61-844}
(Appendix~\ref{app:permnet61-tables}), each together with the permutation it induces on the
orbit.

We have also found other 61-stage architectures, for example one that first separates an
opposite pair of faces from the remaining ring of four faces, then splits that pair, and
finally splits all four ring faces at once ($24 + 7 + 30$ stages), and one that needs only
two phases ($31 + 30$ stages).  These algorithms perform slightly worse when
optimizing for  the worst-case behavior, but slightly better when optimizing
for average-case behavior. For brevity, we
focus on the one above and give only its move
sequences, but all data is available in our repository.\footnote{\url{https://github.com/rokicki/brob-paper-data}}

\begin{lemma}[permnet61]
\label{lem:permnet61}
The permnet61 universal sequence of $L = 24 + 16 + 3\cdot 7 = 61$ stages, each a multi-piece move
sequence with $+2/+X$ parallelization behavior for a constant $X$, selectively solves every
center orbit and therefore solves all centers of an $n \times n \times n$ Rubik's Cube in
$30.5\,n^2/\lg n + o(n^2/\lg n)$ moves; that is, permnet61 has leading constant $\mathbf{30.5}$.
This is the smallest leading constant achieved in this paper.
\end{lemma}
\begin{proof}
Like permnet72 and permnet62, the $24\to8{+}16$ network, followed by the $16\to8{+}8$ network
and by the $8\to4{+}4$ network applied to each of the three resulting groups of $8$, reduces
any orbit to solved; every configuration is selectively solved by some subsequence, verified exhaustively by the validation scripts of Section~\ref{introduction} (the move
sequences are listed in Tables~\ref{tab:permnet61-2416}--\ref{tab:permnet61-844} and are
checkable on a $6\times6\times6$ cube). Each of the $61$ multi-piece stages has $+2/+X$
behavior for a constant $X$. Applying Lemma~\ref{lem:stageconstant} with $L = 61$ gives
leading constant $61/2 = 30.5$.
\end{proof}

Because permnet61 lowers the leading constant to $30.5$, the ratio to the lower bound
$C_{\mathrm{lb}} \approx 12.88$ (Section~\ref{lower}) improves accordingly:
\[
  \lim_{n\to\infty}
  \frac{F_{\mathrm{upper}}(n)}{F_{\mathrm{lower}}(n)}
  \;=\; \frac{30.5}{C_{\mathrm{lb}}}
  \;\approx\; 2.368.
\]
So as $n \to \infty$, in the worst case permnet61 is within $2.368$ times the lower bound.
}

\subsection{Solving non-center pieces}
\label{noncenter}
\new{
Solving the wing cubies ($12(n-2)$ in total for even $n$) and corners ($8$ cubies) does not affect the asymptotic leading constant, because only the $\Theta(n^2)$
center cubies contribute to the leading term (Section~\ref{algorithmicanalysis}).
We nonetheless give algorithms for them---wings in Section~\ref{wings}, corners (and, for odd $n$, middle edges)
in Section~\ref{midedges}---together with an upper-bound model, so that we can
report total worst-case move counts for $n \times n \times n$ Rubik's Cubes of various
sizes (Table~\ref{tab:discrete-ub} in Appendix~\ref{app:discrete}). For even $n$, our solver first solves the corners, then the wing cubies, and finally the centers (Sections~\ref{upper_n2}--\ref{2D_par}),
whose solution dominates the total move count.
}


\subsection{Solving wing cubies}
\label{wings}
{\color{black}
As with the center-solving algorithms of Section~\ref{upper_n2}, there exist
3-cycle algorithms for wing cubies that can be 1D-parallelized. Unlike center
cubies---where four cubies of an orbit share each color---every wing cubie within
an orbit is distinguishable. Hence an orbit of $24$ wings can have
$\binom{24}{3}\cdot 2 = 4048$ distinct 3-cycle cases.

Our solver iterates over the $24$ positions of an orbit in Speffz order
(Figure~\ref{fig:Speffz}) and never revisits a solved position, analogously to the greedy
center solver of Section~\ref{greedy}. Let $i$ be the lowest-index unsolved position. We choose the
three targets of the next 3-cycle so that position~$i$ becomes solved: the second
target is $i$ itself, the first is the current location of the piece that belongs
at~$i$, and the third is the position where the piece currently occupying~$i$
belongs. When this would form a 2-cycle (two pieces are simply swapped, so a single 3-cycle cannot place both), we instead borrow a piece from another even-length cycle, which exists because the orbit permutation is even after the parity fix. Such a borrow step places only position~$i$ and merges the other piece of the 2-cycle with the borrowed cycle into a single cycle of odd length. Every other step places two pieces of the cycle containing~$i$, or all three if that cycle has length~$3$, and so preserves the parity of the cycle's length. To bound the number of operations, we charge them to the cycles of the initial permutation. A cycle of odd length~$\ell$ is never borrowed from and needs $(\ell-1)/2$ operations. Cycles of even length are merged in pairs by borrow steps: if a pair has lengths $\ell_1$ and $\ell_2$, one of them is reduced to a 2-cycle in $(\ell_1-2)/2$ operations and the other to some even length $m \ge 2$ in $(\ell_2-m)/2$ operations, one borrow step merges them into a cycle of length $m+1$, and that cycle is then completed in $m/2$ operations, for $(\ell_1+\ell_2)/2$ operations in total, regardless of when these operations occur. Summing over all cycles, each orbit needs at most $24/2 = 12$ 3-cycles; the bound is attained when the orbit consists of $12$ transpositions.

Because the lowest unsolved position is always solved and, after canonicalization
(lowest index first), remains the lowest target, the resulting universal sequence of
wing 3-cycles is sorted and has length $4048$. By brute-force search we computed,
for each of the $4048$ cases, a 3-cycle algorithm that uses only $2$ slice moves,
so that 1D-parallelization adds exactly $+2$ moves per additional orbit. The $4048$ move sequences are, however, generated by expanding a set of $87$ base wing
move sequences by mirroring, rotating, inverting, and applying conjugation moves; the
base sequences are listed in Table~\ref{tab:wing-base-algs} in Appendix~\ref{app:wings}.

A more refined upper-bound model is derived in Appendix~\ref{app:wings}; because we
focus on asymptotics, the model we present here is kept deliberately simple and consists
of three parts. Fixing parity costs one quarter-turn per wing orbit, i.e.,\ $r =
\tfrac{n}{2}-1$ moves over the $r$ wing orbits (cf.\ the parity discussion at the
beginning of Section~\ref{algorithmicanalysis} and \cite{DemaineEtAlRubik}). During the
\emph{non-saturated} phase of wing solving, we charge each orbit with at most $12$ move
sequences of maximum length $13$, i.e., at most
\[
  13 \cdot 12 \, r \;=\; 156\,r
\]
moves. Once roughly $4048/12 \approx 338$ orbits are reached, we enter the
\emph{saturated} phase: every one of the $4048$ move sequences is charged at most once at
full length, and each of the at most $12$ 3-cycles of every further orbit is accommodated
with $2$ additional slice moves. The cost is then the total length of the universal sequence
without its slice moves, $\sum_c(\ell_c - 2) = 28{,}592$ (Appendix~\ref{app:wings}),
plus $24\,r$ inserted slice moves, i.e., $24\,r + 28{,}592$. Together, the worst-case
wing move count is at most
\[
  r + \min\bigl(156\,r,\; 24\,r + 28{,}592\bigr).
\]
}

\subsection{Solving middle edges and corners}
\label{midedges}
{\color{black}
For even $n$ there are no middle edges, and the corners reduce to a $2 \times 2 \times 2$
cube. For odd $n$, after adjusting for slice parity, the middle edges and corners reduce to
a $3 \times 3 \times 3$ cube, which is solvable in at most $20$
moves~\cite{rokicki2014diameter}. Rather than solving them jointly at this optimum, our
solver treats the two groups separately, for simplicity of implementation and runtime,
using Kociemba's two-phase algorithm with upper bounds of $14$ moves for the corners and,
for odd $n$, $11$ moves for the middle edges. Our implementation therefore uses at most
$14$ moves for this stage when $n$ is even and at most $11 + 14 = 25$ moves when $n$ is
odd, deliberately trading up to five extra moves for speed; the worst-case bound of
Section~\ref{discreteub} charges $20$ moves in both cases, which suffices because the
reduced cube is a $2 \times 2 \times 2$ or, solved jointly, a $3 \times 3 \times 3$.
Both counts are $O(1)$ and hence, like the wings, do not affect the leading term.
}

\subsection{Discrete worst-case upper bound for $n \times n \times n$ Rubik's Cubes up to $n=100$}
\label{discreteub}
{\color{black}
Combining the pieces above yields a concrete worst-case upper bound on the number of
moves to solve an entire $n \times n \times n$ Rubik's Cube. For even $n$, the total is
\[
  \mathrm{total\_ub}(n) \;=\; \underbrace{C(n)}_{\text{centers}}
  \;+\; \underbrace{\mathrm{wings\_ub}(n)}_{\text{wings, Eq.~\eqref{eq:wings-ub}}}
  \;+\; \underbrace{20}_{\text{corners}} ,
\]
where $C(n)$ is the worst-case center move count. For each $n$, $C(n)$ is produced by
our solver, which selects the optimal discrete strip widths---one or two widths differing
by one, because the per-strip cost is convex in the width (Appendix~\ref{choicew}). The wing
term $\mathrm{wings\_ub}(n)$ includes the $\tfrac{n}{2}-1$ parity slice moves, and the
corners reduce to a $2\times2\times2$ cube; we conservatively retain the $20$-move
allowance of the $3\times3\times3$ bound (Section~\ref{midedges}), which also covers the middle edges for odd~$n$. Table~\ref{tab:discrete-ub} in Appendix~\ref{app:discrete}
lists $\mathrm{total\_ub}(n)$ and its breakdown for all even $n \le 100$.
}

\subsection{Running time and space complexity}
\label{complexity}
Our best fixed algorithm, permnet61, uses a universal sequence of $61$ stages and therefore has a leading constant of $30.5$. Now we analyze its running time and space; the analysis applies equally to brobdicube, which selects among a finite library of such networks for each cube size.
The algorithm runs in linear time in the input size. Determining which stages each scrambled orbit participates in (subsequence selection) runs in $O(1)$ time per orbit. For each stage, the corresponding combinatorial rectangles are retrieved and processed in a single pass over the strips, and more specifically over the columns within each strip. Because each element of the cube state is visited a constant number of times, the overall runtime is $O(N)$, where $N$ denotes the size of the cube state (equivalently, $N = \Theta(n^2)$ for cube dimension $n$).

The solver runs stripwise, keeping the extra memory usage (beyond the cube state itself) low. The size of a strip is $n \cdot w = \sqrt{N} \cdot O(\lg n) = O(\sqrt{N}\,\lg N)$, which therefore also bounds the extra space used.

\section{Experimental Results}
\label{experimentalresults}

We include a video of $n=256$ and $n=65{,}536$ solves at the link \url{https://largecubebench.org}, with eight evenly spaced frames of the $n=256$ solve shown in Figure~\ref{fig:video}.

\begin{figure}[t]
  \centering
  \includegraphics[width=\textwidth]{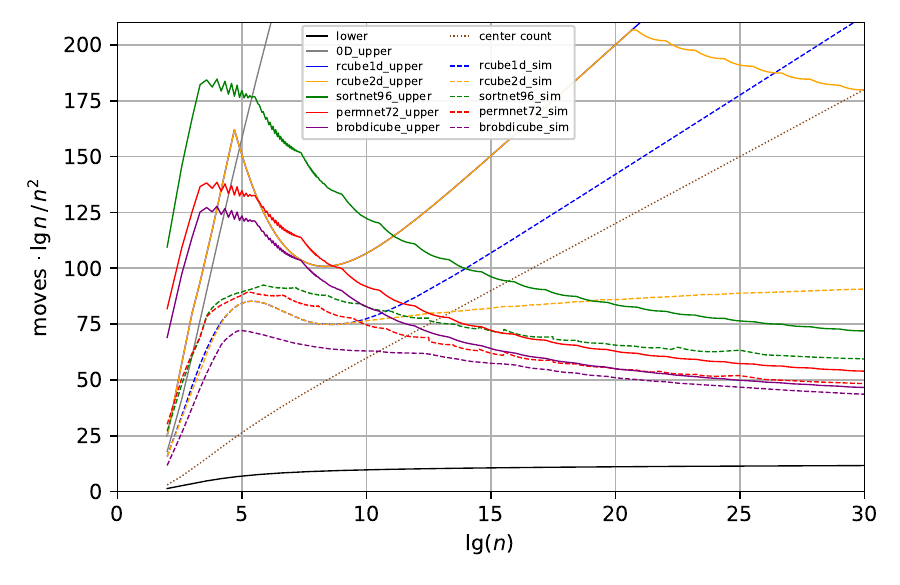}
  \caption{This graph shows how the number of center moves (wings, middle edges and corners excluded) scales for large $n$, normalized by $n^2 / \lg n$. Solid lines show worst-case upper bounds; dashed lines show simulated average-case results; the dotted line is the number of center cubies, $6(n-2)^2$, normalized the same way.}
  \label{fig:main-graph}
\end{figure}

We implemented the algorithms described above and evaluated them experimentally on cube sizes from $n=4$ up to $n=10^9$. For smaller cubes (below roughly $n=1024$) we took numerous samples to obtain a valid average; above $n=1024$, a single run suffices for all algorithms, because the many center orbits are each independently shuffled. Above $n=65{,}536$, we solved only a single strip of width $w$ (at the best value of $w$) and extrapolated that result to the full cube.
For permnet72 we derived an exact formula for the expected-case center move count, validated it against the experimental runs, and used it to generate the reported data. This formula is a refinement of the worst-case analysis presented in this paper, using exact probability calculations for each essential and additional vector (Section~\ref{strip2d}) together with the discrete strip-width optimization of Appendix~\ref{choicew}. For permnet61 the reported averages are measured rather than modeled: whole cubes up to $n = 4096$ and single strips up to $n = 2^{24}$. Beyond that range we use the analogous expected-cost formula derived for permnet61, which agrees with the measured expected center move count to within a relative error of $3 \cdot 10^{-6}$.
To our knowledge, the only existing solver that can reasonably solve cubes beyond roughly $n=256$ is RCube~\cite{RCubeGitHub}, which we reimplemented (RCube1D); other $n \times n \times n$ solvers rely on IDA* search and already require hours of CPU time at $n=256$.

Tables~\ref{tab:maketable5}, \ref{tab:maketable5b}, and~\ref{tab:maketable6} in Appendix~\ref{appendix} present the move-count results of the implemented solvers (RCube1D, RCube2D, sortnet96, permnet72, permnet61).

All of our calculated upper bounds are above our experimental data, which is consistent with the models. Furthermore, in Figure~\ref{fig:main-graph}, it can be seen that the 2D solvers' simulation and upper bound asymptotically match, which confirms our hypothesis that for $n \to \infty$, all combinatorial rectangle patterns come up and there are few essential vectors in a strip of cubies, leading to matching results.
For low $n$, we see that using no parallelization (0D\_upper) initially yields a lower upper move count than RCube (1D), due to its worst case being strongly limited by the number of 3-cycles performed per orbit. For RCube, the worst-case estimates were refined to take into account that for small $n$ only very few of the $240$ stages of its schedule ($15$ two-face swaps of $16$ 3-cycles each) are needed for each orbit; this refinement explains why the RCube upper-bound curves start low, rise sharply for low $n$, and peak at about $n=21$. On the $n^2/\lg n$ scale of Figure~\ref{fig:main-graph}, the upper bounds of the 2D solvers do not start at their highest values: they begin at moderate move counts per cubie, rise to a maximum at around $n = 16$ to $32$, and only then decrease toward their asymptotes. The maximum arises because the upper bound has to account for the worst case, which has far more cases than those that actually occur, while the benefit of parallelization grows only slowly with $n$.
There exist cases where many 3-cycles need to be applied from the universal sequence. In the 2D-variants, the upper bounds decrease toward their asymptote as parallelization increases.
Because the scale is normalized by $n^2 / \lg n$, the difference between RCube1D and our own variant RCube2D is visible in both the upper bound and the simulation. While RCube1D does not manage to stabilize toward a constant move count per cubie on an $n^2 / \lg n$ scale, the 2D variant can. The RCube 2D-variant's upper bound and simulation approach the $y = 120$ asymptote, with the upper bound approaching from above and the simulation from below. For the solvers sortnet96, permnet72 and permnet61 the upper bound calculation approaches the curve of the simulation result quickly as $n \to \infty$, and they approach their horizontal asymptotes of $y=48$, $y=36$ and $y=30.5$ respectively. The approach to the last of these is slow: at $n = 2^{30}$ the permnet61 curve is still at $46.6$ on this scale.

Our algorithms establish new tighter upper bounds for God's number on the
$n \times n \times n$ cube for every even $n \geq 22$: permnet61---our best solver---sets
the tightest bounds throughout that range, while for $n \leq 20$ the 0D approach still
gives the smallest bound (Tables~\ref{tab:maketable5} and~\ref{tab:maketable6}).

\section{Conclusion and Future Work}
\label{conclusionandfuturework}

We presented a new Rubik's Cube algorithm with asymptotic move count $30.5 \cdot \frac{n^2}{\lg n}$, within a factor of about $2.37$ of the lower-bound constant $\approx 12.88$.
A key practical advantage is that the algorithm parallelizes well, enabling fast solves for large $n$. In contrast, BFS, DFS, and A*~\cite{higuita2025algorithmic} can be useful for exhaustive search, but do not scale well to large $n \times n \times n$ cubes.
Future work could improve the leading constant by exploring other mappings of cubie groups into sorting networks, beyond the current six 4-cubie groups, and by developing better 3-cycle algorithms and permutation networks. The same framework also raises questions for $n \times n \times n$ Megaminx and Pyraminx variants, as well as 4D and 5D hypercubes. More broadly, our use of sorting and permutation networks with 3-cycles as the basic operation may apply to other combinatorial algorithms.

\section*{Acknowledgments}

We thank Lucas Garron for developing Twizzle Explorer, which we used throughout
the algorithm development process, and for helpful comments on the manuscript.
We thank Herbert Kociemba for helpful comments on the analysis and the
experiments. We also thank Elias Fischer for proofreading.

\bibliography{main}
\appendix

\section{Omitted Tables}
\label{appendix}

\subsection{Lower Bound and Upper Bound Move counts for different solvers (centers only)}
\label{maketable5a}

See Table~\ref{tab:maketable5}.

\begin{table}[t]
  \centering
  \setlength{\tabcolsep}{4pt}
  \scriptsize

  \begin{tabular}{r| r| r r r r r r}
    \hline
    $n$
    & lower
    & \shortstack{0D\\upper}
    & \shortstack{rcube1d\\upper}
    & \shortstack{rcube2d\\upper}
    & \shortstack{sortnet96\\upper}
    & \shortstack{permnet72\\upper}
    & \shortstack{permnet61\\upper} \\
    \hline
    4 & 11 & \cellcolor{green!25}144 & 200 & 200 & 876 & 655 & 552 \\
    8 & 76 & \cellcolor{green!25}$1.30\cdot10^{3}$ & $1.70\cdot10^{3}$ & $1.70\cdot10^{3}$ & $3.56\cdot10^{3}$ & $2.67\cdot10^{3}$ & $2.43\cdot10^{3}$ \\
    16 & 354 & \cellcolor{green!25}$7.06\cdot10^{3}$ & $8.10\cdot10^{3}$ & $8.10\cdot10^{3}$ & $1.18\cdot10^{4}$ & $8.86\cdot10^{3}$ & $8.18\cdot10^{3}$ \\
    32 & $1.42\cdot10^{3}$ & $3.24\cdot10^{4}$ & $3.09\cdot10^{4}$ & $3.09\cdot10^{4}$ & $3.68\cdot10^{4}$ & $2.76\cdot10^{4}$ & \cellcolor{green!25}$2.53\cdot10^{4}$ \\
    64 & $5.41\cdot10^{3}$ & $1.38\cdot10^{5}$ & $8.33\cdot10^{4}$ & $8.33\cdot10^{4}$ & $1.14\cdot10^{5}$ & $8.58\cdot10^{4}$ & \cellcolor{green!25}$7.84\cdot10^{4}$ \\
    128 & $2.01\cdot10^{4}$ & $5.72\cdot10^{5}$ & $2.50\cdot10^{5}$ & $2.50\cdot10^{5}$ & $3.59\cdot10^{5}$ & $2.69\cdot10^{5}$ & \cellcolor{green!25}$2.46\cdot10^{5}$ \\
    256 & $7.44\cdot10^{4}$ & $2.32\cdot10^{6}$ & $8.28\cdot10^{5}$ & $8.28\cdot10^{5}$ & $1.15\cdot10^{6}$ & $8.66\cdot10^{5}$ & \cellcolor{green!25}$7.90\cdot10^{5}$ \\
    512 & $2.75\cdot10^{5}$ & $9.36\cdot10^{6}$ & $2.97\cdot10^{6}$ & $2.97\cdot10^{6}$ & $3.88\cdot10^{6}$ & $2.91\cdot10^{6}$ & \cellcolor{green!25}$2.62\cdot10^{6}$ \\
    1,024 & $1.02\cdot10^{6}$ & $3.76\cdot10^{7}$ & $1.12\cdot10^{7}$ & $1.12\cdot10^{7}$ & $1.28\cdot10^{7}$ & $9.62\cdot10^{6}$ & \cellcolor{green!25}$8.68\cdot10^{6}$ \\
    2,048 & $3.81\cdot10^{6}$ & $1.51\cdot10^{8}$ & $4.33\cdot10^{7}$ & $4.33\cdot10^{7}$ & $4.39\cdot10^{7}$ & $3.29\cdot10^{7}$ & \cellcolor{green!25}$2.96\cdot10^{7}$ \\
    4,096 & $1.42\cdot10^{7}$ & $6.03\cdot10^{8}$ & $1.71\cdot10^{8}$ & $1.71\cdot10^{8}$ & $1.54\cdot10^{8}$ & $1.16\cdot10^{8}$ & \cellcolor{green!25}$1.03\cdot10^{8}$ \\
    8,192 & $5.34\cdot10^{7}$ & $2.41\cdot10^{9}$ & $6.77\cdot10^{8}$ & $6.77\cdot10^{8}$ & $5.39\cdot10^{8}$ & $4.04\cdot10^{8}$ & \cellcolor{green!25}$3.60\cdot10^{8}$ \\
    16,384 & $2.01\cdot10^{8}$ & $9.66\cdot10^{9}$ & $2.70\cdot10^{9}$ & $2.70\cdot10^{9}$ & $1.91\cdot10^{9}$ & $1.43\cdot10^{9}$ & \cellcolor{green!25}$1.28\cdot10^{9}$ \\
    32,768 & $7.61\cdot10^{8}$ & $3.86\cdot10^{10}$ & $1.08\cdot10^{10}$ & $1.08\cdot10^{10}$ & $6.89\cdot10^{9}$ & $5.17\cdot10^{9}$ & \cellcolor{green!25}$4.58\cdot10^{9}$ \\
    65,536 & $2.89\cdot10^{9}$ & $1.55\cdot10^{11}$ & $4.30\cdot10^{10}$ & $4.30\cdot10^{10}$ & $2.51\cdot10^{10}$ & $1.88\cdot10^{10}$ & \cellcolor{green!25}$1.66\cdot10^{10}$ \\
    131,072 & $1.10\cdot10^{10}$ & $6.18\cdot10^{11}$ & $1.72\cdot10^{11}$ & $1.72\cdot10^{11}$ & $9.15\cdot10^{10}$ & $6.86\cdot10^{10}$ & \cellcolor{green!25}$6.05\cdot10^{10}$ \\
    262,144 & $4.18\cdot10^{10}$ & $2.47\cdot10^{12}$ & $6.87\cdot10^{11}$ & $6.87\cdot10^{11}$ & $3.35\cdot10^{11}$ & $2.51\cdot10^{11}$ & \cellcolor{green!25}$2.21\cdot10^{11}$ \\
    524,288 & $1.60\cdot10^{11}$ & $9.90\cdot10^{12}$ & $2.75\cdot10^{12}$ & $2.75\cdot10^{12}$ & $1.24\cdot10^{12}$ & $9.28\cdot10^{11}$ & \cellcolor{green!25}$8.16\cdot10^{11}$ \\
    1,048,576 & $6.11\cdot10^{11}$ & $3.96\cdot10^{13}$ & $1.10\cdot10^{13}$ & $1.10\cdot10^{13}$ & $4.60\cdot10^{12}$ & $3.45\cdot10^{12}$ & \cellcolor{green!25}$3.02\cdot10^{12}$ \\

    \hline
  \end{tabular}
  \caption{Lower Bound and Upper Bound Move counts for different solvers (centers only)}
  \label{tab:maketable5}
\end{table}

\subsection{Move counts for different solvers (simulations)}
\label{maketable5b}

See Table~\ref{tab:maketable5b}.

\begin{table}[t]
  \centering
  \setlength{\tabcolsep}{4pt}
  \scriptsize

  \begin{tabular}{r| r r r r r}
    \hline
    $n$
    & \shortstack{RCube1D\\sim}
    & \shortstack{RCube2D\\sim}
    & \shortstack{sortnet96\\sim}
    & \shortstack{permnet72\\sim}
    & \shortstack{brobdicube\\sim} \\
    \hline
    4 & 126 & 126 & 217 & 241 & \cellcolor{green!25}94 \\
    8 & $1.01\cdot10^{3}$ & 963 & $1.29\cdot10^{3}$ & $1.32\cdot10^{3}$ & \cellcolor{green!25}801 \\
    16 & $4.66\cdot10^{3}$ & $4.59\cdot10^{3}$ & $5.40\cdot10^{3}$ & $5.27\cdot10^{3}$ & \cellcolor{green!25}$3.98\cdot10^{3}$ \\
    32 & $1.73\cdot10^{4}$ & $1.73\cdot10^{4}$ & $1.83\cdot10^{4}$ & $1.80\cdot10^{4}$ & \cellcolor{green!25}$1.48\cdot10^{4}$ \\
    64 & $5.72\cdot10^{4}$ & $5.72\cdot10^{4}$ & $6.29\cdot10^{4}$ & $6.00\cdot10^{4}$ & \cellcolor{green!25}$4.76\cdot10^{4}$ \\
    128 & $1.85\cdot10^{5}$ & $1.85\cdot10^{5}$ & $2.13\cdot10^{5}$ & $2.01\cdot10^{5}$ & \cellcolor{green!25}$1.56\cdot10^{5}$ \\
    256 & $6.17\cdot10^{5}$ & $6.17\cdot10^{5}$ & $7.23\cdot10^{5}$ & $6.81\cdot10^{5}$ & \cellcolor{green!25}$5.30\cdot10^{5}$ \\
    512 & $2.18\cdot10^{6}$ & $2.18\cdot10^{6}$ & $2.48\cdot10^{6}$ & $2.28\cdot10^{6}$ & \cellcolor{green!25}$1.85\cdot10^{6}$ \\
    1,024 & $8.11\cdot10^{6}$ & $8.02\cdot10^{6}$ & $8.68\cdot10^{6}$ & $7.83\cdot10^{6}$ & \cellcolor{green!25}$6.60\cdot10^{6}$ \\
    2,048 & $3.11\cdot10^{7}$ & $2.96\cdot10^{7}$ & $3.08\cdot10^{7}$ & $2.74\cdot10^{7}$ & \cellcolor{green!25}$2.37\cdot10^{7}$ \\
    4,096 & $1.22\cdot10^{8}$ & $1.10\cdot10^{8}$ & $1.09\cdot10^{8}$ & $9.66\cdot10^{7}$ & \cellcolor{green!25}$8.67\cdot10^{7}$ \\
    8,192 & $4.82\cdot10^{8}$ & $4.13\cdot10^{8}$ & $3.89\cdot10^{8}$ & $3.41\cdot10^{8}$ & \cellcolor{green!25}$3.12\cdot10^{8}$ \\
    16,384 & $1.92\cdot10^{9}$ & $1.55\cdot10^{9}$ & $1.41\cdot10^{9}$ & $1.22\cdot10^{9}$ & \cellcolor{green!25}$1.12\cdot10^{9}$ \\
    32,768 & $7.65\cdot10^{9}$ & $5.88\cdot10^{9}$ & $5.15\cdot10^{9}$ & $4.43\cdot10^{9}$ & \cellcolor{green!25}$4.11\cdot10^{9}$ \\
    65,536 & $3.05\cdot10^{10}$ & $2.23\cdot10^{10}$ & $1.90\cdot10^{10}$ & $1.62\cdot10^{10}$ & \cellcolor{green!25}$1.52\cdot10^{10}$ \\
    131,072 & $1.22\cdot10^{11}$ & $8.46\cdot10^{10}$ & $6.99\cdot10^{10}$ & $5.94\cdot10^{10}$ & \cellcolor{green!25}$5.55\cdot10^{10}$ \\
    262,144 & $4.88\cdot10^{11}$ & $3.23\cdot10^{11}$ & $2.59\cdot10^{11}$ & $2.19\cdot10^{11}$ & \cellcolor{green!25}$2.04\cdot10^{11}$ \\
    524,288 & $1.95\cdot10^{12}$ & $1.23\cdot10^{12}$ & $9.63\cdot10^{11}$ & $8.10\cdot10^{11}$ & \cellcolor{green!25}$7.53\cdot10^{11}$ \\
    1,048,576 & $7.81\cdot10^{12}$ & $4.72\cdot10^{12}$ & $3.60\cdot10^{12}$ & $3.02\cdot10^{12}$ & \cellcolor{green!25}$2.80\cdot10^{12}$ \\

    \hline
  \end{tabular}
  \caption{Move counts for different solvers (simulations, centers only); note that brobdicube uses its choice of optimal network for a given cube size.}
  \label{tab:maketable5b}
\end{table}

\subsection{Normalized move counts of center cubies (moves per cubie)}
\label{maketable6}

See Table~\ref{tab:maketable6}.

\begin{table}[t]
  \centering
  \setlength{\tabcolsep}{4pt}
  \scriptsize
  \newcommand{\heading}[1]{\multicolumn{1}{c}{\rotatebox[origin=c]{90}{\shortstack{\scriptsize #1}}}}

  \begin{tabular}{r| r| r r r r r r | r r r r r}
    \hline
    $n$
    & \heading{lower}
    & \heading{0D\\upper}
    & \heading{rcube1d\\upper}
    & \heading{rcube2d\\upper}
    & \heading{sortnet96\\upper}
    & \heading{permnet72\\upper}
    & \heading{brobdicube\\upper}
    & \heading{RCube1D\\sim}
    & \heading{RCube2D\\sim}
    & \heading{sortnet96\\sim}
    & \heading{permnet72\\sim}
    & \heading{brobdicube\\sim} \\
    \hline
    4             & 0.458
              & \cellcolor{green!40}6.00 & 8.33 & 8.33 & 36.5 & 27.3 & 23.0 & 5.23 & 5.23 & 9.02 & 10.05 & \cellcolor{green!20}3.90 \\
    8             & 0.352
              & \cellcolor{green!40}6.00 & 7.89 & 7.89 & 16.5 & 12.4 & 11.2 & 4.67 & 4.46 & 5.99 & 6.11 & \cellcolor{green!20}3.71 \\
    16            & 0.301
              & \cellcolor{green!40}6.00 & 6.89 & 6.89 & 10.05 & 7.54 & 6.95 & 3.96 & 3.91 & 4.59 & 4.48 & \cellcolor{green!20}3.38 \\
    32            & 0.263
              & 6.00 & 5.72 & 5.72 & 6.81 & 5.11 & \cellcolor{green!40}4.68 & 3.20 & 3.20 & 3.39 & 3.34 & \cellcolor{green!20}2.73 \\
    64            & 0.234
              & 6.00 & 3.61 & 3.61 & 4.96 & 3.72 & \cellcolor{green!40}3.40 & 2.48 & 2.48 & 2.73 & 2.60 & \cellcolor{green!20}2.06 \\
    128           & 0.211
              & 6.00 & 2.62 & 2.62 & 3.77 & 2.83 & \cellcolor{green!40}2.58 & 1.94 & 1.94 & 2.24 & 2.11 & \cellcolor{green!20}1.64 \\
    256           & 0.192
              & 6.00 & 2.14 & 2.14 & 2.98 & 2.24 & \cellcolor{green!40}2.04 & 1.59 & 1.59 & 1.87 & 1.76 & \cellcolor{green!20}1.37 \\
    512           & 0.176
              & 6.00 & 1.90 & 1.90 & 2.49 & 1.86 & \cellcolor{green!40}1.68 & 1.40 & 1.40 & 1.59 & 1.46 & \cellcolor{green!20}1.18 \\
    1,024         & 0.163
              & 6.00 & 1.78 & 1.78 & 2.05 & 1.53 & \cellcolor{green!40}1.38 & 1.29 & 1.28 & 1.39 & 1.25 & \cellcolor{green!20}1.05 \\
    2,048         & 0.152
              & 6.00 & 1.73 & 1.73 & 1.75 & 1.31 & \cellcolor{green!40}1.18 & 1.24 & 1.18 & 1.22 & 1.09 & \cellcolor{green!20}0.945 \\
    4,096         & 0.142
              & 6.00 & 1.70 & 1.70 & 1.53 & 1.15 & \cellcolor{green!40}1.03 & 1.21 & 1.10 & 1.08 & 0.961 & \cellcolor{green!20}0.862 \\
    8,192         & 0.133
              & 6.00 & 1.68 & 1.68 & 1.34 & 1.00 & \cellcolor{green!40}0.896 & 1.20 & 1.03 & 0.967 & 0.848 & \cellcolor{green!20}0.775 \\
    16,384        & 0.125
              & 6.00 & 1.67 & 1.67 & 1.19 & 0.890 & \cellcolor{green!40}0.792 & 1.19 & 0.965 & 0.875 & 0.759 & \cellcolor{green!20}0.698 \\
    32,768        & 0.118
              & 6.00 & 1.67 & 1.67 & 1.07 & 0.802 & \cellcolor{green!40}0.711 & 1.19 & 0.912 & 0.799 & 0.688 & \cellcolor{green!20}0.638 \\
    65,536        & 0.112
              & 6.00 & 1.67 & 1.67 & 0.975 & 0.731 & \cellcolor{green!40}0.645 & 1.19 & 0.864 & 0.736 & 0.631 & \cellcolor{green!20}0.590 \\
    131,072       & 0.106
              & 6.00 & 1.67 & 1.67 & 0.887 & 0.666 & \cellcolor{green!40}0.587 & 1.18 & 0.821 & 0.678 & 0.577 & \cellcolor{green!20}0.538 \\
    262,144       & 0.101
              & 6.00 & 1.67 & 1.67 & 0.813 & 0.609 & \cellcolor{green!40}0.537 & 1.18 & 0.783 & 0.627 & 0.530 & \cellcolor{green!20}0.494 \\
    524,288       & 0.0968
              & 6.00 & 1.67 & 1.67 & 0.750 & 0.562 & \cellcolor{green!40}0.494 & 1.18 & 0.749 & 0.584 & 0.491 & \cellcolor{green!20}0.457 \\
    1,048,576     & 0.0927
              & 6.00 & 1.67 & 1.67 & 0.697 & 0.523 & \cellcolor{green!40}0.458 & 1.18 & 0.716 & 0.546 & 0.458 & \cellcolor{green!20}0.425 \\
    2,097,152     & 0.0888
              & 6.00 & 1.67 & 1.63 & 0.652 & 0.489 & \cellcolor{green!40}0.427 & 1.18 & 0.687 & 0.514 & 0.429 & \cellcolor{green!20}0.397 \\
    4,194,304     & 0.0853
              & 6.00 & 1.67 & 1.53 & 0.612 & 0.459 & \cellcolor{green!40}0.400 & 1.18 & 0.660 & 0.485 & 0.404 & \cellcolor{green!20}0.373 \\
    8,388,608     & 0.0820
              & 6.00 & 1.67 & 1.43 & 0.573 & 0.430 & \cellcolor{green!40}0.375 & 1.18 & 0.636 & 0.459 & 0.380 & \cellcolor{green!20}0.350 \\
    16,777,216    & 0.0790
              & 6.00 & 1.67 & 1.35 & 0.539 & 0.404 & \cellcolor{green!40}0.352 & 1.18 & 0.613 & 0.435 & 0.358 & \cellcolor{green!20}0.329 \\
    33,554,432    & 0.0762
              & 6.00 & 1.67 & 1.27 & 0.509 & 0.382 & \cellcolor{green!40}0.332 & 1.18 & 0.592 & 0.422 & 0.346 & \cellcolor{green!20}0.311 \\
    67,108,864    & 0.0736
              & 6.00 & 1.67 & 1.21 & 0.483 & 0.362 & \cellcolor{green!40}0.314 & 1.18 & 0.571 & 0.393 & 0.322 & \cellcolor{green!20}0.294 \\
    134,217,728   & 0.0712
              & 6.00 & 1.67 & 1.15 & 0.459 & 0.344 & \cellcolor{green!40}0.298 & 1.18 & 0.553 & 0.375 & 0.306 & \cellcolor{green!20}0.280 \\
    268,435,456   & 0.0689
              & 6.00 & 1.67 & 1.09 & 0.437 & 0.328 & \cellcolor{green!40}0.284 & 1.18 & 0.536 & 0.358 & 0.293 & \cellcolor{green!20}0.266 \\
    536,870,912   & 0.0667
              & 6.00 & 1.67 & 1.05 & 0.418 & 0.314 & \cellcolor{green!40}0.271 & 1.18 & 0.519 & 0.343 & 0.280 & \cellcolor{green!20}0.254 \\
    671,088,640   & 0.0661
              & 6.00 & 1.67 & 1.03 & 0.411 & 0.308 & \cellcolor{green!40}0.266 & 1.18 & 0.514 & 0.339 & 0.275 & \cellcolor{green!20}0.250 \\
    805,306,368   & 0.0655
              & 6.00 & 1.67 & 1.01 & 0.406 & 0.304 & \cellcolor{green!40}0.263 & 1.18 & 0.510 & 0.336 & 0.272 & \cellcolor{green!20}0.247 \\
    1,073,741,824 & 0.0647
              & 6.00 & 1.67 & 0.999 & 0.400 & 0.300 & \cellcolor{green!40}0.259 & 1.18 & 0.504 & 0.330 & 0.268 & \cellcolor{green!20}0.242 \\

    \hline
  \end{tabular}
  \caption{Normalized move counts of center cubies (moves per cubie).}
  \label{tab:maketable6}
\end{table}


\clearpage

\section{Choice of $w$}
\label{choicew}

While we have established a closed-form expression $F_{2\mathrm{D}}(n)$ (Section~\ref{2D_par}) for the move count of an $n \times n \times n$ Rubik's Cube, that expression assumes a single uniform strip width $w$ and therefore does not always capture the best discrete choice of strip widths.
For other values of $n$, it is better to choose a
\emph{multiset} of widths $\{w_1, w_2, \ldots, w_k\}$ whose sum equals $Y$, i.e.\
$\sum_{i=1}^{k} w_i = \frac{n}{2} - 1$ for even $n$.

Our optimization problem then becomes
\begin{equation}
    \min_{\substack{w_1,\ldots,w_k \;\in\; \mathbb{Z}^{+} \\[2pt] \sum_{i=1}^{k} w_i \;=\; \frac{n}{2}-1}}
    \quad
    \sum_{i=1}^{k} C_{\mathrm{strip}}(n,\, w_i),
    \label{eq:opt}
\end{equation}
where the cost function per strip is
\begin{equation}
    C_{\mathrm{strip}}(n,\, w_i) \;=\; 4\!\left(2^{w_i}-1\right) + 2\!\left(\tfrac{n}{2}-1\right) + 2^{w_i}\cdot w_i.
    \label{eq:cost}
\end{equation}

We approach this optimization problem with a dynamic programming algorithm to generate the multiset of specific discrete widths across the quadrant that sum up to the overall width $n/2 - 1$ while minimizing the total worst-case move count.

The total cost across all strips is then
\begin{equation}
  C_{\mathrm{total}}(n) \;=\; \sum_{i=1}^{k} C_{\mathrm{strip}}(n,\, w_i).
  \label{eq:total}
\end{equation}

Table~\ref{tab:comparison} compares the DP-optimized total cost $C_{\mathrm{total}}$ with the
closed-form expression $F_{2\mathrm{D}}(n)$ for selected values of $n$.

\begin{table}[t]
\centering
\caption{DP-optimal cost $C_{\mathrm{total}}$ vs.\ closed-form $F_{2\mathrm{D}}(n)$.
         The notation $w^c$ denotes width $w$ used $c$ times.}
\label{tab:comparison}
\begin{tabular}{r r r r l}
\toprule
$n$ & $Y$ & DP cost & $F_{2\mathrm{D}}(n)$ & Optimal multiset \\
\midrule
    66 &   32 &      1\,244 &      5\,892 & $\{2^1,\, 3^{10}\}$ \\
   100 &   49 &      2\,472 &      7\,794 & $\{3^{15},\, 4^1\}$ \\
   130 &   64 &      3\,852 &     10\,223 & $\{3^{20},\, 4^1\}$ \\
   258 &  128 &     12\,160 &     24\,585 & $\{4^{32}\}$ \\
   500 &  249 &     38\,724 &     65\,259 & $\{4^{61},\, 5^1\}$ \\
   514 &  256 &     40\,704 &     68\,096 & $\{4^{64}\}$ \\
 1\,000 &  499 &    128\,040 &    196\,045 & $\{4^1,\, 5^{99}\}$ \\
 1\,026 &  512 &    134\,120 &    204\,423 & $\{5^{100},\, 6^2\}$ \\
 5\,000 & 2\,499 &  2\,285\,514 &  3\,001\,119 & $\{7^{357}\}$ \\
10\,000 & 4\,999 &  8\,142\,692 & 10\,191\,274 & $\{7^{713},\, 8^1\}$ \\
\bottomrule
\end{tabular}
\end{table}

\section{Discrete Worst-Case Upper Bounds for $n \le 100$}
\label{app:discrete}
{\color{black}
Table~\ref{tab:discrete-ub} lists the discrete worst-case upper bound
$\mathrm{total\_ub}(n) = C(n) + \mathrm{wings\_ub}(n) + 20$ (Section~\ref{discreteub})
for solving an entire $n \times n \times n$ Rubik's Cube, for all even $n \le 100$,
together with its center, wing, and corner components. The center column $C(n)$ is
the worst-case center move count produced by our solver using the optimal discrete strip
widths (Appendix~\ref{choicew}); the wing column is $\mathrm{wings\_ub}(n)$ from
Equation~\eqref{eq:wings-ub}; and the ``E{+}C'' column is the conservative $20$-move allowance for the
corners (Sections~\ref{midedges} and~\ref{discreteub}); for even $n$ there are no middle edges.

\begin{table}[htbp]
  \centering
  \scriptsize
  \setlength{\tabcolsep}{5pt}
  \begin{tabular}{r r r r r l}
    \toprule
    $n$ & Centers & Wings & E{+}C & Total & Optimal multiset \\
    \midrule
    4 & 552 & 157 & 20 & 729 & $\{1^{1}\}$ \\
    6 & 1{,}356 & 314 & 20 & 1{,}690 & $\{1^{2}\}$ \\
    8 & 2{,}428 & 471 & 20 & 2{,}919 & $\{1^{3}\}$ \\
    10 & 3{,}768 & 628 & 20 & 4{,}416 & $\{1^{4}\}$ \\
    12 & 5{,}108 & 785 & 20 & 5{,}913 & $\{1^{1},\, 2^{2}\}$ \\
    14 & 6{,}448 & 942 & 20 & 7{,}410 & $\{2^{3}\}$ \\
    16 & 8{,}175 & 1{,}099 & 20 & 9{,}294 & $\{1^{1},\, 2^{3}\}$ \\
    18 & 9{,}638 & 1{,}256 & 20 & 10{,}914 & $\{2^{4}\}$ \\
    20 & 11{,}733 & 1{,}413 & 20 & 13{,}166 & $\{1^{1},\, 2^{4}\}$ \\
    22 & 13{,}326 & 1{,}570 & 20 & 14{,}916 & $\{2^{5}\}$ \\
    24 & 15{,}811 & 1{,}727 & 20 & 17{,}558 & $\{1^{1},\, 2^{5}\}$ \\
    26 & 17{,}534 & 1{,}884 & 20 & 19{,}438 & $\{2^{6}\}$ \\
    28 & 20{,}409 & 2{,}041 & 20 & 22{,}470 & $\{1^{1},\, 2^{6}\}$ \\
    30 & 22{,}262 & 2{,}198 & 20 & 24{,}480 & $\{2^{7}\}$ \\
    32 & 25{,}279 & 2{,}355 & 20 & 27{,}654 & $\{2^{6},\, 3^{1}\}$ \\
    34 & 27{,}510 & 2{,}512 & 20 & 30{,}042 & $\{2^{8}\}$ \\
    36 & 30{,}669 & 2{,}657 & 20 & 33{,}346 & $\{2^{7},\, 3^{1}\}$ \\
    38 & 33{,}278 & 2{,}802 & 20 & 36{,}100 & $\{2^{9}\}$ \\
    40 & 36{,}579 & 2{,}947 & 20 & 39{,}546 & $\{2^{8},\, 3^{1}\}$ \\
    42 & 39{,}566 & 3{,}092 & 20 & 42{,}678 & $\{2^{10}\}$ \\
    44 & 43{,}009 & 3{,}237 & 20 & 46{,}266 & $\{2^{9},\, 3^{1}\}$ \\
    46 & 46{,}374 & 3{,}382 & 20 & 49{,}776 & $\{2^{11}\}$ \\
    48 & 49{,}584 & 3{,}527 & 20 & 53{,}131 & $\{2^{1},\, 3^{7}\}$ \\
    50 & 52{,}710 & 3{,}672 & 20 & 56{,}402 & $\{3^{8}\}$ \\
    52 & 56{,}298 & 3{,}817 & 20 & 60{,}135 & $\{2^{2},\, 3^{7}\}$ \\
    54 & 59{,}566 & 3{,}962 & 20 & 63{,}548 & $\{2^{1},\, 3^{8}\}$ \\
    56 & 62{,}912 & 4{,}107 & 20 & 67{,}039 & $\{3^{9}\}$ \\
    58 & 66{,}942 & 4{,}252 & 20 & 71{,}214 & $\{2^{2},\, 3^{8}\}$ \\
    60 & 70{,}418 & 4{,}397 & 20 & 74{,}835 & $\{2^{1},\, 3^{9}\}$ \\
    62 & 73{,}846 & 4{,}542 & 20 & 78{,}408 & $\{3^{10}\}$ \\
    64 & 78{,}444 & 4{,}687 & 20 & 83{,}151 & $\{2^{2},\, 3^{9}\}$ \\
    66 & 82{,}014 & 4{,}832 & 20 & 86{,}866 & $\{2^{1},\, 3^{10}\}$ \\
    68 & 85{,}626 & 4{,}977 & 20 & 90{,}623 & $\{3^{11}\}$ \\
    70 & 90{,}702 & 5{,}122 & 20 & 95{,}844 & $\{2^{2},\, 3^{10}\}$ \\
    72 & 94{,}444 & 5{,}267 & 20 & 99{,}731 & $\{2^{1},\, 3^{11}\}$ \\
    74 & 98{,}174 & 5{,}412 & 20 & 103{,}606 & $\{3^{12}\}$ \\
    76 & 103{,}782 & 5{,}557 & 20 & 109{,}359 & $\{2^{2},\, 3^{11}\}$ \\
    78 & 107{,}654 & 5{,}702 & 20 & 113{,}376 & $\{2^{1},\, 3^{12}\}$ \\
    80 & 111{,}496 & 5{,}847 & 20 & 117{,}363 & $\{3^{13}\}$ \\
    82 & 117{,}582 & 5{,}992 & 20 & 123{,}594 & $\{2^{2},\, 3^{12}\}$ \\
    84 & 121{,}554 & 6{,}137 & 20 & 127{,}711 & $\{2^{1},\, 3^{13}\}$ \\
    86 & 125{,}526 & 6{,}282 & 20 & 131{,}828 & $\{3^{14}\}$ \\
    88 & 132{,}008 & 6{,}427 & 20 & 138{,}455 & $\{3^{13},\, 4^{1}\}$ \\
    90 & 136{,}234 & 6{,}572 & 20 & 142{,}826 & $\{2^{1},\, 3^{14}\}$ \\
    92 & 140{,}336 & 6{,}717 & 20 & 147{,}073 & $\{3^{15}\}$ \\
    94 & 146{,}966 & 6{,}862 & 20 & 153{,}848 & $\{3^{14},\, 4^{1}\}$ \\
    96 & 151{,}694 & 7{,}007 & 20 & 158{,}721 & $\{2^{1},\, 3^{15}\}$ \\
    98 & 155{,}926 & 7{,}152 & 20 & 163{,}098 & $\{3^{16}\}$ \\
    100 & 162{,}704 & 7{,}273 & 20 & 169{,}997 & $\{3^{15},\, 4^{1}\}$ \\

    \bottomrule
  \end{tabular}
  \caption{Discrete worst-case move-count upper bounds for solving an entire
    $n \times n \times n$ Rubik's Cube (even $n \le 100$), split into center, wing, and
    corner (``E{+}C'') contributions. Centers use the optimal discrete strip widths
    (Appendix~\ref{choicew}); wings use Equation~\eqref{eq:wings-ub}; corners
    use the conservative $20$-move allowance (for even $n$ there are no middle edges). Our solver chooses the strip widths per
    stage; the ``Optimal multiset'' column therefore gives the strip-width partition of
    $S=\tfrac{n}{2}-1$ (Appendix~\ref{choicew}) used by the largest number of stages, where
    $w^{c}$ denotes width $w$ used $c$ times.}
  \label{tab:discrete-ub}
\end{table}
}

\section{Wing Move Sequences and Refined Wing Upper Bound}
\label{app:wings}
{\color{black}
Table~\ref{tab:wing-base-algs} lists the $87$ base wing 3-cycle move sequences from
which all $4048$ wing 3-cycle cases are generated by mirroring, rotating, inverting, and
applying conjugation moves (Section~\ref{wings}). The total move lengths $\ell_c$ of
the expanded algorithm set follow the histogram in Table~\ref{tab:wing-histogram}.
}

\begin{table}[t]
  \centering
  \color{black}
  \begin{tabular}{l|rrrrr|r}
    \toprule
    Move length $\ell_c$ & 8 & 9 & 10 & 12 & 13 & total \\
    \midrule
    Number of cases & 2112 & 912 & 448 & 384 & 192 & 4048 \\
    \bottomrule
  \end{tabular}
  \caption{Histogram of move lengths over the $4048$ wing 3-cycle cases.
    Here $\sum_c \ell_c = 36{,}688$ and $\sum_c (\ell_c - 2) = 28{,}592$.}
  \label{tab:wing-histogram}
\end{table}

\begin{table}[p]
  \centering
  \color{black}
  \scriptsize
  \setlength{\tabcolsep}{3pt}
  \begin{tabular}{r l | r l}
    \toprule
    \# & base move sequence & \# & base move sequence \\
    \midrule
    1 & \texttt{F D 2B' D' F' D 2B D'} & 45 & \texttt{F D 2B D' F D 2B' D' F2} \\
    2 & \texttt{F' L' 2B L F L' 2B' L} & 46 & \texttt{F' L' 2B' L F' L' 2B L F2} \\
    3 & \texttt{F D 2B2 D' F' D 2B2 D'} & 47 & \texttt{F D 2B' D' F D 2B D' F2} \\
    4 & \texttt{F' L' 2B2 L F L' 2B2 L} & 48 & \texttt{F D 2F D' F D 2F' D' F2} \\
    5 & \texttt{F D 2B D' F' D 2B' D'} & 49 & \texttt{F D 2B D' F2 D 2B' D' F} \\
    6 & \texttt{F' L' 2B' L F L' 2B L} & 50 & \texttt{F' L' 2B' L F2 L' 2B L F'} \\
    7 & \texttt{F D 2F' D' F' D 2F D'} & 51 & \texttt{F D F 2D F' D' F 2D' F2} \\
    8 & \texttt{F' L' 2F L F L' 2F' L} & 52 & \texttt{F' L' F' 2L' F L F' 2L F2} \\
    9 & \texttt{F D 2F2 D' F' D 2F2 D'} & 53 & \texttt{F D2 F 2D F' D2 F 2D' F2} \\
    10 & \texttt{F' L' 2F2 L F L' 2F2 L} & 54 & \texttt{F' L2 F' 2L' F L2 F' 2L F2} \\
    11 & \texttt{F D2 F' 2D F D2 F' 2D'} & 55 & \texttt{F D' F 2D F' D F 2D' F2} \\
    12 & \texttt{F' L2 F 2L' F' L2 F 2L} & 56 & \texttt{F' L F' 2L' F L' F' 2L F2} \\
    13 & \texttt{F D2 F' 2D2 F D2 F' 2D2} & 57 & \texttt{F D' F 2D2 F' D F 2D2 F2} \\
    14 & \texttt{F' L2 F 2L2 F' L2 F 2L2} & 58 & \texttt{F' L F' 2L2 F L' F' 2L2 F2} \\
    15 & \texttt{F D2 F' 2D' F D2 F' 2D} & 59 & \texttt{F 2D F D2 F' 2D' F D2 F2} \\
    16 & \texttt{F' L2 F 2L F' L2 F 2L'} & 60 & \texttt{F 2D F U F' 2D' F U' F2} \\
    17 & \texttt{F D2 F' 2U F D2 F' 2U'} & 61 & \texttt{F' 2L' F' R' F 2L F' R F2} \\
    18 & \texttt{F' L2 F 2R' F' L2 F 2R} & 62 & \texttt{F 2D2 F D F' 2D2 F D' F2} \\
    19 & \texttt{F D' 2B2 D F' D' 2B2 D} & 63 & \texttt{F 2D2 F U F' 2D2 F U' F2} \\
    20 & \texttt{F' L 2B2 L' F L 2B2 L'} & 64 & \texttt{F B D F' 2D' F D' F' 2D B'} \\
    21 & \texttt{F D2 F' 2U2 F D2 F' 2U2} & 65 & \texttt{F B D F' 2U F D' F' 2U' B'} \\
    22 & \texttt{F' L2 F 2R2 F' L2 F 2R2} & 66 & \texttt{F B D' F' 2D' F D F' 2D B'} \\
    23 & \texttt{F D' 2F2 D F' D' 2F2 D} & 67 & \texttt{F' B' L F 2L F' L' F 2L' B} \\
    24 & \texttt{F' L 2F2 L' F L 2F2 L'} & 68 & \texttt{F B 2D2 F' D F 2D2 F' D' B'} \\
    25 & \texttt{F D2 F' 2U' F D2 F' 2U} & 69 & \texttt{F' B' 2L2 F L' F' 2L2 F L B} \\
    26 & \texttt{F' L2 F 2R F' L2 F 2R'} & 70 & \texttt{F B' D F' 2D' F D' F' 2D B} \\
    27 & \texttt{F D' F' 2D2 F D F' 2D2} & 71 & \texttt{F' B L' F 2L F' L F 2L' B'} \\
    28 & \texttt{F' L F 2L2 F' L' F 2L2} & 72 & \texttt{F D R' 2D2 R D' R' 2D2 R F'} \\
    29 & \texttt{F D' F' 2D' F D F' 2D} & 73 & \texttt{F D R' 2U2 R D' R' 2U2 R F'} \\
    30 & \texttt{F' L F 2L F' L' F 2L'} & 74 & \texttt{F D2 L' 2D' L D2 L' 2D L F'} \\
    31 & \texttt{F D' F' 2U F D F' 2U'} & 75 & \texttt{F' L2 D 2L D' L2 D 2L' D' F} \\
    32 & \texttt{F' L F 2R' F' L' F 2R} & 76 & \texttt{F D B2 L' 2F L B2 L' 2F' L D' F'} \\
    33 & \texttt{F D' F' 2U2 F D F' 2U2} & 77 & \texttt{F D' L F' D 2L D' F L' D F' 2L'} \\
    34 & \texttt{F' L F 2R2 F' L' F 2R2} & 78 & \texttt{F' L D' F L' 2D' L F' D L' F 2D} \\
    35 & \texttt{F 2D F' D2 F 2D' F' D2} & 79 & \texttt{F D' L F' D 2L2 D' F L' D F' 2L2} \\
    36 & \texttt{F' 2L' F L2 F' 2L F L2} & 80 & \texttt{F D' L 2F2 L' D F' L D' 2F2 D L'} \\
    37 & \texttt{F 2D F' U2 F 2D' F' U2} & 81 & \texttt{F' L D' 2F2 D L' F D' L 2F2 L' D} \\
    38 & \texttt{F' 2L' F R2 F' 2L F R2} & 82 & \texttt{F D' L 2F' L' D F' L D' 2F D L'} \\
    39 & \texttt{F 2D2 F' D2 F 2D2 F' D2} & 83 & \texttt{F' L D' 2F D L' F D' L 2F' L' D} \\
    40 & \texttt{F' 2L2 F L2 F' 2L2 F L2} & 84 & \texttt{F D L' F 2D' F' L D' F L' 2D L F2} \\
    41 & \texttt{F 2D2 F' U2 F 2D2 F' U2} & 85 & \texttt{F' L' D F' 2L F D' L F' D 2L' D' F2} \\
    42 & \texttt{F' 2L2 F R2 F' 2L2 F R2} & 86 & \texttt{F D L' F 2U F' L D' F L' 2U' L F2} \\
    43 & \texttt{F 2D' F' U2 F 2D F' U2} & 87 & \texttt{F' L' D F' 2R' F D' L F' D 2R D' F2} \\
    44 & \texttt{F' 2L F R2 F' 2L' F R2} & & \\
    \bottomrule
  \end{tabular}
  \caption{The $87$ base wing 3-cycle move sequences. All $4048$ wing 3-cycle move
    sequences are generated from these by mirroring, rotating, inverting, and applying
    conjugation moves (Section~\ref{wings}).}
  \label{tab:wing-base-algs}
\end{table}

{\color{black}
Now we refine the simple worst-case wing bound of Section~\ref{wings}. As before,
fixing parity costs one quarter-turn per wing orbit, i.e.,\ $r = \tfrac{n}{2}-1$ moves
over the $r$ wing orbits.
For the 3-cycles, rather than the rough linear bound $13 \cdot 12\, r$ (each of the
$\le 12$ 3-cycles per orbit charged at the maximum length $13$), we use a tighter
\emph{ramp} of decreasing slope. Order the $4048$ algorithms by decreasing length.
Each additional orbit contributes at most $12$ new 3-cycles, which in the worst
case are charged to the longest algorithms not yet used, each applied at its full
length. After $m = 12r$
such charges, the worst-case 3-cycle cost is the sum $G(m)$ of the $m$ largest
lengths. With the length multiset $\{13^{192}, 12^{384}, 10^{448}, 9^{912},
8^{2112}\}$ (Table~\ref{tab:wing-histogram}), $G$ is piecewise linear with
breakpoints at the cumulative counts $192, 576, 1024, 1936, 4048$ and slope
decreasing from $13$ to $8$. Once $m \ge 4048$ (equivalently $r \ge 338$), the charging
has exhausted the multiset: the outer-turn contribution is capped at $\sum_c(\ell_c - 2)
= 28{,}592$, and each further orbit adds only the $2$ parallelized slice moves per
3-cycle (slope $24$). This is a property of the worst-case charging, not a claim that any
particular solve uses all $4048$ algorithms. Because $G$ is concave, we can express:
\begin{equation}
  \label{eq:wings-ub}
  \mathrm{wings\_ub}(n) = r + \min\!\big(
      156\,r,\;
      144\,r + 192,\;
      120\,r + 1344,\;
      108\,r + 2368,\;
      96\,r + 4304,\;
      24\,r + 28{,}592
    \big),
\end{equation}
where $r = \tfrac{n}{2} - 1$ and the slopes $156, 144, 120, 108, 96$ equal
$13\cdot 12,\, 12\cdot 12,\, 10\cdot 12,\, 9\cdot 12,\, 8\cdot 12$. This ramp lies
everywhere below the linear bound $13 \cdot 12\, r$ and approaches the saturated
line $24\,r + 28{,}592$ for large $n$. Table~\ref{tab:discrete-ub}
(Appendix~\ref{app:discrete}) uses this model for the wing contribution to the total
worst-case upper bounds.
}

\section{Universal Sequences of permnet62}
\label{app:permnet62-tables}
Tables~\ref{tab:permnet62-2412} and~\ref{tab:permnet62-124a} list the universal
sequences of the permnet62 networks discussed in Section~\ref{beyond3cycles}.

\begin{table}[p]
  \centering \footnotesize
  \setlength{\tabcolsep}{4pt}
  \renewcommand{\arraystretch}{1.15}
  \begin{tabular}{r l l}
    \toprule
    \# & permutation & representative move sequence \\
    \midrule
    1 & $(1,19,5)+(9,13,21)$ & \texttt{2R B2 3U B2 2R' B2 3U' B2} \\
    2 & $(1,12,23,11,8,16,7)$ & \texttt{B' D2 F2 2R F2 R' 3D' F2 2R' F2 R 2U' R' 3D R 2U D2 B} \\
    3 & $(0,9,15,19,4,20,12)$ & \texttt{B' F' U 2F U' 3L U 2F' U' B2 2U B2 3L' B2 2U' B2 F B} \\
    4 & $(0,7,22,15)+(6,10,12,16)$ & \texttt{L R U2 F2 2U F2 3R U2 F2 2U' F2 2B' U2 3R' U2 2B R' L'} \\
    5 & $(0,5,21,10,13,18,14)$ & \texttt{L F2 2B D' 3L' D F2 2B' 2D F2 D' 3L D F2 2D' L'} \\
    6 & $(1,18,21,14,17,4,8)$ & \texttt{B2 D2 D 2R' D' 3B D R2 2R 2U R2 D' 3B' R2 2U' R2 D2 B2} \\
    7 & $(2,9,19,18,6,23,13)$ & \texttt{R F B2 2L' D' 3R' D 2L B2 2D' B2 D' 3R D B2 2D F' R'} \\
    8 & $(0,15,23,4)+(1,17,22,10)$ & \texttt{L' F U' 2L B2 3D B2 2L' R2 2B R2 B2 3D' R2 2B' R2 B2 U F' L} \\
    9 & $(1,14,15,7,11,20,19)$ & \texttt{R U L2 2F L2 3D' L2 2F' L2 B 2U' B' 3D B 2U B' U' R'} \\
    10 & $(0,20,23,8,15,17,5)$ & \texttt{B D U2 2F U2 3L U2 2F' U2 B 2R' B' 3L' B 2R B' D' B'} \\
    11 & $(3,4,11,21,16,22,12)$ & \texttt{L2 B D R F2 2B L' 3U' L F2 2B' 2L F2 L' 3U L F2 2L' R' D' B' L2} \\
    12 & $(2,16,3,19,22,8,6)$ & \texttt{L B' D' U2 2B' D' 3L' D 2B F 2R2 F' D' 3L D F 2R2 F' U2 D B L'} \\
    13 & $(2,19,22,16,12,10,4)$ & \texttt{L2 B D' F U L 2U L' 3F L 2U' L' D2 2L' D2 3F' D2 2L D2 U' F' D B' L2} \\
    14 & $(0,4,11,3,13,21,7)$ & \texttt{L' D 2B2 2L' D' R' 3F' R D 2L 2B2 D' R' 3F R L} \\
    15 & $(1,15,8,12,18,20,6)$ & \texttt{L2 R U' B2 2U B2 3L D B2 2U' 2B B2 D' 3L' D 2B' D' U R' L2} \\
    16 & $(3,17,13)+(4,21,10)$ & \texttt{L F' U2 B2 2R B2 3U' B2 2R' B2 3U U2 F L'} \\
    17 & $(2,7,8)+(13,17,22)$ & \texttt{B' F2 3R U2 2B' U2 3R' U2 2B U2 F2 B} \\
    18 & $(1,18,6)+(11,15,23)$ & \texttt{R' F' 2D L2 3B L2 2D' L2 3B' L2 F R} \\
    19 & $(1,10,5)+(13,22,17)$ & \texttt{F' U' 3U L2 2B L2 3U' L2 2B' L2 U F} \\
    20 & $(2,18,7)+(8,12,20)$ & \texttt{L U2 3L' B2 2U B2 3L B2 2U' B2 U2 L'} \\
    21 & $(2,6,18)+(10,23,14)$ & \texttt{U' L2 3B L2 2D L2 3B' L2 2D' U} \\
    22 & $(0,14,18)+(4,10,20)$ & \texttt{L2 2D F2 3L' F2 2D' F2 3L F2 L2} \\
    23 & $(1,14,10)+(5,16,23)$ & \texttt{L' B2 3B' L2 2D L2 3B L2 2D' L2 B2 L} \\
    24 & $(5,10,9)$ & \texttt{3U L 2D' L' 3U' L 2D L'} \\
    25 & $(1,4,8)+(13,17,21)$ & \texttt{L F 3U' B2 2R' B2 3U B2 2R B2 F' L'} \\
    26 & $(3,19,15)+(4,23,11)$ & \texttt{L' B2 2L' B2 3D B2 2L B2 3D' L} \\
    \bottomrule
  \end{tabular}
  \caption{Universal sequence of the $24\to12{+}12$ permutation network of permnet62 ($26$ stages). For each stage we give its representative move sequence and the permutation it induces on the $24$ orbit positions (labeled $0$--$23$); the multi-piece stages act as products of disjoint cycles. All sequences are verifiable on a $6\times6\times6$ cube.}
  \label{tab:permnet62-2412}
\end{table}

\begin{table}[p]
  \centering \footnotesize
  \setlength{\tabcolsep}{4pt}
  \renewcommand{\arraystretch}{1.15}
  \begin{tabular}{r l l}
    \toprule
    \# & permutation & representative move sequence \\
    \midrule
    1 & $(0,8,9,1,17)$ & \texttt{B2 2R2 B' D' 3L2 D B 2R' B' D' 3L2 D B 2R' B2} \\
    2 & $(1,10,16)+(3,17,9)$ & \texttt{F U' 2B2 3R D U 2L D' U' 3R' D U 2L' D' U' 2B2 U F'} \\
    3 & $(2,8,3)+(11,16,19)$ & \texttt{B F' 2L D' 3R2 D 2L' F 2L2 F' D' 3R2 D F 2L2 B'} \\
    4 & $(0,1,3)+(10,18,19)$ & \texttt{F2 U F 2F2 D 2R2 D' F' 3L2 F D 2R2 D' F' 3L2 2F2 U' F2} \\
    5 & $(0,10,18)+(3,11,19)$ & \texttt{F2 U 2R F 2R2 F' D' 3L2 D F 2R2 F' D' 3L2 D 2R' U' F2} \\
    6 & $(0,8,17)+(2,18,11)$ & \texttt{F' B' 2B2 3R D U 2L D' U' 3R' D U 2L' D' U' 2B2 B F} \\
    7 & $(0,11,10,3,16)$ & \texttt{B U' 2L' D' B' 3R2 B D 2L' D' B' 3R2 B D 2L2 U B'} \\
    8 & $(1,8)+(18,19)$ & \texttt{F 2R' F' U' 3L U F 2R F' U' 3L' U} \\
    9 & $(0,10,9)$ & \texttt{F' 3L' U 2R U' 3L U 2R' U' F} \\
    10 & $(8,18,19)$ & \texttt{F 2R2 F' 3L2 F 2R2 F' 3L2} \\
    11 & $(3,16,17)$ & \texttt{2L U' 3R' U 2L' U' 3R U} \\
    12 & $(2,10,11)$ & \texttt{U 2L' U' 3R U 2L U' 3R'} \\
    13 & $(0,19,18)$ & \texttt{3L U 2R' U' 3L' U 2R U'} \\
    14 & $(11,17,16)$ & \texttt{F' 3R2 F 2L2 F' 3R2 F 2L2} \\
    15 & $(3,10,11)$ & \texttt{2L' U' 3R U 2L U' 3R' U} \\
    16 & $(1,18,19)$ & \texttt{2R' U' 3L U 2R U' 3L' U} \\
    17 & $(0,9,19)$ & \texttt{F' 3L F 2R2 F' 3L' F 2R2} \\
    18 & $(2,16,11)$ & \texttt{3R B 2L2 B' 3R' B 2L2 B'} \\
    \bottomrule
  \end{tabular}
  \caption{Universal sequence of the first $12\to4{+}4{+}4$ permutation network of permnet62 ($18$ stages). It acts on one of the two $12$-cubie groups produced by the $24\to12{+}12$ network, namely the positions $\{0,1,2,3,8,9,10,11,16,17,18,19\}$; the second $12\to4{+}4{+}4$ network is the face-relabeled mirror image acting on the complementary group and is omitted.}
  \label{tab:permnet62-124a}
\end{table}

\section{Universal Sequences of permnet61}
\label{app:permnet61-tables}
Tables~\ref{tab:permnet61-2416}--\ref{tab:permnet61-844} list the universal sequences of the
three permnet61 networks discussed in Section~\ref{beyond3cycles}.

\begin{table}[htbp]
  \centering \scriptsize
  \setlength{\tabcolsep}{4pt}
  \renewcommand{\arraystretch}{1.15}
  \begin{tabular}{r l l}
    \toprule
    \# & permutation & representative move sequence \\
    \midrule
    1 & $(3,15,21,12,18,20,7)$ & \texttt{R D B U 2B2 2R U' L' 3F' L U 2R' 2B2 U' L' 3F L B' D' R'} \\
    2 & $(2,17,5,11,12,22,18)$ & \texttt{B' R' D2 2F D2 F' 3R' D2 2F' D2 F 2L2 F' 3R F 2L2 R B} \\
    3 & $(1,11,21,17)+(4,10,15,18)$ & \texttt{R' F' B 2R' D2 3B L2 D2 2R 2D D2 L2 3B' D2 L2 2D' L2 B' F R} \\
    4 & $(1,18,21,10)+(2,14,20,6)$ & \texttt{L' B' D' R F U 2L' F2 R2 3U F2 R2 2L 2F R2 F2 3U' R2 2F' F2 U' F' R' D B L} \\
    5 & $(3,7,23,15)+(6,11,12,19)$ & \texttt{B D' L F U R U2 2U F2 3R U2 F2 2U' 2B' F2 U2 3R' F2 U2 2B R' U' F' L' D B'} \\
    6 & $(1,7,21,13)+(6,9,14,17)$ & \texttt{F U L' B D' R 2F' D2 B2 3R D2 B2 2F 2D B2 D2 3R' B2 2D' D2 R' D B' L U' F'} \\
    7 & $(0,11,20,19)+(5,18,13,8)$ & \texttt{R' D B L' U F 2R' U2 L2 3F U2 L2 2R 2U L2 U2 3F' L2 2U' U2 F' U' L B' D' R} \\
    8 & $(0,12,20,4)+(1,10,23,16)$ & \texttt{R' B' U L F' D 2R' F2 L2 3D F2 L2 2R 2F L2 F2 3D' L2 2F' F2 D' F L' U' B R} \\
    9 & $(2,7,23,14)+(3,16,22,11)$ & \texttt{B R' U L2 2L F2 3U L2 F2 2L' 2B' F2 L2 3U' F2 L2 2B U' R B'} \\
    10 & $(0,5,21,15)+(6,9,14,19)$ & \texttt{B' R' F D L D2 2D F2 3L D2 F2 2D' 2B F2 D2 3L' F2 D2 2B' L' D' F' R B} \\
    11 & $(0,6,20,14)+(3,8,21,16)$ & \texttt{B' L U F R' D R2 2R F2 3D R2 F2 2R' 2B' F2 R2 3D' F2 R2 2B D' R F' U' L' B} \\
    12 & $(0,12,13)+(6,7,22)$ & \texttt{2B D' U' 3F' D U 2B' D' U' 3F D U} \\
    13 & $(2,3,20,14,7)$ & \texttt{U' L D' 3B2 D R L 2F R' L' D' 3B2 D R L 2F' R' L' L' U} \\
    14 & $(3,4,10)+(13,16,21)$ & \texttt{R' F2 L2 3F' L2 2U L2 3F L2 2U' F2 R} \\
    15 & $(0,12,6)$ & \texttt{3F2 L 2B L' 3F2 L 2B' L'} \\
    16 & $(3,6,11)+(12,17,21)$ & \texttt{D2 2L' D2 3F D2 2L D2 3F'} \\
    17 & $(1,14,19)+(4,9,23)$ & \texttt{D2 2R' D2 3B D2 2R D2 3B'} \\
    18 & $(2,18,5)+(8,15,20)$ & \texttt{2F U2 3L U2 2F' U2 3L' U2} \\
    19 & $(1,4,9)$ & \texttt{D 2R D' 3B' D 2R' D' 3B} \\
    20 & $(5,17,20)$ & \texttt{R 2F R' 3U' R 2F' R' 3U} \\
    21 & $(2,23,3)$ & \texttt{D' 3R2 D 2L2 D' 3R2 D 2L2} \\
    22 & $(3,21)+(11,17)$ & \texttt{D2 2L D2 3F2 D2 2L' D2 3F2} \\
    23 & $(1,8,4)+(14,23,16)$ & \texttt{R2 2U R2 3B R2 2U' R2 3B'} \\
    24 & $(6,22,13)$ & \texttt{2B R' 3F2 R 2B' R' 3F2 R} \\
    \bottomrule
  \end{tabular}
  \caption{Universal sequence of the $24\to8{+}16$ permutation network of permnet61 ($24$ stages).
  It acts on all $24$ orbit positions and splits them into a group of $8$ (faces $U$, $L$; positions $0$--$7$) and a group of $16$ (faces $F$, $R$, $B$, $D$; positions $8$--$23$). For each stage we give its representative move sequence and the permutation it
  induces on the orbit positions (labeled $0$--$23$ in Speffz order); the multi-piece stages act as
  products of disjoint cycles. All sequences are verifiable on a $6\times6\times6$ cube.}
  \label{tab:permnet61-2416}
\end{table}

\begin{table}[htbp]
  \centering \footnotesize
  \setlength{\tabcolsep}{4pt}
  \renewcommand{\arraystretch}{1.15}
  \begin{tabular}{r l l}
    \toprule
    \# & permutation & representative move sequence \\
    \midrule
    1 & $(0,10,12,3,4,9,7)$ & \texttt{2R' 2U2 L' 3F L 2U2 2R 2B L' F' 2R' 3F' 2R F L 2B'} \\
    2 & $(0,4,5,12,9,3,10)$ & \texttt{F L R U' R 2L2 2F' R' F' 3U F R 2F R' 2L 2D F' 3U' F 2D' 2L U R' L' F'} \\
    3 & $(0,9,1,7,3,15,11)$ & \texttt{R2 2R2 D 2B' R2 3D R2 2B D' R2 2R' U2 2R 3D' 2R' U2 2R'} \\
    4 & $(5,9,12)+(7,10,11)$ & \texttt{L2 2B R 2L2 3U' L 2B 2D' L' 3U L 2D 2B' R' 2L2 L' 2B' L2} \\
    5 & $(0,15,10,1,3,6,4)$ & \texttt{U L F 2F U 2R 3D 2R' U' L2 2F' 2R L2 3D' 2R' L2 2F L2 2F' F' L' U'} \\
    6 & $(5,7,9)+(10,11,12)$ & \texttt{2B2 2L2 R 3U R 2F2 2D R' 3U' R 2D' 2F2 2L2 R2 2B2} \\
    7 & $(0,12)+(3,7)$ & \texttt{R 2B' R' L' D' 3F' D R L 2B R' L' D' 3F D L} \\
    8 & $(1,11)+(5,15)$ & \texttt{L2 2R 3D L2 2F2 L2 3D' L2 2F2 2R'} \\
    9 & $(2,11,15)$ & \texttt{3D L 2F L' 3D' L 2F' L'} \\
    10 & $(3,12,8)$ & \texttt{3F L 2U' L' 3F' L 2U L'} \\
    11 & $(0,7,12)$ & \texttt{2B L' 3F2 L 2B' L' 3F2 L} \\
    12 & $(1,5,14)$ & \texttt{3B U 2F U' 3B' U 2F' U'} \\
    13 & $(1,4,9)$ & \texttt{D 2R D' 3B' D 2R' D' 3B} \\
    14 & $(5,14,15)$ & \texttt{2F2 L' 3B2 L 2F2 L' 3B2 L} \\
    15 & $(2,6,10)$ & \texttt{B 2D B' 3R' B 2D' B' 3R} \\
    16 & $(7,12,13)$ & \texttt{2B2 L' 3F2 L 2B2 L' 3F2 L} \\
    \bottomrule
  \end{tabular}
  \caption{Universal sequence of the $16\to8{+}8$ permutation network of permnet61 ($16$ stages).
  It acts on the $16$-group produced by the $24\to8{+}16$ network, splitting it into faces $\{F,R\}$ and $\{B,D\}$, and leaves the other $8$ positions untouched. The network is shown in a representative orientation on faces $U$, $L$, $F$, $R$; the solver applies it to faces $F$, $R$, $B$, $D$ by relabeling faces. For each stage we give its representative move sequence and the permutation it
  induces on the orbit positions in this orientation (labeled $0$--$15$ in Speffz order); the multi-piece stages act as
  products of disjoint cycles. All sequences are verifiable on a $6\times6\times6$ cube.}
  \label{tab:permnet61-1688}
\end{table}

\begin{table}[htbp]
  \centering \footnotesize
  \setlength{\tabcolsep}{4pt}
  \renewcommand{\arraystretch}{1.15}
  \begin{tabular}{r l l}
    \toprule
    \# & permutation & representative move sequence \\
    \midrule
    1 & $(1,2,4)$ & \texttt{L 2F' L' 3B L 2F L' 3B'} \\
    2 & $(2,7,6)$ & \texttt{U 3F U 2B' U' 3F' U 2B U' U'} \\
    3 & $(1,6)+(2,7)$ & \texttt{L2 2F2 L 2F L' D' 3B D L 2F' L' D' 3B' D 2F2 L2} \\
    4 & $(0,6,7)$ & \texttt{2B' U' 3F U 2B U' 3F' U} \\
    5 & $(1,2,4)$ & \texttt{L 2F' L' 3B L 2F L' 3B'} \\
    6 & $(1,2,5)$ & \texttt{2F' L' 3B L 2F L' 3B' L} \\
    7 & $(3,6,7)$ & \texttt{U 2B' U' 3F U 2B U' 3F'} \\
    \bottomrule
  \end{tabular}
  \caption{Universal sequence of the $8\to4{+}4$ permutation network of permnet61 ($7$ stages).
  It acts on one group of $8$ and is applied to each of the three groups of $8$ in turn. The network is shown in a representative orientation on faces $U$, $L$; the solver applies it to faces $\{F,R\}$ and $\{B,D\}$ by relabeling faces. For each stage we give its representative move sequence and the permutation it
  induces on the orbit positions in this orientation (labeled $0$--$7$ in Speffz order); the multi-piece stages act as
  products of disjoint cycles. All sequences are verifiable on a $6\times6\times6$ cube.}
  \label{tab:permnet61-844}
\end{table}

\section{Full Network Diagrams}
\label{app:networks}
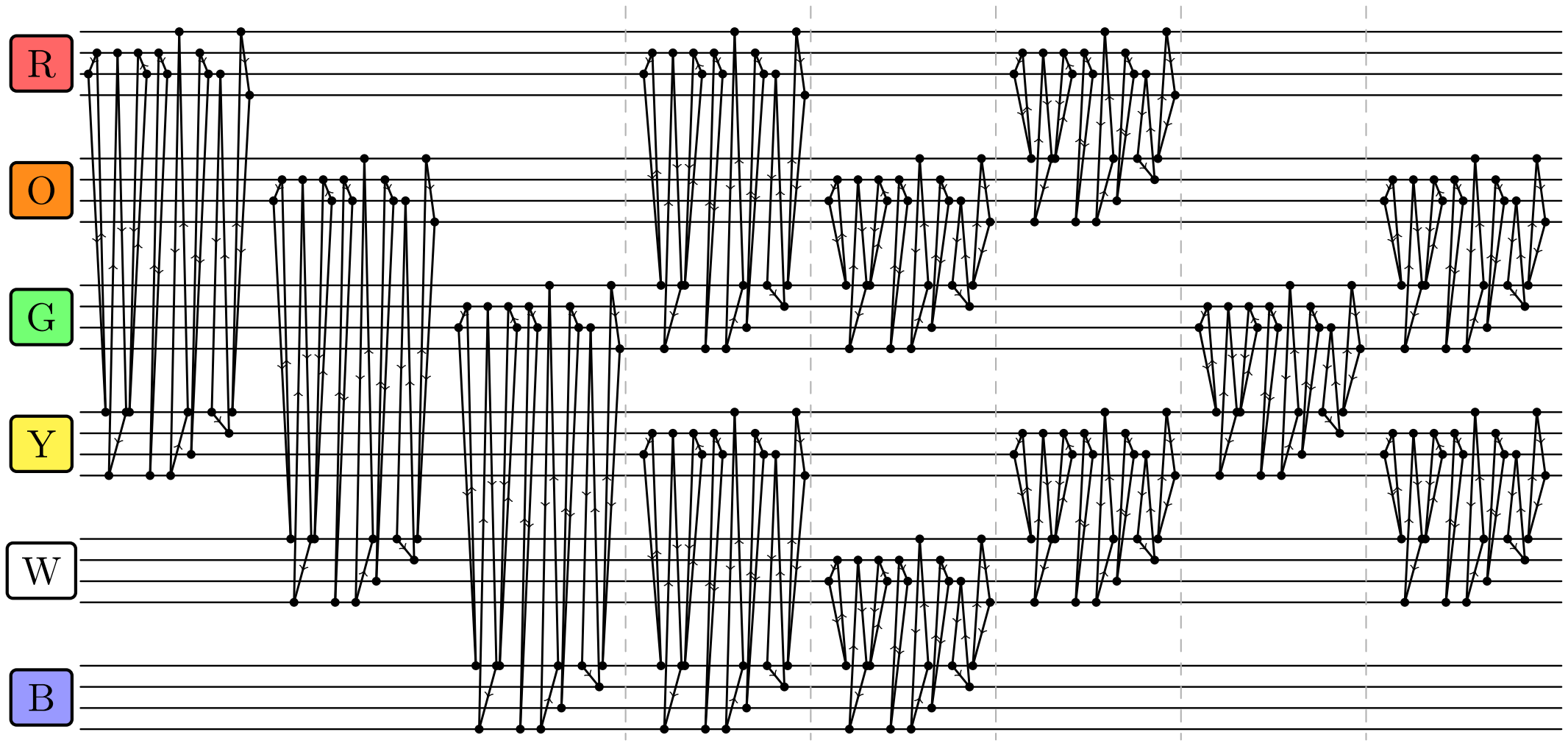
\begin{figure}[p]
    \centering
    \begin{tikzpicture}[x=0.185cm,y=0.19cm,line cap=round,line join=round]
        \tikzset{
            wire/.style={black, line width=0.45pt},
            cyc/.style={black, line width=0.55pt,
                postaction={decorate, decoration={markings,
                    mark=at position 0.5 with {\arrow[scale=0.6]{>}}}}},
            dot/.style={circle, fill=black, inner sep=0.8pt},
            fbox/.style={draw=black, thick, rounded corners=1.5pt,
                minimum width=0.55cm, minimum height=0.45cm, font=\small},
            phasesep/.style={gray!60, dashed, line width=0.4pt}
        }

        %

        \foreach \fb/\fcol/\fname in {
            23/faceR/R, 17/faceO/O, 11/faceG/G, 5/faceY/Y, -1/faceW/W, -7/faceB/B
        } {
            \foreach \w in {1,2,3,4} {
                \pgfmathsetmacro{\wy}{\fb - \w + 1}
                \draw[wire] (0,\wy) -- (72,\wy);
            }
            \pgfmathsetmacro{\midwy}{\fb - 1.5}
            \node[fbox, fill=\fcol] at (-1.9,\midwy) {\fname};
        }

        \foreach \px in {26.5, 35.5, 44.5, 53.5, 62.5} {
            \draw[phasesep] (\px,24.2) -- (\px,-10.5);
        }


        \foreach \xo/\ab/\bb in {
             0/23/ 5,   
             9/17/-1,   
            18/11/-7,   
            27/23/11,   
            27/ 5/-7,   
            36/17/11,   
            36/-1/-7,   
            45/23/17,   
            45/ 5/-1,   
            54/11/ 5,   
            63/17/11,   
            63/ 5/-1    
        } {
            \foreach \lx/\la/\lb/\lc/\ls in {
                0.8/2/3/5/ 1,
                1.8/2/5/8/-1,
                2.8/2/5/3/ 1,
                3.8/2/3/8/-1,
                4.8/1/8/5/ 1,
                5.8/2/3/7/-1,
                6.8/3/5/6/ 1,
                7.8/1/4/5/-1
            } {
                \pgfmathsetmacro{\xpos}{\xo + \lx}
                \pgfmathsetmacro{\dx}{0.42*\ls}
                \pgfmathsetmacro{\yA}{ifthenelse(\la<=4, \ab-\la+1, \bb-\la+5)}
                \pgfmathsetmacro{\yB}{ifthenelse(\lb<=4, \ab-\lb+1, \bb-\lb+5)}
                \pgfmathsetmacro{\yC}{ifthenelse(\lc<=4, \ab-\lc+1, \bb-\lc+5)}
                \path (\xpos,\yA) coordinate (A)
                      (\xpos-\dx,\yB) coordinate (B)
                      (\xpos+\dx,\yC) coordinate (C);
                \draw[cyc] (A)--(B);
                \draw[cyc] (B)--(C);
                \draw[cyc] (C)--(A);
                \node[dot] at (A) {};
                \node[dot] at (B) {};
                \node[dot] at (C) {};
            }
        }

    \end{tikzpicture}
    \caption{Full combined network of the sortnet96 algorithm, showing all 96 3-cycles of the universal sequence on 24 wires
    ($6\text{ faces} \times 4\text{ stickers each}$). The 6 layers of the 12-comparator
    sorting network (Figure~\ref{fig:optimal-sorting-network}) are separated by dashed lines.
    In each layer, every comparator pair is expanded into the 8-3-cycle permutation network of Figure~\ref{fig:new-permutation-network}: local wires $0$--$3$ map to the higher-ranked
    face's stickers and local wires $4$--$7$ map to the lower-ranked face's stickers.}
    \label{fig:full-combined-network}
\end{figure}

\begin{figure}[p]
  \centering
  \begin{tikzpicture}[x=0.25cm,y=0.265cm,line cap=round,line join=round]
    \tikzset{
      wire/.style={black, line width=0.45pt},
      cyc/.style={black, line width=0.55pt,
          postaction={decorate, decoration={markings,
              mark=at position 0.5 with {\arrow[scale=0.6]{>}}}}},
      dot/.style={circle, fill=black, inner sep=0.8pt},
      phasesep/.style={gray!60, dashed, line width=0.4pt}
    }

    \foreach \i in {1,...,12} {
      \draw[wire] (0,\i) -- (55,\i);
    }
    \foreach \i in {13,...,24} {
      \pgfmathsetmacro{\wy}{\i+3}
      \draw[wire] (0,\wy) -- (55,\wy);
    }

    \foreach \i in {1,...,12} {
      \pgfmathtruncatemacro{\lbl}{\i-1}
      \node[left, font=\tiny] at (-0.2,\i) {\lbl};
    }
    \foreach \i in {13,...,24} {
      \pgfmathsetmacro{\wy}{\i+3}
      \pgfmathtruncatemacro{\lbl}{\i-1}
      \node[left, font=\tiny] at (-0.2,\wy) {\lbl};
    }

    \draw[phasesep] (33.5, 0.3) -- (33.5, 27.7);
    \node[font=\scriptsize, align=center] at (16.5, 28.5)
        {24-to-12 permutation network \;(32 3-cycles)};
    \node[font=\scriptsize, align=center] at (44, 28.5)
        {12-to-4 networks \;(20 3-cycles each)};

    \foreach \k/\la/\lb/\lc/\ls in {
        1/11/18/14/ 1,
        2/ 2/13/18/-1,
        3/ 5/14/22/ 1,
        4/ 7/16/22/-1,
        5/12/23/13/ 1,
        6/11/15/19/-1,
        7/ 7/23/15/ 1,
        8/12/20/19/-1,
        9/10/23/19/ 1,
       10/10/13/16/-1,
       11/ 3/15/16/ 1,
       12/ 9/14/20/-1,
       13/ 9/23/18/ 1,
       14/ 7/20/13/-1,
       15/ 9/22/21/ 1,
       16/12/14/17/-1,
       17/12/13/18/ 1,
       18/12/15/20/-1,
       19/11/23/15/ 1,
       20/ 3/18/22/-1,
       21/ 5/18/15/ 1,
       22/ 2/15/14/-1,
       23/ 7/13/21/ 1,
       24/11/16/17/-1,
       25/ 8/18/14/ 1,
       26/ 2/18/19/-1,
       27/12/18/24/ 1,
       28/ 9/18/16/-1,
       29/10/15/16/ 1,
       30/ 1/22/13/-1,
       31/ 6/13/16/ 1,
       32/ 4/24/16/-1
     } {
      \pgfmathsetmacro{\xpos}{\k + 0.3}
      \pgfmathsetmacro{\dx}{0.36*\ls}
      \pgfmathsetmacro{\wyA}{ifthenelse(\la<=12, \la, \la+3)}
      \pgfmathsetmacro{\wyB}{ifthenelse(\lb<=12, \lb, \lb+3)}
      \pgfmathsetmacro{\wyC}{ifthenelse(\lc<=12, \lc, \lc+3)}
      \path (\xpos,\wyA) coordinate (A)
            (\xpos-\dx,\wyB) coordinate (B)
            (\xpos+\dx,\wyC) coordinate (C);
      \draw[cyc] (A)--(B);
      \draw[cyc] (B)--(C);
      \draw[cyc] (C)--(A);
      \node[dot] at (A) {};
      \node[dot] at (B) {};
      \node[dot] at (C) {};
    }

    \foreach \yoff in {0, 15} {
      \foreach \k/\la/\lb/\lc/\ls in {
          1/ 1/ 2/12/ 1,    2/ 3/ 4/12/-1,    3/ 1/ 8/ 7/ 1,    4/ 2/ 7/ 3/-1,
          5/ 2/ 8/ 5/ 1,    6/ 7/12/11/-1,    7/ 3/ 4/ 7/ 1,    8/ 2/ 3/ 6/-1,
          9/ 1/ 9/10/ 1,   10/ 5/11/12/-1,   11/ 7/ 8/10/ 1,   12/ 6/10/ 9/-1,
         13/ 1/ 2/ 9/ 1,   14/ 3/10/11/-1,   15/ 5/ 9/10/ 1,   16/ 3/ 4/ 9/-1,
         17/ 1/11/12/ 1,   18/ 3/ 4/ 7/-1,   19/ 2/ 8/ 3/ 1,   20/ 1/ 5/ 6/-1
      } {
        \pgfmathsetmacro{\xpos}{33.8 + \k}
        \pgfmathsetmacro{\dx}{0.36*\ls}
        \pgfmathsetmacro{\wyA}{\la + \yoff}
        \pgfmathsetmacro{\wyB}{\lb + \yoff}
        \pgfmathsetmacro{\wyC}{\lc + \yoff}
        \path (\xpos,\wyA) coordinate (A)
              (\xpos-\dx,\wyB) coordinate (B)
              (\xpos+\dx,\wyC) coordinate (C);
        \draw[cyc] (A)--(B);
        \draw[cyc] (B)--(C);
        \draw[cyc] (C)--(A);
        \node[dot] at (A) {};
        \node[dot] at (B) {};
        \node[dot] at (C) {};
      }
    }

  \end{tikzpicture}
  \caption{Full combined permutation network of the permnet72 algorithm on 24 wires.
  Left of the dashed line: the 32 3-cycles of the 24-to-12 network.
  Right: the 20 3-cycles of the 12-to-4 network, applied independently to the bottom group (wires~0--11) and the top group (wires~12--23).
  Each directed triangle represents one 3-cycle.}
  \label{fig:full-perm-network}
\end{figure}
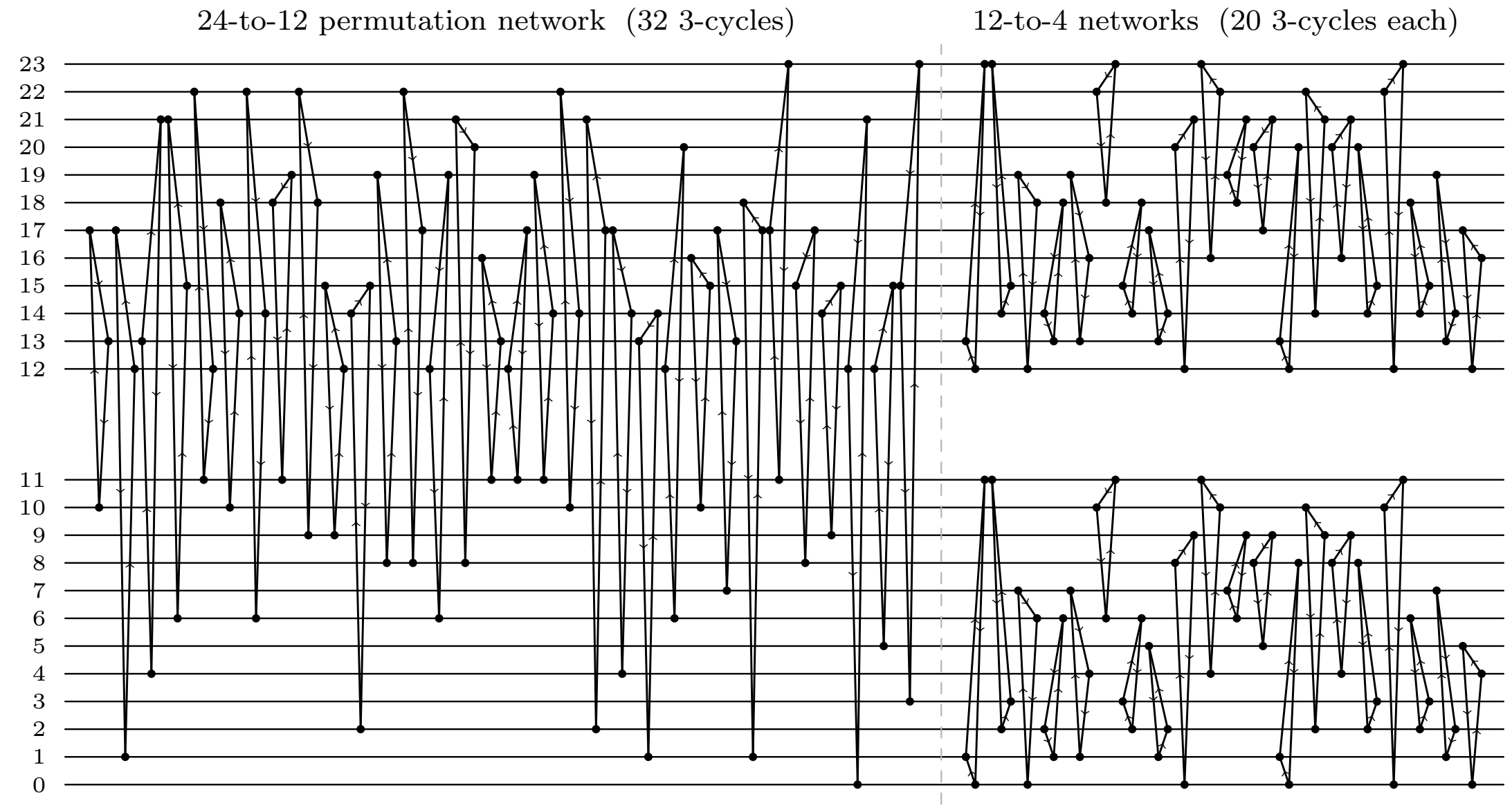

For completeness, we give the full expansions of the two universal sequences discussed in Section~\ref{algorithmicanalysis}: the 96-3-cycle sorting/permutation network (Figure~\ref{fig:full-combined-network}) and the 72-3-cycle permnet72 network (Figure~\ref{fig:full-perm-network}).

\end{document}